\documentclass[english,twocolumn]{aastex62}
\usepackage{amssymb}
\usepackage{graphicx}

\makeatletter

\providecommand{\tabularnewline}{\\}

\usepackage{amsmath}
\usepackage{colortbl}

\definecolor{lgray}{gray}{0.8}

\makeatother

\usepackage{babel}
\begin{document}

\title{Cosmic Reionization May Still Have Started Early and Ended Late:
Confronting Early Onset with CMB Anisotropy and 21 cm Global Signals}

\author{Kyungjin Ahn}

\affiliation{Department of Earth Sciences, Chosun University, Gwangju 61452, Korea}

\author{Paul R. Shapiro}

\affiliation{Department of Astronomy, University of Texas at Austin, Austin, TX
78712, USA}
\begin{abstract}
The global history of reionization was shaped by the relative amounts
of starlight released by three halo mass groups: atomic-cooling halos
(ACHs) with virial temperatures $T_{{\rm vir}}>10^{4}\,{\rm K}$,
either (1) massive enough to form stars even after reionization (HMACHs,
$\gtrsim10^{9}\,M_{\odot}$) or (2) less-massive (LMACHs), subject
to star formation suppression when overtaken by reionization, and
(3) ${\rm H_{2}}$-cooling minihalos (MHs) with $T_{{\rm vir}}<10^{4}\,{\rm K}$,
whose star formation is predominantly suppressed by the ${\rm H_{2}}$-dissociating
Lyman-Werner (LW) background. Our previous work showed that including
MHs caused two-stage reionization -- early rise to $x\lesssim0.1$,
driven by MHs, followed by a rapid rise, late, to $x\sim1$, driven
by ACHs -- with a signature in CMB polarization anisotropy predicted
to be detectable by the Planck satellite. Motivated by this prediction,
we model global reionization semi-analytically for comparison with
Planck CMB data and the EDGES global 21cm absorption feature, for
models with: (1) ACHs, no feedback; (2) ACHs, self-regulated; and
(3) ACHs and MHs, self-regulated. Model (3) agrees well with Planck
E-mode polarization data, even with a substantial tail of high-redshift
ionization, beyond the limit proposed by the Planck Collaboration
(2018). No model reproduces the EDGES feature. For model (3), $\left|\delta T_{b}\right|\lesssim60\,{\rm mK}$
across the EDGES trough, an order of magnitude too shallow, and absorption
starts at higher $z$ but is spectrally featureless. Early onset reionization
by Population III stars in MHs is compatible with current constraints,
but only if the EDGES interpretation is discounted or else other processes
we did not include account for it.
\end{abstract}


\section{Introduction\label{sec:Introduction}}

Cosmic reionization is commonly believed to have commenced with the
birth of first stars in the Universe and ended when all the intergalactic
medium (IGM) became ionized due to the net production of ionizing-photons
surpassing the number of neutral hydrogen atoms. This whole epoch
marks the epoch of reionization (EoR) in the history of the Universe.
There exist several observational constraints, including (1) the quasar
Gunn-Peterson trough (\citealt{2001AJ....122.2850B,Fan2002}): reionization
is likely to have ended at $z\gtrsim6$, (2) the temporal evolution
of the 21 cm background at redshift $z\sim6$ (\citealt{Bowman2010}):
a sudden reionization scenario is ruled out, (3) the polarization
anisotropy of the cosmic microwave background (CMB)(\citealt{PlanckCollaboration2018}):
the optical depth to the CMB photon, $\tau_{{\rm es}}$, is $0.054_{-0.0081}^{+0.0070}$($1\sigma$),
(4) the sudden change of the Ly$\alpha$-emitter (LAE) population
(\citealt{Pentericci2011}): the global ionized fraction, $x$, at
$z\simeq7$ is at least $\gtrsim0.1$, (5) the quasar proximity effect
(\citealt{Calverley2011}): the metagalactic ionization rate $\Gamma$ at $z\simeq6$
is about $10^{-13}\,{\rm s^{-1}}$. These constrains are nevertheless
insufficient to constrain the full history of reionization.

Constraining the cosmic reionization process by observations requires
theoretical modelling, of course. The simplest approach is to calculate
the history of reionization using semi-analytical, one-zone models
(see e.g. \citealt{1996ApJ...461...20H,Haiman1997a,2003ApJ...595....1H,2006MNRAS.371..867F}).
The more difficult approach is to simulate the process numerically
(see e.g. \citealt{Iliev2007,Trac2007,2007ApJ...657...15K,Mesinger2011}).
These methods are mutually complementary. For example, the former
allows a very fast exploration of the parameter space, while the latter
enables simulating the 3-dimensional (3D) structure of the process.
If one is interested only in the averaged quantities, such as the
global ionized fraction $x$ and the kinetic temperature of the intergalactic
medium (IGM) $T_{k}$, the one-zone calculation can provide surprisingly
reliable estimates very easily. Therefore the one-zone calculation
has been used quite extensively. This method has become very useful
in reionization-parameter estimation effort, while one has to be wary
about the fact that such an estimation is limited to specific reionization
models that are assumed to be fully described by those parameters.

Some of the semi-analytical one-zone modellings used to have common
``conventions'': (1) only atomic-cooling halos, with virial temperature
$T_{{\rm vir}}\gtrsim10^{4}\,{\rm K}$, are considered as radiation
sources, (2) the star formation rate (SFR) is proportional to the
growth rate $df_{{\rm coll}}/dt$ of the halo-collapsed fraction $f_{{\rm coll}}$,
and (3) feedback effects are neglected (\citealt{2001ARA&A..39...19L,2006MNRAS.371..867F,Pritchard2006}).
We categorize those models that follow these conventions as the vanilla
model. First, main justification for the first convention is from
the fact that stars inside less-massive halos, or minihalos (with
$T_{{\rm vir}}\lesssim10^{4}\,{\rm K}$ and mass limited down to the
Jeans mass of the intergalactic medium), are susceptible to the Lyman-Werner
feedback effect and the Jeans-mass filtering. The Lyman-Werner feedback
strongly regulates the amount of ${\rm H}_{2}$ inside minihalos,
the main cooling agent in the primordial environment, by photo-dissociation.
The Jeans-mass filtering can happen if minihalos are exposed to the
hydrogen-ionizing radiation field, either from inside or from outside.
Therefore, their contribution was believed to be negligible. Second,
justification for SFR $\propto df_{{\rm coll}}/dt$ is not solid,
and this assumption is equivalent of null duty cycle of star formation
(Section \ref{subsec:Gamma_F_dF}). Third, neglecting the feedback
effect is a conventional simplification, while there may exist atomic-cooling
halos in the low-mass end that are affected by the Jeans-mass filtering
if embedded in photoionized regions (Section \ref{subsec:SRI}). 

Many numerical simulations inherited the convention of neglecting
minihalos, usually in those using a large ($\gtrsim$ a few tens of
comoving Mpc) simulation box. While the justification is the strong
LW feedback effect, a practical reason is the numerical resolution
limit that becomes worse as the size of the simulation box becomes
larger. Therefore, large-scale simulations of reionization usually
suffers from this limit. To overcome this limit and include the impact
of minihalo stars, a sub-grid treatment of missing halos (\citealt{Ahn2015,Nasirudin2020})
was included in a large simulation box by \citet{Ahn2012}. They used
an empirical, deterministic bias of minihalo population for any given
density environment, to successfully populate minihalos over the domain
of calculation. Using theoretical predictions on the formation of
first stars inside minihalos, together with the impact of the LW feedback,
they could then cover the fully dynamic range of halos. They found
that (1) minihalo stars cannot finish reionization but can ionize
the universe up to $\sim20\,\%$ depending on the mass of Population
III (Pop III) stars, (2) the resulting reionization history is composed
of the early staggered stage with slow growth of $\left\langle x\right\rangle $
and the late rapid stage with rapid growth of $\left\langle x\right\rangle $,
(3) the reionization is finished by photons from atomic-cooling halos,
and (4) there exist degeneracy in reionization histories that can
result in the same $\tau_{{\rm es}}$ and $z_{{\rm reion}}$.

There exists an interesting hint from the recent large-scale CMB polarization
observation that is related to the result by \citet{Ahn2012}. Constraining
the full history of reionization has become available only recently,
through observation of the large-scale CMB polarization anisotropy.
Having such constraints used to be impossible with the given quality
of CMB polarization data before, and thus only a two-parameter constraint
on reionization with $\tau_{{\rm es}}$ and the reionization redshift
$z_{{\rm reion}}$ had been available (e.g. \citealt{Hinshaw2013,PlanckCollaboration2013}).
The Planck 2015 data (\citealt{Planck2015}) first allowed constraining
the history of reionization beyond this two-parameter constraint,
and it was claimed that the Planck data favored a type of reionization
histories composed of (1) the early, slow growth of $x$ for a large
range of redshift $z\sim30-10$ and (2) the late, rapid growth of
$x$ for a short range of redshift $z\sim10-6$ (\citealt{Miranda2017};
\citealt{Heinrich2018}). While criticisms on this claim appeared due
to the limitation of having only the Planck Low Frequency Instrument
(LFI) data (\citealt{Millea2018}), this ``two-stage'' reionization
histories had been indeed predicted by \citet{Ahn2012}, in which
the early stage is dominated by strongly self-regulated formation
of stars inside minihalos, presumably Pop III stars, and the late
stage by stars inside atomically-cooling halos, presumably Population
II (Pop II) stars, with weaker (inside low-mass atomic cooling halos,
LMACHs) or even no modulation (inside high-mass atomic cooling halos,
HMACHs) on star formation. The most refined, non-parametric constraint
comes from the Planck 2018 observation that also includes the High
Frequency Instrument (HFI) data (\citealt{PlanckCollaboration2018,Millea2018}),
and the constraint bears qualitative similarity with but some quantitative
difference from the analyses by \citet{Miranda2017} and \citet{Heinrich2018}.
First, the two-stage reionization is still favored at $\lesssim2\sigma$
level. Second, a ``significant'' amount of ionization at $z\gtrsim15$
is disfavored, with maximum of $\sim10\%$ ionized fraction allowed
at $z\sim15$ at 1$\sigma$ level. Even though \citet{PlanckCollaboration2018}
and \citet{Millea2018} stress the latter finding and even claim that
the early stage of reionization dominated by self-regulated Pop III
stars is strongly disfavored, their constraint indeed allows very
extended ionization histories reaching to $z\sim25$ and $z\sim30$
for the 1$\sigma$ and the $2\sigma$ constraint, respectively, in
addition to showing a two-stage reionization feature. Therefore, we
take the analysis on the Planck 2018 data (\citealt{PlanckCollaboration2018,Millea2018})
as a mild proof for the two-stage reionization, which will be  tested
quantitatively in this paper.

Meanwhile, observing the hydrogen 21 cm line background is believed
to provide a direct probe of the the Dark Ages and the epoch of reionization.
The deepest ($z=14-26$) observation so far, in terms of the sky-averaged
global 21 cm background, has been delivered by the Experiment to Detect
the Global Epoch of Reionization (EDGES), and the detection of $\sim500\,{\rm mK}$
continuum absorption of the hydrogen 21-cm signal against the smooth
background around $z\sim17$ was claimed (\citealt{Bowman2018}).
It is hard to grasp such a large absorption in the standard picture,
compared to the maximum $\sim200\,{\rm mK}$ depth allowed in the
$\Lambda$CDM universe. The overall shape of the signal over the redshift
range is also incompatible with usual model predictions. This ``conflict''
stimulated many resolutions, which fall into roughly three categories:
(1) the analysis of removing a smooth galactic foreground and obtaining
another smooth absorption signal is unreliable (\citealt{Hills2018}),
(2) non-standard models beyond $\Lambda$CDM should be considered,
such as large baryon-dark matter interaction (\citealt{Tashiro2014};
\citealt{Barkana2018}) and abnormally large expansion rate at high
redshift (\citealt{Hill2018}), and (3) the contribution of yet unknown
sources to the radio continuum (the ``excess radio background'')
but still in the standard $\Lambda$CDM framework, may generate such
a large absorption signal (e.g. \citealt{Feng2018}; \citealt{EwallWice2018}).

Motivated by the hint on the two-stage reionization from the Planck
2018 data and the compelling EDGES observation, we explore the possibility
for the two-stage reionization to leave any characteristic signature
in the CMB and the global 21 cm background. Toward this end, we use
semi-analytical one-zone models to cover a wide range of reionization
scenarios. We cast reionization models into three categories, from
the vanilla model without feedback effect to progressively sophisticated
ones, expanding the star-hosting halo species and considering feedback
effects that regulate star formation in those halos. We then investigate
how the history of reionization and the evolution of other relevant
radiation fields shape the CMB the 21cm background, with special focus
on finding whether the two-phase reionization models that carry the
high-redshift ionization tail, constructed by the LW-regulated star
formation in MHs, will leave any imprint on the CMB and the 21 cm
background. While it is worthwhile to include the resolutions, alternative
to the standard picture, for the deep absorption signal of the EDGES
in the parameter estimation effort (\citealt{Mirocha2018,Mirocha2019,Mebane2020,Qin2020_tt_I,Qin2021_tt_II}),
we limit our study to the standard $\Lambda$CDM framework without
the excess radio background and other alternatives but instead explore
any new possible observational predictions. We indeed find out very
unique and novel features of the two-phase ionization models both
in the CMB and the global $\delta T_{{\rm b}}$, opening up exciting
observational prospects.

The paper is organized as follows. In Section \ref{sec:Method}, we
describe the details of the one-zone model, including the model categories
and the numerical method. In Section \ref{sec:Result}, we present
the estimates on the ionized fraction,  the CMB polarization
  anisotropy and the 21 cm background for 
various reionization models, with special focus on the estimate from
the strongly self-regulated models. 
We summarize and discuss the result in Section \ref{sec:Discussion}.
Throughout this paper, we use cMpc to denote ``comoving megaparsec'',
or the comoving length in units of Mpc.

\section{Method\label{sec:Method}}

We take a simple one-zone model for studying the global evolution
of the physical states of IGM and backgrounds including the 21 cm
signal. Two crucial parameters are $x$ and $T_{k}$, whose evolutions
will be governed by physical properties, the evolution and the feedback
of radiation sources. Therefore, the main equations for reionization
models are the rate equations for $x$ and $T_{k}$.

The ionizing-photon production rate per baryon (PPR) is the source
term that increases $x$, or the source term in the rate equation
$dx/dt$ where $t$ is the cosmic time. PPR is the main parameter
whose variance leads to the variance in the histories of reionization
among different models. PPR during EoR is quite uncertain and will
be parameterized by the species of host halos, spectral shape and the
star formation duty cycle. PPR is also affected by various feedback
effects. These will be the main subjects to be described in Sections
\ref{subsec:Gamma_F_dF} - \ref{subsec:Model-variance}.

The recombination rate is the sink term in the ionization equation.
The recombination rate does not have the model variance as large as
PPR. Nevertheless, there exists some uncertainty in the recombination
rate mainly due to the difficulty in quantifying the clumping factor.
We conservatively take a combination of previous studies that quantified
the clumping factor.

A few radiation background fields are crucial in determining $T_{k}$
and the 21 cm spin temperature $T_{s}$. Variance in $T_{k}$ is caused
by variance in the heating efficiency, and variance in $T_{s}$ by
variance in Lyman-resonance-line backgrounds. Therefore, the model
variance will be reflected also in the 21 cm background, which will
be described in Sections \ref{subsec:Background} - \ref{subsec:Back_21cm}.
We defer the description of the used spectral energy distributions
(SED) of the stellar radiation, which cause the variance in $T_{s}$
even at a similar level of the ionization state, to Sections
\ref{subsec:Result-reion-history} and \ref{subsec:Result-21cm}.

Throughout this paper, we use cosmological parameters reported by
\citet{PlanckCollaboration2018}: $h=0.6732$, $\Omega_{m,\,0}=0.3144$,
$\Omega_{b,\,0}=0.04939$, $n_{s}=0.966$, and $\sigma_{8}=0.812$,
which are the present Hubble parameter in the unit of $100\,{\rm km\,s^{-1}\,Mpc}$,
the present matter density in units of the critical density, the present
baryon density in units of the critical density, the power index of
the primordial curvature perturbation, and the present variance of
matter density at the filtering scale of $8h^{-1}\,{\rm Mpc}$. We
use the mass function by \citet{Sheth1999} when calculating the halo-collapsed
fraction.

\subsection{Star formation rate density and duty cycle\label{subsec:Gamma_F_dF}}

It is a usual approximation that PPR is proportional to the star formation
rate density (SFRD: star formation rate per comoving volume, in units
of $M_{\odot}\,{\rm s^{-1}\,cMpc^{-3}}$). Rigorously, this becomes
true only if $t_{*}$, the lifetime of stars which we take as a constant
for a given stellar species (e.g. Pop II stars), is infinitesimal
compared to the ionization time, i.e. $x/(dx/dt)$. Even more rigorously,
for PPR at time $t$ to be the ionization rate at $t$, the photon
travel time inside H II regions should also be very small. In this
paper, we take this usual assumption, PPR $\propto$SFRD, that has
been used extensively in semi-analytical modelling. The ionization
rate $\xi$ (in the unit of ${\rm s}^{-1}$) will then be given by
\begin{equation}
\xi=f_{{\rm esc}}{\rm PPR}=f_{{\rm esc}}N_{{\rm ion}}\frac{{\rm SFRD}}{m_{b}n_{b,\,0}},\label{eq:Gamma_PDR_SFRD}
\end{equation}
where $f_{{\rm esc}}$, $N_{{\rm ion}}$, $m_{b}$, $n_{b,\,0}$ are
the ionizing-photon escape fraction out of halos, the total number
of ionizing photons emitted per stellar baryon during the lifetime
of a star, the baryon mass, and the average comoving baryon number
density, respectively.

We also modify the typical assumption of semi-analytical calculations
to incorporate non-zero duty cycle of star formation. Let us first
briefly review the typical assumption in semi-analytical calculations:
PPR and SFRD are assumed proportional to $df_{{\rm coll}}/dt$, the
growth rate of the halo-collapsed fraction $f_{{\rm coll}}$, such
that
\begin{align}
\xi(t)&=\xi_{{\rm a}}(t)=f_{*}f_{{\rm esc}}\frac{N_{{\rm
      ion}}}{t_{*}}\int_{t-t_{*}}^{t}dt'\,\frac{df_{{\rm coll}}}{dt'}\nonumber\\
&\simeq f_{*}f_{{\rm esc}}N_{{\rm ion}}\frac{df_{{\rm coll}}}{dt}(t)=f_{\gamma}\frac{df_{{\rm coll}}}{dt}(t),
\label{eq:Gamma_typical}
\end{align}
where $\xi_{{\rm a}}$ denotes the ionization rate due to newly ``accreted''
matter, $f_{*}$ is the star formation efficiency in such an episode
(e.g. \citealt{2001ARA&A..39...19L,2006MNRAS.371..867F}), and $f_{\gamma}\equiv f_{*}f_{{\rm esc}}N_{{\rm ion}}$
is the proportionality coefficient between $\xi$ and $df_{{\rm coll}}/dt$.
The second equality in equation (\ref{eq:Gamma_typical}) is a rigorous
form that considers only the newly accreted matter, whose maximal
lookback time in the integration should be $t_{*}$. Therefore, we
can denote this assumption as the ``mass-accretion dominated star
formation'' scenario (MADSF). In the limit of infinitesimal $t_{*}$
or slowly-varying $f_{{\rm coll}}$, the last approximation becomes
valid. Because $df_{{\rm coll}}$ in equation (\ref{eq:Gamma_typical})
is the amount of matter that has been newly accreted into halos per
time interval $dt$, considering $\xi_{{\rm a}}(t)$ only is identical
to having null duty cycle such that any gas that has once formed stars
can never form stars again. Note also that $df_{{\rm coll}}$ does
not include halos whose mass has increased by a merger event at $t$,
even though in nature ``wet merger'' events occur quite frequently.
Because this null duty cycle assumption is somewhat extreme, we need
to consider a more general case of non-zero duty cycle.

We quantity the non-zero duty cycle as the time fraction that stars
emit radiation, $f_{{\rm DC}}\equiv t_{*}/(t_{*}+t_{{\rm dorm}})$,
where $t_{{\rm dorm}}$ is the duration that a stellar baryon remains
dormant after the death of the host star until a new star formation
episode occurs. The same account on $t_{{\rm dorm}}$ has been addressed
by \citet{Mirocha2018}. Both $t_{*}$ and $t_{{\rm dorm}}$ are average
quantities, and consequently $f_{{\rm DC}}$ will be a global parameter.
Ergodicity is assumed such that this temporal duty cycle is the same
as the spatially averaged fraction of gas that has once resided inside
a star and now resides in a post-generation star. In this work, we
further assume that $f_{{\rm DC}}$ is constant over time for a given
stellar species. With non-zero $f_{{\rm DC}}$, we obtain the ionization
rate $\xi_{{\rm d}}$ due to ``duty cycle'':
\begin{align}
\xi_{{\rm d}}(t)&=f_{*}f_{{\rm esc}}f_{{\rm DC}}f_{{\rm coll}}(t-t_{*})\frac{N_{{\rm ion}}}{t_{*}}\nonumber\\&\simeq f_{*}f_{{\rm esc}}f_{{\rm DC}}f_{{\rm coll}}(t)\frac{N_{{\rm ion}}}{t_{*}}=f_{*,{\rm d}}f_{{\rm esc}}f_{{\rm coll}}(t)\frac{N_{{\rm ion}}}{t_{*}},
\label{eq:Gamma_withDC}
\end{align}
where only ``old'' gas at lookback time $t_{*}$ is allowed to re-generate
stars and the approximation holds when $t_{*}$ is treated infinitesimal.
The last equality absorbs $f_{{\rm DC}}$ into the effective $f_{*}$
of \emph{halo gas}, $f_{*,{\rm d}}$. Equation (\ref{eq:Gamma_withDC})
obviously does not account for any newly accreted gas. Note also that
equation (\ref{eq:Gamma_withDC}) holds due to ergodicity: $f_{*,{\rm d}}$
is a spatially averaged quantity over many galaxies, which will be
identical to the time average of star formation episodes on a single
galaxy. We denote a scenario based on $\xi=\xi_{{\rm d}}$ as the
``all-halo replenished star formation'' scenario (AHRSF).

The most generic form for $\xi$ should implement both contributions
from old gas and newly accreted gas, or combining the episodes of
MADSF and AHRSF, such that
\begin{equation}
\xi=\xi_{{\rm d}}+\xi_{{\rm a}}=f_{*,{\rm d}}f_{{\rm esc}}f_{{\rm coll}}\frac{N_{{\rm ion}}}{t_{*}}+f_{*,{\rm a}}f_{{\rm esc}}N_{{\rm ion}}\frac{df_{{\rm coll}}}{dt},\label{eq:Gamma_generic}
\end{equation}
where $f_{*}$ of \emph{newly accreted gas} is denoted by $f_{*,{\rm a}}$,
to be distinguished from $f_{*,{\rm d}}$. $f_{*,{\rm a}}=0$ corresponds
to the case where newly accreted gas waits longer than $t_{*}$ to
start star formation activity, and practically there is no theoretical
constraint on the relative strengths of $f_{*,{\rm a}}$ over $f_{*,{\rm d}}$.
With equation (\ref{eq:Gamma_generic}), both cases of $\xi\propto f_{{\rm coll}}/t_{*}$
and $\xi\propto df_{{\rm coll}}/dt$ can be accommodated in terms
of special cases of this generic form with \{$f_{*,{\rm d}}\ne0$,
$f_{*,{\rm a}}=0$\} and \{$f_{*,{\rm d}}=0$, $f_{*,{\rm a}}\ne0$\},
respectively. Usually, the relation $\xi\propto f_{{\rm coll}}/t_{*}$
has been used in numerical radiation transfer simulations (e.g. \citealt{Iliev2007})
and the relation $\xi\propto df_{{\rm coll}}/dt$ in semi-analytical
calculations (e.g. \citealt{2001ARA&A..39...19L,2006MNRAS.371..867F}).
In this paper, we only consider these two special cases, denoting
the former by ``F'' and the latter by ``dF'' as the nomenclature
for star formation scenarios. Because we use $f_{*,{\rm a}}$ and
$f_{*,{\rm d}}$ mutually exclusively for these special cases, we
drop the subscripts ``a'' (accretion) and ``d'' (duty cycle) from
$f_{*}$ for simplicity.

As long as cosmology is fixed, the evolution of the volume ionized
fraction will be uniquely determined for given $f_{\gamma}$ in MADSF
and for given $g_{\gamma}\equiv f_{*}f_{{\rm esc}}N_{{\rm ion}}/(t_{*}/10\,{\rm Myr})$
in AHRSF, because $df_{{\rm coll}}/dt$ and $f_{{\rm coll}}$ are
determined solely by cosmological parameters. Therefore, reionization
models with MADSF and AHRSF will be parameterized by
$f_{\gamma}$ (e.g. \citealt{2006MNRAS.371..867F}) and
$g_{\gamma}$ (e.g. \citealt{Ahn2012}), respectively.
Of course, $f_{\gamma}$ and $g_{\gamma}$ need not be constant in
  time. $f_{{\rm esc}}$, $f_{*}$, $t_{*}$ and $N_{\rm ion}$ can be
  time-varying individually
  and collectively (in terms of $f_{\gamma}$ and
  $g_{\gamma}$). Allowing $f_{\gamma}$ and $g_{\gamma}$ to change in
  time can yield even more variants of reionization scenarios,
  especially by reshaping $dx/dz$. For simplicity, we do not explore
  this possibility. Nevertheless,
  future CMB observations will become more accurate and
  models inferred from the data could not be matched well by the
  constant values of $f_{\gamma}$ and $g_{\gamma}$. In such a case,
  time-varying $f_{\gamma}$ and $g_{\gamma}$ may have to be
  considered in our models.

\subsection{Model variance\label{subsec:Model-variance}}

We consider three types of models: (1) the vanilla model, (2) the
self-regulated model type I (SRI), and (3) the self-regulated model
type II (SRII). From type (1) to type (3), these models become progressively
sophisticated in physical processes considered: the vanilla model
does not consider any feedback, SRI considers the photo-heating feedback,
and SRII considers both photo-heating and LW feedback effects. In
each model, we allow both cases of $\xi\propto df_{{\rm coll}}/dt$
and $\xi\propto f_{{\rm coll}}/t_{*}$ that were described in Section
\ref{subsec:Gamma_F_dF}. In all these models the source term is
the ionization rate
and the sink term is the recombination rate, such that the change
rate of the global (volume) ionization fraction $x$ is
\begin{equation}
\frac{dx}{dt}=\xi-\alpha Cn_{{\rm e}}x,\label{eq:dxvol_dt}
\end{equation}
where $\alpha$ is the hydrogen recombination coefficient (in units
of ${\rm cm}^{3}\,{\rm s}^{-1}$), $C$ is the average clumping factor,
and $n_{{\rm e}}=n_{{\rm H}}+n_{{\rm He}}$ is the (proper) number
density of electrons inside H II regions (e.g. \citealt{2006MNRAS.371..867F,Iliev2007}).
In this work, we use the case B recombination coefficient for $\alpha$,
or $\alpha=\alpha_{{\rm B}}(T=10^{4}\,{\rm K})=2\times10^{-13}\,{\rm cm^{3}}{\rm s}^{-1}$,
and adopt a fitting formula for $C$ given by
\begin{equation}
C={\rm max(3,\,17.6\exp[-0.1z+0.0011z^{2}])},\label{eq:clumping}
\end{equation}
which is a conservative combination of work by \citet{2005ApJ...624..491I},
\citet{Pawlik2009} and \citet{So2014}. While it is possible that
numerical resolution limit of previous numerical simulations may have
led to underestimation of $C$ (\citealt{Mao2020}), we do not consider
this possibility in this paper. The model variance is mainly caused
by the variance in $\xi$ in equation (\ref{eq:dxvol_dt}), which
will be described in the following subsections.

Physical parameters governing $\xi$, such as $f_{{\rm esc}}$, $f_{*}$,
$N_{{\rm ion}}$, etc., would depend on halo properties. One of the
crucial halo properties is the virial temperature $T_{{\rm vir}}$.
The natural borderline between MHs and ACHs is the temperature endpoint
of the atomic line cooling (if dominated by hydrogen Ly$\alpha$ line
cooling), $\sim10^{4}\,{\rm K}$. ACHs have $T_{{\rm vir}}>10^{4}\,{\rm K}$,
and MHs have $T_{{\rm vir}}<10^{4}\,{\rm K}$. This classification
can be cast into the mass criterion
\begin{equation}
M\gtrless10^{8}\,h^{-1}M_{\odot}\,\left[1.98\,\left\{ \Omega_{m}+\Omega_{\Lambda}(1+z)^{-3}\right\} \,\left(\frac{1+z}{10}\right)\right]^{-\frac{3}{2}},\label{eq:M_10000}
\end{equation}
where we assume the mean molecular weight $\mu=0.6$ for fully ionized
gas (\citealt{2001PhR...349..125B}). One important subtlety is that
this redshift-dependent mass criterion is not commonly respected in
numerical RT simulations, because the minimum mass of halos resolvable
in accompanying N-body or uniform-grid simulations tends to be constant
in time\footnote{If the particle-splitting scheme or the adaptive mesh refinement scheme
is used, one may in principle recover the time-varying mass criterion
for MH and ACH determination. }. Therefore, many RT simulations take a constant-mass classification
scheme, such that ACHs and MHs correspond to halos with $M>M_{{\rm th}}$
and $M<M_{{\rm th}}$, respectively, with a constant mass threshold
$M_{{\rm th}}$. When a simulation domain is increased to a few $\sim10$
cMpc, the numerical resolution limit quickly reaches $\sim 10^{8}\,M_{\odot}$,
a common value of $M_{{\rm th}}$ for such classification. For example,
in many large-volume simulations MHs are ignored and all halos that
are numerically resolved, or e.g. those with $M>10^{8}\,M_{\odot}$,
are taken as ACHs. However, this scheme loses track of those halos
with $T_{{\rm vir}}<10^{4}\,{\rm K}$ and $M<10^{8}\,M_{\odot}$ at
$z\gtrsim18$. Therefore, one should be wary of the negligence when
the high-redshift ($z\gtrsim20$) astrophysics, for example the early
stage of cosmic reionization, is investigated with such a scheme.
It may ignore not only MHs but also a substantial amount of ACHs,
as long as we believe in the constant-temperature criterion as a natural
distinction.

In order to have a fair comparison of the semi-analytical models to
the previous numerical RT simulation results based on this constant-mass
criterion for ACH/MH classification, we adopt a constant
$M_{{\rm th}}$ ($=10^{8}\,M_{\odot}$) in this work. 
Comparison
will be made to numerical simulations that (1) further split ACHs
into low-mass and high-mass species with a constant-mass classification
(Section \ref{subsec:SRI}) and (2) cover the full dynamic range of
halos, namely MHs, low-mass ACHs and high-mass ACHs but again with
a constant-mass classification (Sec \ref{subsec:SRII}). 

\subsubsection{Vanilla model\label{subsec:Vanilla}}

This model assumes a very simplified form for PPR: PPR is typically
assumed to be proportional $df_{{\rm coll}}/dt$, and no feedback
effect on star formation is considered. The essential ingredient is
not the relation $\xi\propto df_{{\rm coll}}/dt$ but the lack of
feedback, and thus we also allow the relation $\xi\propto f_{{\rm coll}}/t_{*}$.
We therefore have
\begin{equation}
\xi=\xi_{{\rm a}}=\sum_{i}f_{*}^{(i)}f_{{\rm esc}}^{(i)}N_{{\rm ion}}^{(i)}\frac{df_{{\rm coll}}^{(i)}}{dt}=\sum_{i}f_{\gamma}^{(i)}\frac{df_{{\rm coll}}^{(i)}}{dt}\label{eq:Gamma_vanilla_dF}
\end{equation}
for $\xi\propto df_{{\rm coll}}/dt$ (``dF'') and
\begin{equation}
\xi=\xi_{{\rm d}}=\sum_{i}f_{*}^{(i)}f_{{\rm esc}}^{(i)}f_{{\rm coll}}^{(i)}\frac{N_{{\rm ion}}^{(i)}}{t_{*}^{(i)}}\label{eq:Gamma_vanilla_F}
\end{equation}
for $\xi\propto f_{{\rm coll}}/t_{*}$ (``F''). These equations
are generalized from equations (\ref{eq:Gamma_typical}) and (\ref{eq:Gamma_withDC})
with summation to accommodate different halo species $i$, with the
superscript ``($i$)'' denoting physical quantities of the halo
species $i$. The simplicity of equations (\ref{eq:Gamma_vanilla_dF})
and (\ref{eq:Gamma_vanilla_F}) is the essence of vanilla models that
ignore any feedback effects.

\subsubsection{Self-Regulated Model Type I\label{subsec:SRI}}

The work by \citet{Iliev2007} is among the first 3D RT simulations
of self-regulated reionization based on multi-species halo stars,
but with radiation sources restricted to ACHs. This simulation starts
with classifying ACHs into two mass categories, namely (1) the high-mass
atomic-cooling halos (HMACH) and (2) the low-mass atomic-cooling halos
(LMACH), with $M\gtrsim10^{9}\,M_{\odot}$ and $10^{8}\,M_{\odot}\lesssim M\lesssim10^{9}\,M_{\odot}$,
respectively. Again, even though it is more natural to classify halos
in terms of $T_{{\rm vir}}$, to make a direct comparison to \citet{Iliev2007}
we adopt this convention in this work. While the accurate boundary
does not exist, LMACHs defined this way roughly correspond to those
halos that are subject to the ``Jeans-mass filtering'': if these
halos are formed inside regions that have been already ionized, accretion
of baryonic gas will not be efficient enough to form stars inside
due to the high temperature ($T\gtrsim10^{4}\,{\rm K}$) of the regions
(\citealt{1992MNRAS.256P..43E}; \citealt{1994ApJ...427...25S}; \citealt{1996ApJ...465..608T};
\citealt{1997ApJ...478...13N}; \citealt{1998MNRAS.296...44G}; \citealt{2000ApJ...542..535G};
\citealt{2004ApJ...601..666D}). The suppression of star formation
is likely to be not as abrupt as assumed in \citet{Iliev2007} but
rather gradual in halo mass (\citealt{1992MNRAS.256P..43E}; \citealt{1997ApJ...478...13N};
\citealt{2004ApJ...601..666D}). Nevertheless, in this work we simply
adopt the self-regulation scheme of \citet{Iliev2007}.

We denote such a type as the type I self-regulated model (SRI). Because
HMACHs are unaffected by the feedback and LMACHs are suppressed inside
H II regions, the ionization rate will be given by 
\begin{equation}
\xi=f_{*}^{{\rm H}}f_{{\rm esc}}^{{\rm H}}N_{{\rm ion}}^{{\rm H}}\frac{df_{{\rm coll}}^{{\rm H}}}{dt}+f_{*}^{{\rm L}}f_{{\rm esc}}^{{\rm L}}N_{{\rm ion}}^{{\rm L}}\frac{df_{{\rm coll}}^{{\rm L}}}{dt}(1-x^{\eta})\label{eq:Gamma_SRI_dF}
\end{equation}
for $\xi\propto df_{{\rm coll}}/dt$ (``dF'') and
\begin{equation}
\xi=f_{*}^{{\rm H}}f_{{\rm esc}}^{{\rm H}}f_{{\rm coll}}^{{\rm H}}\frac{N_{{\rm ion}}^{{\rm H}}}{t_{*}^{{\rm H}}}+f_{*}^{{\rm L}}f_{{\rm esc}}^{{\rm L}}f_{{\rm coll}}^{{\rm L}}\frac{N_{{\rm ion}}^{{\rm L}}}{t_{*}^{{\rm L}}}(1-x^{\eta})\label{eq:Gamma_SRI_F}
\end{equation}
for $\xi\propto f_{{\rm coll}}/t_{*}$ (``F''), where superscripts
H and L denote HMACH and LMACH respectively, and $0<\eta<1$. The
reason why such an amplified suppression term $(1-x^{\eta})$ is used
instead of $(1-x)$ is that LMACHs are clustered more strongly inside
H II regions than in neutral regions. The value $\eta=0.1$ is the
empirical one found in 3D numerical simulations by
\citet{Iliev2007}, such that $x^{\eta}$ roughly estimates the mass
  fraction of LMACHs occupied by H II regions when the global
  ionization fraction is $x$. 
  The effect of such a small value of $\eta$ is to make star
  formation in LMACHs, and consequnetly the reionization history, more
  strongly regulated than the unrealistic case with $\eta=1$ where
  LMACHs are uniformly distributed in space. 
The assumption for the Jeans-mass filtering in equation (\ref{eq:Gamma_SRI_F})
is that even those halos that have collapsed earlier cannot host new
star-formation episodes. Even though one can generalize $\xi$
into a combined form of equations (\ref{eq:Gamma_SRI_dF}) and (\ref{eq:Gamma_SRI_F}),
as we mentioned in Section \ref{subsec:Gamma_F_dF} we restrict our
models to these two categories of ``dF'' and ``F''.

\subsubsection{Self-Regulated Model Type II\label{subsec:SRII}}

The work by \citet{Ahn2012} is a unique 3D RT simulation of reionization
in that (1) a full dynamic range of halos, from MHs to HMACHs, is
treated as radiation sources, (2) both the Jeans-mass filtering and
the LW feedback are considered, and (3) the simulation box is large
($\sim150\,{\rm Mpc}$) enough to provide a reliable statistical significance.
This simulation does not neglect MHs as most other large-box simulations
do. MHs are subject both to the Jeans-mass filtering and the LW feedback.
Jeans-mass filtering of MHs is obvious due to the smallness of the
the virial temperature ($T\lesssim10^{4}\,{\rm K}$). The LW feedback
occurs due to the fact that ${\rm H}_{2}$ is the main cooling agent
in the primordial environment, and MHs are usually formed first in
the primordial environment. Stars born in this environment will be
Pop III stars. Even inside MHs, after a few episodes of star formation
the chemical environment can gain metallicity beyond the critical value
$Z\simeq10^{-3}$. However, dynamical feedback from supernova explosion
inside MHs is believed to be very destructive (\citealt{Yoshida2007,Greif2007}),
such that it may take longer than e.g. the halo merger time for post-generation
star-formation episodes to occur in the same MH. It is possible that
the supernova feedback in the massive MHs, which remained neutral
even after photoionization from stars inside, could have been confined
inside the halo and led to the next episode of star formation (\citealt{Whalen2008}).
However, the dominant contribution to the number of MHs is from the
least massive ones, and so it is appropriate to assume the destructive
feedback. If one assumed the most destructive feedback effect of the
first episode of star formation inside MHs, then it would be equivalent
to assuming that only the newly forming MHs form stars. This assumption
was taken in \citet{Ahn2012}, which we implement in our modelling
here as well.

We denote such a type as the type II self-regulated model (SRII).
the ionization rate will be given by
\begin{align}
\xi&=f_{*}^{{\rm H}}f_{{\rm esc}}^{{\rm H}}N_{{\rm ion}}^{{\rm
    H}}\frac{df_{{\rm coll}}^{{\rm H}}}{dt}+f_{*}^{{\rm L}}f_{{\rm
    esc}}^{{\rm L}}N_{{\rm ion}}^{{\rm L}}\frac{df_{{\rm coll}}^{{\rm
      L}}}{dt}(1-x^{\eta})\nonumber\\
&+\frac{f_{{\rm esc}}^{{\rm M}}M_{{\rm III}}N_{{\rm ion}}^{{\rm M}}}{\mu m_{{\rm H}}n_{b,0}}\frac{dn^{{\rm M}}}{dt}\left[1-\min\left\{ {\rm max}\left(x_{{\rm LW}},x\right),\,1\right\} \right]
\label{eq:Gamma_mine_dF}
\end{align}
for $\xi\propto df_{{\rm coll}}/dt$ (``dF'') and
\begin{align}
\xi&=f_{*}^{{\rm H}}f_{{\rm esc}}^{{\rm H}}f_{{\rm coll}}^{{\rm
    H}}\frac{N_{{\rm ion}}^{{\rm H}}}{t_{*}^{{\rm H}}}+f_{*}^{{\rm
    L}}f_{{\rm esc}}^{{\rm L}}f_{{\rm coll}}^{{\rm L}}\frac{N_{{\rm
      ion}}^{{\rm L}}}{t_{*}^{{\rm L}}}(1-x^{\eta})\nonumber\\
&+\frac{f_{{\rm esc}}^{{\rm M}}M_{{\rm III}}N_{{\rm ion}}^{{\rm M}}}{\mu m_{{\rm H}}n_{b,0}}\frac{dn^{{\rm M}}}{dt}\left[1-\min\left\{ {\rm max}\left(x_{{\rm LW}},x\right),\,1\right\} \right]\label{eq:Gamma_mine_F}
\end{align}
for $\xi\propto f_{{\rm coll}}/t_{*}$ (``F''), where the superscript
M denotes MH, $M_{{\rm III}}$ is the mass of Pop III stars per MH,
$n^{{\rm M}}$ is the comoving number density of MHs, $n_{b,0}\simeq2\times10^{-7}\,{\rm cm^{-3}}$is
the comoving number density of baryons, $\mu=1.22$ is the mean molecular
weight of MH gas, and $x_{{\rm LW}}$ is the LW intensity $J_{{\rm LW}}$
normalized by the threshold intensity $J_{{\rm LW,th}}$ given by
\begin{equation}
x_{{\rm LW}}=J_{{\rm LW}}/J_{{\rm LW,th}},\label{eq:xLW}
\end{equation}
to implement the suppression of star formation inside minihalos (see
equations \ref{eq:Gamma_mine_dF} and \ref{eq:Gamma_mine_F}) in a
similar fashion with \citet{Ahn2012}.

Because we only take newly-forming MHs as sources, the last term in
equations (\ref{eq:Gamma_mine_dF}) and (\ref{eq:Gamma_mine_F}) are
identical. Having $\xi\propto dn^{{\rm M}}/dt$ instead of $\xi\propto df_{{\rm coll}}^{{\rm M}}/dt$
for MHs is to accommodate the tendency found in numerical simulations
of Pop III star formation: Pop III stars under the primordial environment
will form mostly in isolation (\citealt{2000ApJ...540...39A,2002ApJ...564...23B})
or as a few binary systems at most (\citealt{Turk2009,Stacy2010}),
and the total mass of Pop III stars in MHs is determined by the atomic
physics (\citealt{Hirano2014,Hirano2015}) and is not strongly dependent
on the mass of MHs.
  The actual mass of Pop III stars can vary substantially according to
  physical properties, such as the angular momentum, the local LW
  intensity and the mass accretion rate of their host MHs
  (\citealt{Hirano2014,Hirano2015}), and
  thus $M_{\rm III}$ should be taken as the average mass of Pop III
  stars per MH.

  We limit the MH mass to  $10^{5} \, M_{\odot}\le M \le 10^{8}$ and
  use the Sheth-Tormen mass function (\citealt{Sheth1999}) based on the
  typical linear matter density perturbation obtained from the linear
  Boltzmann solver CAMB (\citealt{Lewis2000}) to calculate
  $n^{\rm  M}$. Our choice of the minimum mass of MHs,
  $M_{\rm min}=10^{5}\,M_{\odot}$, is also used in
  \citet{Ahn2012} and is indeed reasonble due to the following reasons.
  This value is about 1/2 of the Jeans mass of the
  neutral IGM when the baryon-dark matter streaming velocity is
  considered (\citealt{Tseliakhovich2011}).
  The actual minimum mass to host Pop III stars could be
  somewhat larger than this Jeans mass and also redshift-dependent
  (see e.g. Fig. 1 of \citealt{Glover2013}, and also
  \citealt{Hirano2015}).
  Even though some claim that
  $M_{\rm min}\gtrsim$ a few $10^{6}\,M_{\odot}$ and thus the impact
  of MH stars
  on cosmic reionization is negligible (e.g. \citealt{Kimm2017}),
  other high-resolution numerical simulations
  (e.g. \citealt{Hirano2015}) find that massive Pop III stars are hosted
  mostly by halos in the mass range
  $M_{\rm min}\simeq [10^{5},\,10^{6}]\,M_{\odot}$.

  We also note that the ionization history $x(t)$ would depend
  on $M_{\rm min}$ (or similarly on $M_{\rm III}$) more weakly than
  $J_{\rm LW,th}$ and $f_{\rm esc}$, and therefore determining an
  accurate value 
  of $M_{\rm min}$ would not be too crucial.
  It is because 
  star formation in MHs, during the time when  MH stars
  are the dominant radiation sources, is regulated in a way to maintain
  $J_{\rm LW} \simeq J_{\rm LW,th}$ (\citealt{Ahn2012}; see also
  Section \ref{subsec:Result-reion-history}). 
  If $M_{\rm min}$ had been larger than our fiducial value and
  thus $n^{\rm M}$ had been smaller, then MH stars would have produced
  less ionizing and LW radiation in the beginning and drive resulting
  suppression   weaker (or $(1-x_{\rm LW})$ larger). In this case, because of
  reduced suppression, MH star formation will soon be expedited
  until $J_{\rm LW}$ reaches $J_{\rm LW,th}$ and produce an
  $x(t)$ evolution similar to the fiducial case thereafter.
  Similarly, any additional change in
  $n^{\rm M}$ due to the baryon-dark matter streaming effect is likely
  to be unimportant in determing $x(t)$.

The way we implement the LW
feedback as a multiplicative factor $(1-x_{{\rm LW}})$ in equations
(\ref{eq:Gamma_mine_dF}) and (\ref{eq:Gamma_mine_F}) roughly follows
the work of \citet{Yoshida2003} and \citet{OShea2008}, where they
find a gradual increase in $M_{{\rm min}}$, the minimum mass of halos
that can form stars, as $J_{{\rm LW}}$ increases if $J_{{\rm LW}}<J_{{\rm LW,\,th}}$.
\citet{OShea2008} also find that when $J_{{\rm LW}}\ge J_{{\rm LW,th}}\sim0.1\times10^{-21}\,{\rm erg\,s^{-1}\,cm^{-2}\,Hz^{-1}\,sr^{-1}}$,
$M_{{\rm min}}$ and $T_{{\rm vir,\,min}}$ (the minimum virial temperature
of halos that can form stars) jump to those of atomic-cooling halos
such that star formation in MHs are fully suppressed. \citet{Yoshida2003}
find practically the same result, namely the full suppression of star
formation inside MHs when $J_{{\rm LW}}\ge J_{{\rm LW,th}}\sim0.1\times10^{-21}\,{\rm erg\,s^{-1}\,cm^{-2}\,Hz^{-1}\,sr^{-1}}$
based on a combination of semi-analytical analysis and numerical simulation.
Therefore, the exact functional form of the suppression is not important
once $J_{{\rm LW}}$ reaches $J_{{\rm LW,th}}$ and the increasing
number of MHs afterwards try to produce more photons but fail to do
so due to the self-regulation. As long as the condition that MHs become
fully devoid of star formation when the condition $J_{{\rm LW}}\ge J_{{\rm LW,th}}$
is met, the formalism will correctly predict the self-regulation by
LW feedback. This is exactly how the LW feedback is implemented in
equations (\ref{eq:Gamma_mine_dF}) -- (\ref{eq:xLW}). Other work
(e.g. \citealt{Mirocha2018,Qin2021_tt_II}) implementing the LW feedback
on MH stars and forecasting the 21 cm background do not usually take
this approach, which will be discussed in detail in Section
\ref{subsec:Result-21cm}.

 The self-regulation of MH stars is predominantly
  governed by the LW feedback. In all the SR II models
  we tested (see the detailed model parameters in
  section \ref{subsec:Result-reion-history}),
  $x_{{\rm LW}}>x$ and thus the regulation factor
  $(1-{\rm min}\{ {\rm max}(x_{{\rm LW}},x),\,1\})$ in equations
  (\ref{eq:Gamma_mine_dF}) and (\ref{eq:Gamma_mine_F}) are practically
  identical  to $(1-{\rm min}\{x_{{\rm LW}},\,1\})$. We also note that we
  do not use biased suppression of star formation for MH stars, as
  quantified by $\eta$ 
  for LMACHs in equations (\ref{eq:Gamma_SRI_dF}) --
  (\ref{eq:Gamma_mine_F}). This is because star formation inside MHs
  is likely to occur much more diffusively than that inside ACHs. This
  tendency is indeed observed in the numerical simulation by 
  \citet{Ahn2012}: ionized regions generated by MH stars are almost
  uniformly spread in space, in constrast to those by ACH stars
  (Fig. 2 of \citealt{Ahn2012}).
  Therefore, we simply take an unbiased regulation
  factor $(1-{\rm min}\{x_{{\rm LW}},\,1\})$ for MH stars.

We take $N_{{\rm ion}}^{{\rm M}}=50000$, which is a value suitable
for very massive stars ($M_{*}\gtrsim100\,M_{\odot}$). As was noted
in \citet{Fialkov2013}, quantifying suppression of SFR by LW intensity
is not very straightforward when $J_{{\rm LW}}<J_{{\rm LW,th}}$,
if e.g. one considers the temporal evolution of LW intensity during
halo formation. Our suppression scheme described by equations (\ref{eq:Gamma_mine_dF})
-- (\ref{eq:xLW}) could instead perfectly mimic the full suppression
by any threshold intensity $J_{{\rm LW,th}}$, and our ignorance of
any other details is parameterized by the value of $J_{{\rm LW,th}}$.
How $J_{{\rm LW}}$ is evaluated is described in Section \ref{subsec:Back_LW}.

\subsection{Background Radiation and Feedback\label{subsec:Background}}

There are a few radiation backgrounds that determine the reionization
history and the 21 cm background. Any unprocessed background intensity
(${\rm erg\,s^{-1}\,cm^{-2}\,Hz^{-1}\,sr^{-1}}$) at observing frequency
$\nu$ and redshift $z$ is given by
\begin{equation}
J_{\nu}=\frac{c}{4\pi}(1+z)^{3}\int_{z}^{\infty}\frac{h\nu'\mathcal{N}_{\nu'}(z')}{(1+z')H(z')}{\rm e}^{-\tau_{\nu}}dz',\label{eq:Jnu_general}
\end{equation}
where $h$ is the Planck constant, $\mathcal{N}_{\nu'}(z')$ is the
photon-number luminosity density (${\rm s^{-1}\,Hz^{-1}\,cMpc^{-3}}$)
at source frequency $\nu'$ and redshift $z'$, and $\tau_{\nu}$
is the optical depth from $z'$ to $z$ at $\nu$. The soft-UV, LW,
and X-ray backgrounds are all unprocessed types and thus given by
equation (\ref{eq:Jnu_general}). The Lyman alpha background is a
processed type, and thus is not given by equation (\ref{eq:Jnu_general});
an appropriate description will be given instead in Section \ref{subsec:Back_Lya}.

\subsubsection{H-ionization by UV background\label{subsec:Back_UV}}

Ionization of IGM by the UV background is believed to be very inhomogeneous,
and the corresponding ``patchy reionization'' scenario is widely
accepted. Unless a rather extreme scenario of X-ray-dominated reionization
is assumed, patchy reionization will naturally occur in the Universe.
Well-defined H II regions, almost fully ionized inside and connecting
sharply with neutral IGM outside, will be created by UV sources in
patchy reionization. In such scenarios, it is difficult to specify
the background UV intensity by equation (\ref{eq:Jnu_general}) in
our one-zone model. 

The UV background is usually quantified in terms of the metagalactic
H-ionizing rate per baryon, commonly denoted by $\Gamma$ (in the
unit of ${\rm s}^{-1}$), which is currently well constrained for
the post-reionization epoch (e.g. \citealt{Bolton2007,Calverley2011}).
$\Gamma$ is a quantity that is determined after H-ionizing photons
emitted from galaxies (and quasars) are filtered and reprocessed as
they propagate through IGM and dense gas clumps. Because we do not
accurately model the clumping factor and we use practically a one-zone
model, it is difficult to calculate $\Gamma$ that is a processed
quantity linked to IGM properties such as the photon mean free path
$\lambda_{{\rm mfp}}$. Instead, we can use a more transparent quantity,
the UV-photon emissivity, which is blind to any physical properties
of the IGM. The UV-photon emissivity is defined as the H-ionizing
photon production rate per comoving volume (in the unit of ${\rm s}^{-1}\,{\rm cMpc}^{-3}$),
or equivalently $\dot{N}_{{\rm ion}}\equiv\xi n_{b,\,0}$, which can
be linked to $\Gamma$ by the relation $\Gamma\propto\lambda_{{\rm mfp}}\dot{N}_{{\rm ion}}$
(e.g. \citealt{Bolton2007} report $\dot{N}_{{\rm ion}}\simeq10^{50.5}\,{\rm s}^{-1}\,{\rm cMpc}^{-3}$
at $z=6$). We simply check whether our models produce a reasonable
value of $\dot{N}_{{\rm ion}}$ in Section \ref{subsec:Result-reion-history}.

\subsubsection{${\rm H}_{2}$-dissociation by Lyman-Werner background\label{subsec:Back_LW}}

The frequency-averaged, global LW intensity is given by

\begin{align}
  J_{{\rm LW}}(z) & =\left\langle \frac{c}{4\pi}(1+z)^{3}\int_{z}^{\infty}\frac{h\nu'\mathcal{N}_{\nu'}(z')}{(1+z')H(z')}{\rm e}^{-\tau_{\nu}}dz'\right\rangle _{\nu}\nonumber \\
  & =\frac{c}{4\pi}(1+z)^{3}\int_{z}^{\infty}
  \frac{\left\langle h\nu'\mathcal{N}_{\nu'}(z')\right\rangle_{\nu'}}
       {(1+z')H(z')}f_{{\rm mod}}(z,\,z')dz',
  \label{eq:JLW}
\end{align}
where $\left\langle \,\,\,\,\right\rangle _{\nu}$ and $\left\langle \,\,\,\,\right\rangle _{\nu'}$
are frequency-averages in the observed band at $z$ and in the emitted
band at $z'$ respectively, $\nu'$ is limited below the Lyman limit
(``LL'', $h\nu'<13.6\,{\rm eV}$), and $f_{{\rm mod}}$ is the ``picket-fence
modulation factor'' that accounts for the trimming of bands of radiation
from a source at $z'$ due to the redshifting of continuum into Lyman
resonance lines \citep{Ahn2009}. As seen in equation (\ref{eq:JLW}),
$f_{{\rm mod}}$ replaces the attenuation factor ${\rm e}^{-\tau_{\nu}}$
and is given approximately by
\begin{align}
  &f_{{\rm mod}}(z,\,z')\nonumber\\
  &=\begin{cases}
  1.7\exp\left[-\left(\frac{r_{{\rm cMpc}}}{116.29\alpha}\right)^{0.68}\right]-0.7 & {\rm if}\,\,\frac{r_{{\rm cMpc}}}{\alpha}\le97.39\\
  0 & {\rm if\,\,otherwise},
  \end{cases}
  \label{eq:fmod}
\end{align}
where $\alpha\equiv(h/0.7)^{-1}(\Omega_{m}/0.27)^{-1/2}[(1+z')/21]^{-1/2}$
and $r_{{\rm cMpc}}$ is the comoving distance that light has traveled
from $z'$ to $z$, in units of cMpc. $f_{{\rm mod}}$ is useful when
calculating the inhomogeneity of LW background if inhomogeneous source
distribution is given \citep{Ahn2009}, while in this work only serves
as the relative weight that sources at $z'$ contributes to $J_{{\rm LW}}(z)$. 

Because $J_{{\rm LW}}$ is simply the frequency-averaged intensity,
actual ${\rm H}_{2}$-dissociation rates by individual LW lines should
be further implemented. This could be achieved by some multiplication
factor weighted by line-wise dissociation rates, which would change
the effective weight of source-contribution from a smooth form ($f_{{\rm mod}}$)
to a discrete form ($f_{{\rm LW}}$ in \citealt{Fialkov2013}), or
by interpreting $J_{{\rm LW,th}}$ as the threshold intensity weighted
by the same line-wise dissociation rates. We take the latter option
in this work. We also note that in this one-zone model, equation (\ref{eq:JLW})
is equivalent to the LW intensity that is averaged over the sawtooth-modulated
spectrum \citep{Haiman1997}. Then, suppression of SFR by dissociation
of ${\rm H}_{2}$ is implemented in the form of equations (\ref{eq:Gamma_mine_dF})
-{}- (\ref{eq:xLW}).

\subsubsection{Ionization and heating by X-ray background\label{subsec:Back_X}}

Global X-ray intensity determines the heating rate and the ionization
rate of IGM outside H II regions (or ``bulk IGM'' as in \citealt{Mirocha2014}).
X-ray photon-number intensity is given by

\begin{equation}
N_{\nu}(z)=\frac{J_{\nu}(z)}{h\nu}=\frac{c}{4\pi}(1+z)^{2}\int_{z}^{z_{f}}\frac{\mathcal{N}_{\nu'}(z')}{H(z')}{\rm e}^{-\tau_{\nu}}dz',\label{eq:Nnu_X}
\end{equation}
where we used equation (\ref{eq:Jnu_general}) and the fact that $\nu=\nu'(1+z)/(1+z')$.
Once $N_{\nu}(z)$ is known, we can calculate the photo-ionization
rate
\begin{equation}
\xi_{{\rm X}}(z)= 4\pi n_{{\rm HI}}(z)\int_{\nu_{{\rm min}}}^{\nu_{{\rm max}}}N_{\nu}(z)\sigma_{\nu,{\rm HI}}d\nu,\label{eq:Gamma_bulk}
\end{equation}
the secondary ionization rate
\begin{equation}
{\tilde{\xi}}_{{\rm {\rm X}}}(z)= 4\pi n_{{\rm HI}}(z)\int_{\nu_{{\rm min}}}^{\nu_{{\rm max}}}N_{\nu}(z)\sigma_{\nu,{\rm HI}}\frac{h\nu-h\nu_{{\rm th}}}{h\nu}d\nu,\label{eq:gamma_2ndary_bulk}
\end{equation}
and the heating rate (${\rm erg\,s^{-1}}$)
\begin{equation}
\epsilon_{{\rm X}}(z)= 4\pi n_{{\rm HI}}(z)\int_{\nu_{{\rm min}}}^{\nu_{{\rm max}}}N_{\nu}(z)\sigma_{\nu,{\rm HI}}\left(h\nu-h\nu_{{\rm th}}\right)d\nu,\label{eq:gamma_2ndary_bulk-1}
\end{equation}
where $h\nu_{{\rm th}}=13.6\,{\rm eV}$ is the hydrogen Lyman-limit
energy, and $\sigma_{\nu,{\rm HI}}$ is the photo-ionization cross-Section
of the hydrogen atom at frequency $\nu$. The ionization rate equation
for the bulk IGM is then given by
\begin{equation}
\frac{dx_{{\rm b}}}{dt}=\left(\xi_{{\rm X}}+{\tilde{\xi}}_{{\rm X}}\right)(1-x_{{\rm b}})-\alpha Cx_{{\rm e}}^{2}n_{{\rm H}},\label{eq:ion_rate_bulk}
\end{equation}
where $x_{{\rm b}}$ is used to denote the ionized fraction of the
bulk IGM and to be distinguished from the volume ionized fraction
$x$. Even though this will affect the volume ionization rate (equ.
\ref{eq:dxvol_dt}) as well, we take the approximation that $1-x_{{\rm b}}\simeq1$
until the reionization ends. This holds true for reionization scenarios
we consider in this work. The energy rate equation is given by
\begin{equation}
\frac{3}{2}\frac{d}{dt}\left(\frac{k_{{\rm B}}T_{k}n_{{\rm b}}}{\mu}\right)=\epsilon_{{\rm X}}(z)+\epsilon_{{\rm comp}}-\mathcal{C},\label{eq:energy_rate}
\end{equation}
where $k_{{\rm B}}$ is the Boltzmann constant, $T_{k}$ is the kinetic
temperature of gas, $n_{{\rm b}}$ is the proper number density of
baryons, $\epsilon_{{\rm comp}}$ is the Compton heating (when $T_{{\rm K}}<T_{{\rm CMB}}$,
and cooling when $T_{{\rm K}}>T_{{\rm CMB}}$) rate and $\mathcal{C}$
is the cooling rate. We only include the adiabatic cooling by cosmic
expansion, which is dominant over the recombination cooling and the
collisional exctitation+ionization cooling. 

\subsubsection{Lyman Alpha background\label{subsec:Back_Lya}}

Hydrogen Ly$\alpha$ background is crucial in determining the 21 cm
background by decoupling the spin temperature from the CMB temperature
through Ly$\alpha$ pumping process, or the Wouthuysen-Field mechanism
\citep{Wouthuysen1952,Field1958}. The photon-number intensity of
Ly$\alpha$ background is given by
\begin{equation}
N_{\alpha}=\frac{c}{4\pi}(1+z)^{2}\sum_{n=2}^{n_{{\rm max}}}f_{{\rm rec}}(n)\int_{z}^{z'_{n}}dz'\frac{\mathcal{N}_{\nu'}(z')}{H(z')},\label{eq:Nalpha}
\end{equation}
where $n_{{\rm max}}$ (=23) is the effective maximum principal quantum
number of Lyman resonances, $f_{{\rm rec}}(n)$ is the probability
for a Ly$n$ photon (Ly1$\equiv$Ly$\alpha$, Ly2$\equiv$Ly$\beta$,
Ly3$\equiv$Ly$\gamma$, etc.) to be converted to a Ly$\alpha$ photon,
and $z'_{n}$ is the redshift satisfying (\citealt{Pritchard2006})
\begin{equation}
\frac{1+z'_{n}}{1+z}=\frac{1-(n+1)^{-2}}{1-n^{-2}}.\label{eq:zn}
\end{equation}

Ly$\alpha$ background can also be generated by the collisional excitation
of H atoms induced by energetic electrons generated by the X-ray background
(e.g. \citealt{Ahn2014}). However, we do not include this mechanism
here because it is usually negligible when the X-ray efficiency is
not extremely high. As seen in Section \ref{subsec:Back_21cm}, we
only impose a minimal level of X-ray background in this work.

\subsection{21 cm Background\label{subsec:Back_21cm}}

The spin temperature $T_{s}$ is a parameter representing the ratio
of up to down states of the hyperfine structure \citep{Field1958}:
\begin{equation}
\frac{n_{1}}{n_{0}}=3\exp\left[-\frac{T_{*}}{T_{s}}\right],\label{eq:Ts_def}
\end{equation}
where $n_{1}$ and $n_{0}$ are the number of hydrogen atoms in the up
(triplet) state and the down (singlet) state, respectively, and $T_{*}=0.0628\,{\rm K}$
is the energy difference of the two states in terms of temperature.
While in the absence of Lyman resonance photons $T_{s}$ is driven
to the CMB temperature $T_{{\rm CMB}}$ radiatively, the absorption
and re-emission of Lyman resonance photons can drive $T_{s}$ to the
color temperature of Lyman lines. The dominant radiative coupling
is by the Ly$\alpha$ photons, and the repeated scattering of Ly$\alpha$
photons against thermalized gas brings the Ly$\alpha$ color temperature
$T_{\alpha}$ into $T_{k}$. The mechanical pumping by collision,
separately, drives $T_{s}$ into $T_{k}$. Then the spin temperature
becomes
\begin{equation}
T_{s}^{-1}=\frac{T_{{\rm CMB}}^{-1}+(x_{c}+x_{\alpha})T_{k}^{-1}}{1+x_{c}+x_{\alpha}}\label{eq:Ts}
\end{equation}
where $x_{c}$ is the collisional coupling coefficient and $x_{\alpha}$
is the Ly$\alpha$ pumping coefficient. $x_{c}$ is given by
\begin{equation}
x_{c}=\frac{4\kappa(1-0)n_{{\rm H}}T_{*}}{3A_{10}T_{{\rm CMB}}},\label{eq:xc}
\end{equation}
where $A_{10}=2.85\times10^{-15}\,{\rm s}^{-1}$ is the spontaneous
emission coefficient, and $\kappa(1-0)$ is the collisional deexcitation
coefficient defined and tabulated as a function of $T_{k}$ in \citet{2005ApJ...622.1356Z}.
$x_{\alpha}$ is given by \citep{Pritchard2006}
\begin{align}
  x_{\alpha}&=\frac{16\pi^{2}T_{*}e^{2}
    f_{\alpha}}{27A_{10}T_{{\rm CMB}}m_{e}c}S_{\alpha}N_{\alpha} \nonumber\\
&=\frac{S_{\alpha}N_{\alpha}}{1.165\times10^{-10}
  \left[(1+z)/20\right]\,{\rm cm^{-2}\,s^{-1}\,Hz^{-1}\,sr^{-1}}},
\label{eq:xalpha}
\end{align}
where $m_{e}$ is the electron mass and $S_{\alpha}$ is a correction
factor of order of unity that accounts for the distortion of the line
profile by thermalized atoms and peculiar motion \citep{2004ApJ...602....1C,2006MNRAS.367..259H,Chuzhoy2006}.
In this paper, we adopt the functional form of $S_{\alpha}$ suitable
for comoving gas without peculiar motion \citep{Chuzhoy2006}:
\begin{equation}
S_{\alpha}=\frac{\exp\left[-0.37(1+z)^{1/2}T_{k}^{-2/3}\right]}{1+0.4T_{k}^{-1}}.\label{eq:Salpha}
\end{equation}

\section{Result\label{sec:Result}}

\subsection{Reionization history\label{subsec:Result-reion-history}}

We cover a limited but representative set of parameters in calculating
reionization histories. For the vanilla model, we use parameters
that are sampled similarly to \citet{2006MNRAS.371..867F} and \citet{Bernardi2015}.
For SRI, we use parameters including those of \citet{Iliev2007},
which first suggested the self-regulation scheme of the model. For
SRII, we use parameters including those of \citet{Ahn2012}, which
provides the physical basis of the model. Parameters and some characteristics
of reionization are listed in Tables \ref{tab:params} -- \ref{tab:paramsII300}.
For the vanilla model and SRI, we accommodate both dF and F star formation
scenarios (Section \ref{subsec:Gamma_F_dF}). The resulting $x(z)$'s
are plotted in figures \ref{fig:SR1} -- \ref{fig:SR2-1}, and overlaid
on the 68\% and 95\% constraints from the Planck Legacy Data (``PLD'':
\citealt{PlanckCollaboration2018}). Note that in these figures we
show the ionized volume fraction in terms of the electron fraction
$x_{{\rm e}}\equiv\left\langle n_{{\rm e}}\right\rangle /\left\langle n_{{\rm H}}\right\rangle $,
with $\left\langle \,\,\,\,\right\rangle $ being the volume average,
such that $x_{{\rm e}}=(1+\left\langle n_{{\rm He}}\right\rangle /\left\langle n_{{\rm H}}\right\rangle )x=1.079\,x$
if helium atoms are assumed singly ionized in H II regions. The PLD
constraints are in fact on $x_{{\rm e}}$, and they are shown as shaded
regions in figures \ref{fig:SR1} -- \ref{fig:SR2-1}.

\begin{table*}[t!]
\begin{tabular}{ccccccccccccc}
\hline 
Model & $f_{\gamma}$ & $g_{\gamma}$ & $f_{\gamma}^{{\rm H}}$ & $f_{\gamma}^{{\rm L}}$ & $g_{\gamma}^{{\rm H}}$ & $g_{\gamma}^{{\rm L}}$ & $z_{{\rm begin}}$ & $z_{{\rm end}}$ & $\Delta z_{3-97}$ & $\tau_{{\rm es}}$ & $\tau_{{\rm es}}(15-30)$ & $\chi^{2}/\nu$\tabularnewline
\hline 
\cellcolor{lgray}V-L\_dF & 19.62 & $\ldots$ & $\ldots$ & $\ldots$ & $\ldots$ & $\ldots$ & 15.87 & 5.50 & 8.18 & 0.05998 & 3.97E-4 & \cellcolor{lgray}0.81\tabularnewline
V-M1\_dF & 23.55 & $\ldots$ & $\ldots$ & $\ldots$ & $\ldots$ & $\ldots$ & 16.32 & 6.21 & 7.98 & 0.06638 & 5.11E-4 & 0.86\tabularnewline
V-M2\_dF & 34.88 & $\ldots$ & $\ldots$ & $\ldots$ & $\ldots$ & $\ldots$ & 17.01 & 7.27 & 7.66 & 0.07647 & 7.57E-4 & 1.19\tabularnewline
V-H\_dF & 56.69 & $\ldots$ & $\ldots$ & $\ldots$ & $\ldots$ & $\ldots$ & 17.85 & 8.52 & 7.33 & 0.08899 & 1.23E-3 & 2.31\tabularnewline
V-L\_F & $\ldots$ & 0.4077 & $\ldots$ & $\ldots$ & $\ldots$ & $\ldots$ & 12.21 & 5.49 & 5.06 & 0.04794 & 2.73E-5 & 0.86\tabularnewline
\cellcolor{lgray}V-M1\_F & $\ldots$ & 0.7042 & $\ldots$ & $\ldots$ & $\ldots$ & $\ldots$ & 13.00 & 6.25 & 5.10 & 0.05555 & 4.71E-5 & \cellcolor{lgray}0.81\tabularnewline
V-M2\_F & $\ldots$ & 1.472 & $\ldots$ & $\ldots$ & $\ldots$ & $\ldots$ & 14.06 & 7.32 & 5.11 & 0.06651 & 9.85E-5 & 0.87\tabularnewline
V-H\_F & $\ldots$ & 3.140 & $\ldots$ & $\ldots$ & $\ldots$ & $\ldots$ & 15.13 & 8.43 & 5.10 & 0.07852 & 2.10E-4 & 1.32\tabularnewline
\cellcolor{lgray}SRI-L0\_dF & $\ldots$ & $\ldots$ & 26.02 & 0 & $\ldots$ & $\ldots$ & 12.82 & 5.81 & 5.56 & 0.05436 & 2.01E-5 & \cellcolor{lgray}0.82\tabularnewline
SRI-LL\_dF & $\ldots$ & $\ldots$ & 26.02 & 91.93 & $\ldots$ & $\ldots$ & 16.04 & 6.05 & 7.96 & 0.06380 & 4.14E-4 & 0.83\tabularnewline
SRI-LH\_dF & $\ldots$ & $\ldots$ & 26.02 & 919.3 & $\ldots$ & $\ldots$ & 19.11 & 6.82 & 10.41 & 0.09311 & 2.77E-3 & 2.93\tabularnewline
SRI-HL\_dF & $\ldots$ & $\ldots$ & 91.93 & 91.93 & $\ldots$ & $\ldots$ & 16.22 & 8.25 & 6.15 & 0.08057 & 4.60E-4 & 1.46\tabularnewline
SRI-HH\_dF & $\ldots$ & $\ldots$ & 91.93 & 919.3 & $\ldots$ & $\ldots$ & 19.12 & 8.78 & 8.54 & 0.10083 & 2.81E-3 & 4.54\tabularnewline
SRI-0H\_dF & $\ldots$ & $\ldots$ & 0 & 919.3 & $\ldots$ & $\ldots$ & 19.11 & $\ldots$ & $\ldots$ & 0.07659 & 2.75E-3 & 1.80\tabularnewline
SRI-L0\_F & $\ldots$ & $\ldots$ & $\ldots$ & $\ldots$ & 0.8673 & 0 & 10.81 & 5.78 & 3.84 & 0.04756 & 1.53E-6 & 0.87\tabularnewline
\cellcolor{lgray}SRI-LL\_F & $\ldots$ & $\ldots$ & $\ldots$ & $\ldots$ & 0.8673 & 8.673 & 14.41 & 6.24 & 6.49 & 0.06140 & 1.16E-4 & \cellcolor{lgray}0.81\tabularnewline
SRI-LH\_F & $\ldots$ & $\ldots$ & $\ldots$ & $\ldots$ & 0.8673 & 86.73 & 17.09 & 6.93 & 8.34 & 0.08816 & 8.63E-4 & 2.22\tabularnewline
SRI-HL\_F & $\ldots$ & $\ldots$ & $\ldots$ & $\ldots$ & 6.938 & 8.673 & 14.59 & 8.27 & 4.82 & 0.07562 & 1.25E-4 & 1.16\tabularnewline
SRI-HH\_F & $\ldots$ & $\ldots$ & $\ldots$ & $\ldots$ & 6.938 & 86.73 & 17.10 & 8.79 & 6.74 & 0.09310 & 8.71E-4 & 2.95\tabularnewline
SRI-0H\_F & $\ldots$ & $\ldots$ & $\ldots$ & $\ldots$ & 0 & 86.73 & 17.09 & 4.10 & 10.12 & 0.08597 & 8.62E-4 & 1.98\tabularnewline
\hline 
\end{tabular}

\caption{Model parameters and resulting characteristics of the vanilla and
the SRI. Notes: SRI-0H\_dF does not reach the end of reionization,
not even $x=0.97$, until z=0. Highlighted in gray are those that
fit the observed $C_{l}^{{\rm EE}}$ of the PLD best.\label{tab:params}}
\end{table*}

\begin{figure*}\centering
  
\includegraphics[width=0.45\textwidth]{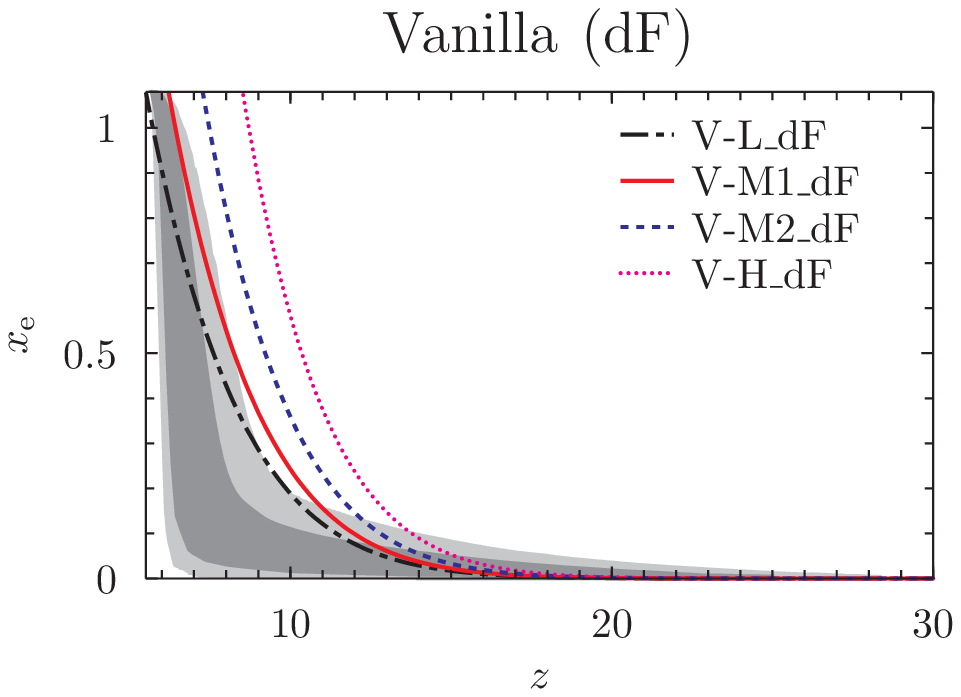}\hspace{0.3cm}\includegraphics[width=0.45\textwidth]{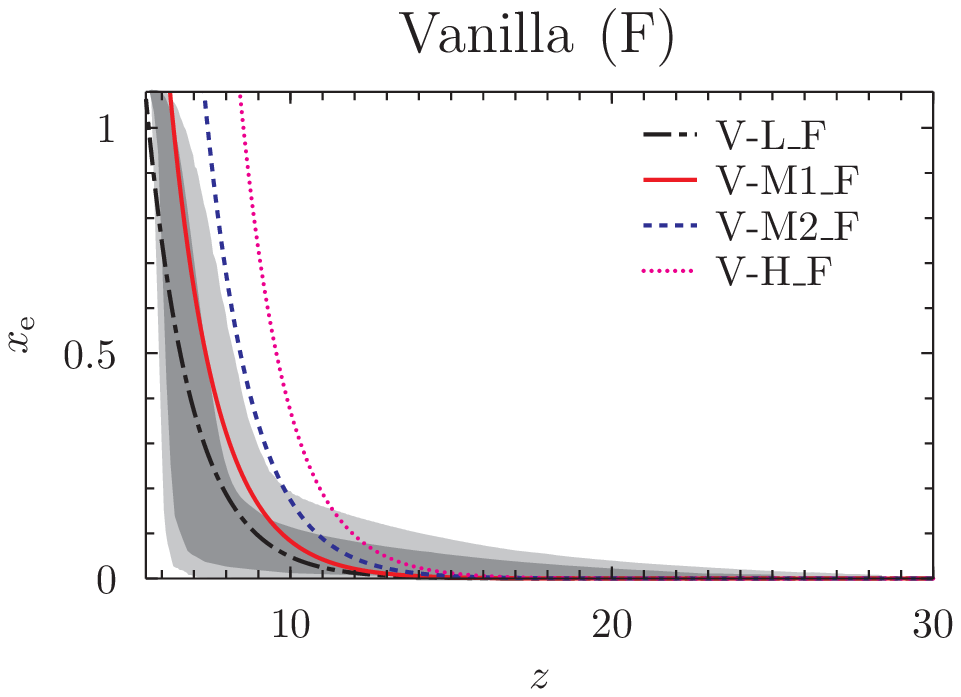}

\caption{Evolution of the global ionized fraction in vanilla (``V'') reionization
models with varying $f_{\gamma}$. Model specifications are shown
in legends, and are listed in Table \ref{tab:params}. (left) Vanilla
models when ``dF'' scheme is used, and (right) vanilla models when
``F'' scheme is used. Throughout Figs \ref{fig:typical} -- \ref{fig:SR2-1},
the ionized fraction is in terms of $x_{{\rm e}}\equiv n_{{\rm e}}/n_{{\rm H}}=1.079\,x$
which reaches the maximum value of 1.079, based on assuming that all
helium atoms inside H II regions are singly ionized before the epoch
of helium reionization occurring at $z\simeq3.5$.\label{fig:typical}}
\end{figure*}

\begin{figure*}\centering
  
\includegraphics[width=0.45\textwidth]{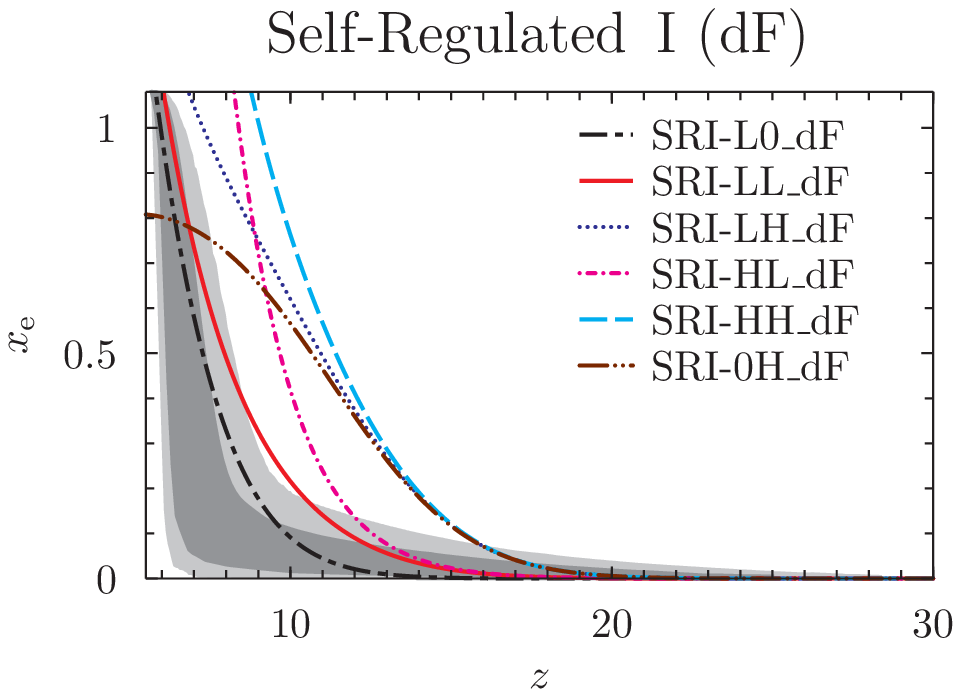}\hspace{0.3cm}\includegraphics[width=0.45\textwidth]{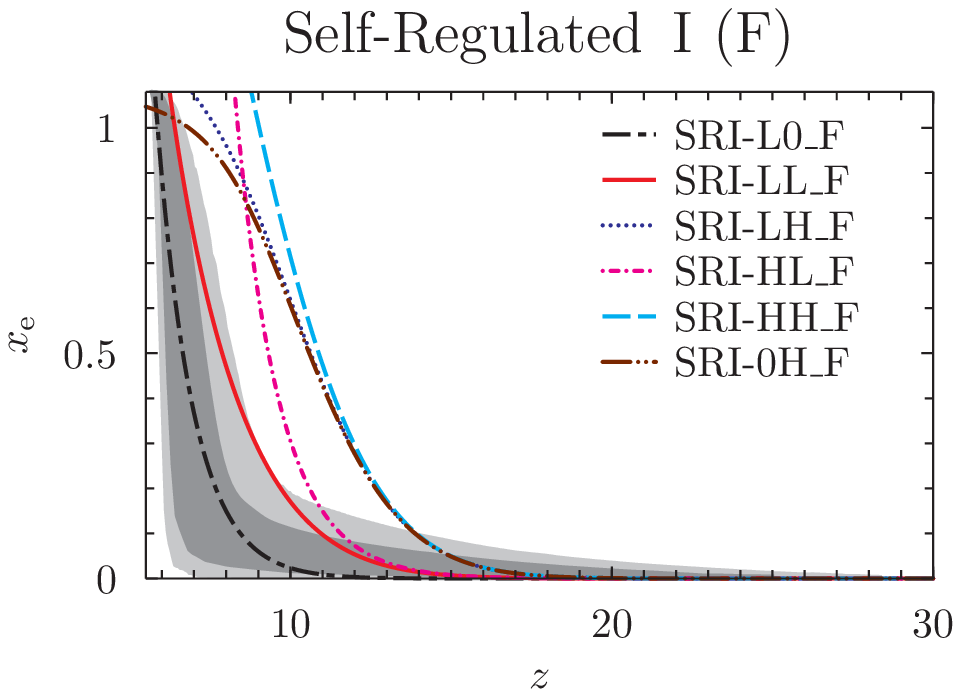}

\caption{Evolution of the global ionized fraction in the model category ``self-regulated
I''. The model variance is specified by the set of $\left\{ g_{\gamma}^{{\rm H}},\,g_{\gamma}^{{\rm L}}\right\} $
and are specified in legends, with the nomenclature ``SI-$g_{\gamma}^{{\rm H}}$$g_{\gamma}^{{\rm L}}$''
but $g_{\gamma}$'s with letters ``0'' for null, ``L'' for low
and ``H'' for high. These models are listed in Table \ref{tab:params}.\label{fig:SR1}}
\end{figure*}

All models are described by simple ordinary differential equations
(ODEs), and thus can be easily integrated with ODE solvers. We start
numerical integration from $z=40$, when the contribution of any type
of halos to reionization and heating is believed to be negligible.
The initial value of $x$ is set to an arbitrarily small value, because
the volume occupied by H II regions at $z=40$ must be negligible.
Ionization rate equations to solve are not stiff, and we use a 4th-order,
adaptive Runge-Kutta integrator with both the relative tolerance and
the absolute tolerance of $x$ set to $10^{-5}$. For SRII, we need
an extra effort to calculate $J_{{\rm LW}}(t)$ at any time $t$,
because $J_{{\rm LW}}(t)$ regulates SFR inside MHs at $t$ via $x_{{\rm LW}}$
and impacts PPR (equations \ref{eq:Gamma_mine_dF} and \ref{eq:Gamma_mine_F}).
Therefore, we calculate $x(t)$ and $J(t)$ at each incrementally
increasing time step, by integrating equation \ref{eq:dxvol_dt} with
equation \ref{eq:Gamma_mine_F} (or equation \ref{eq:Gamma_mine_dF}
if dF assumed) and using equations (\ref{eq:xLW}) and (\ref{eq:JLW}).

\begin{figure*}\centering
  
\includegraphics[width=0.4\paperwidth]{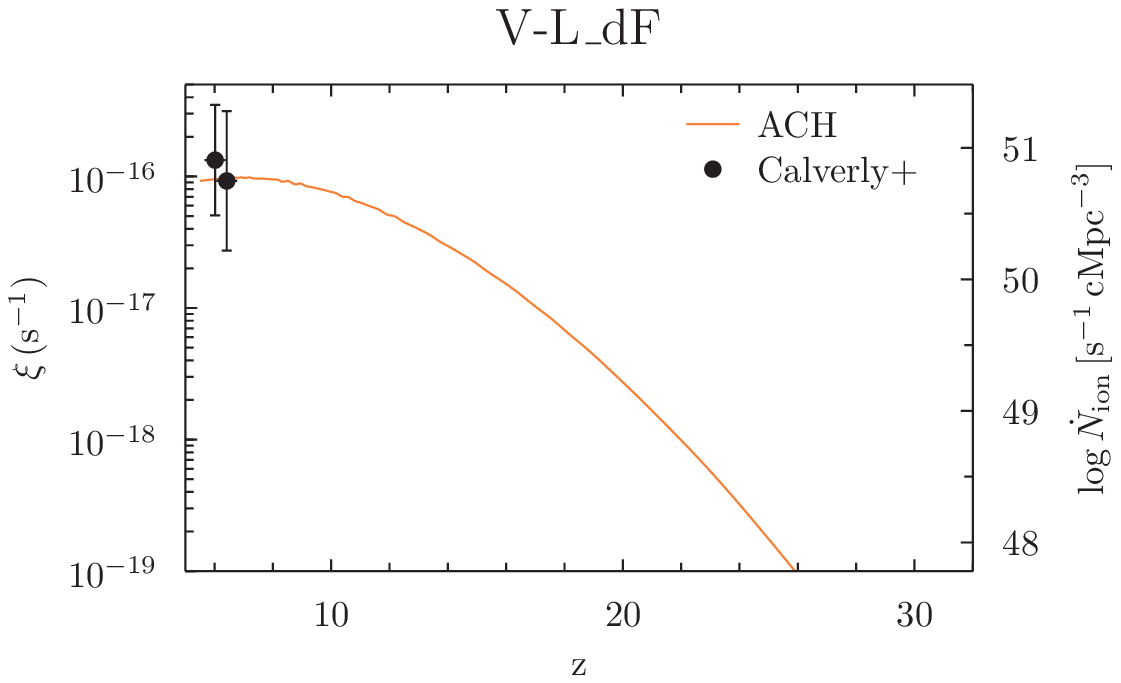}\hspace{0.3cm}\includegraphics[width=0.4\paperwidth]{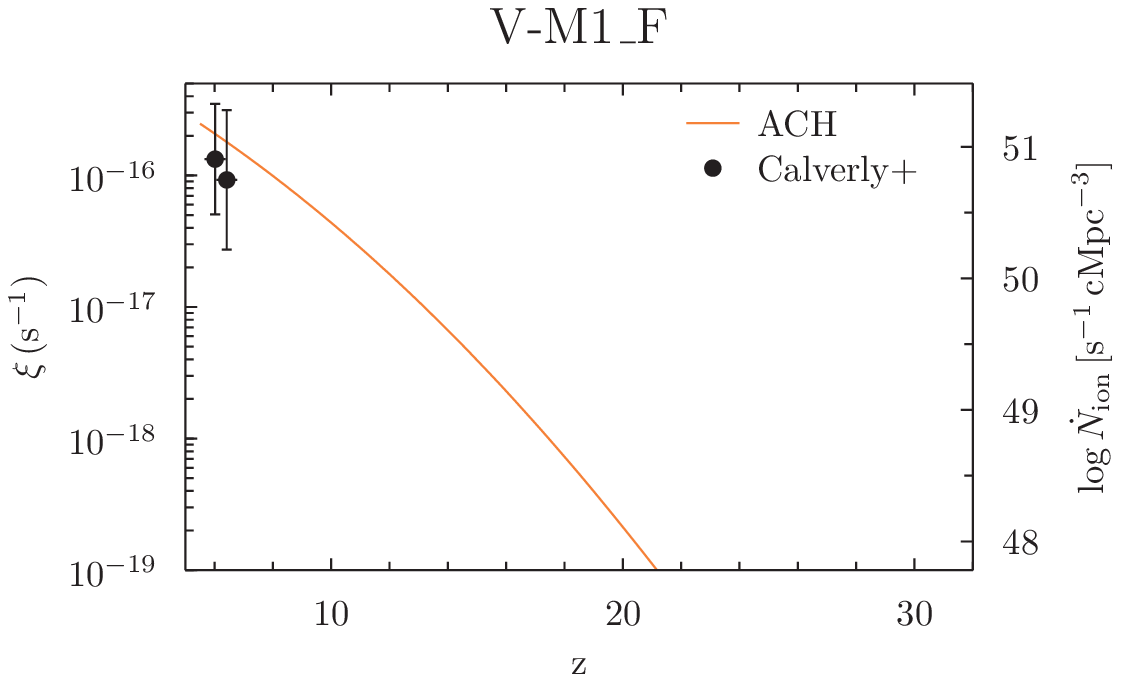}
\vspace{0.3cm}

\includegraphics[width=0.4\paperwidth]{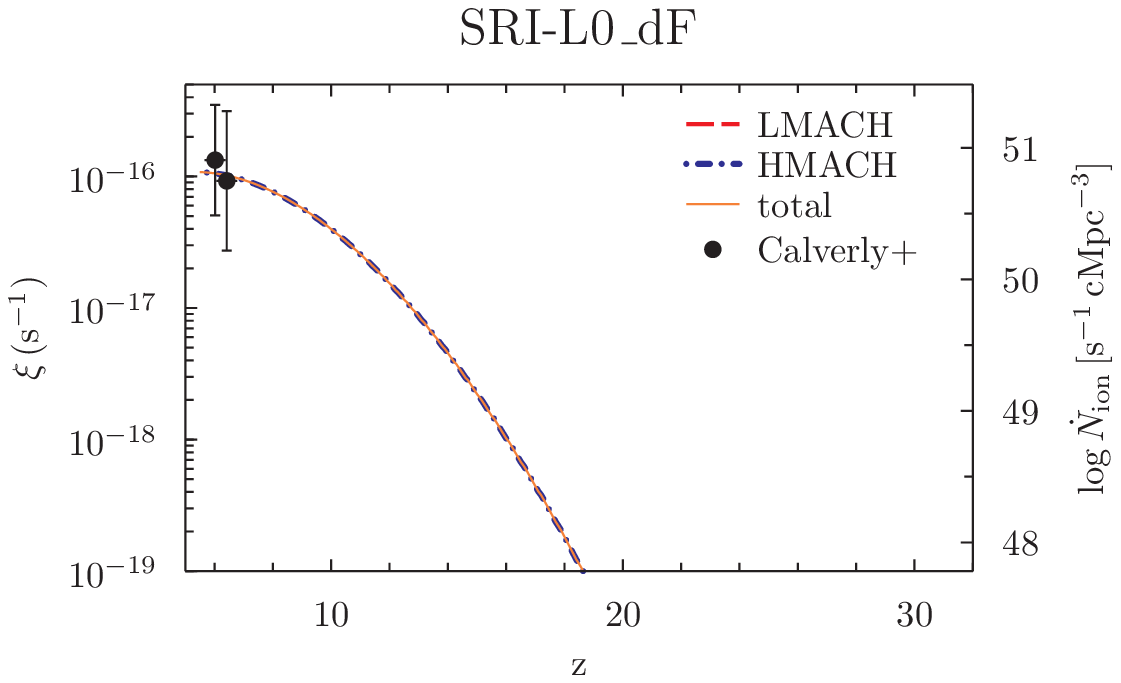}\hspace{0.3cm}\includegraphics[width=0.4\paperwidth]{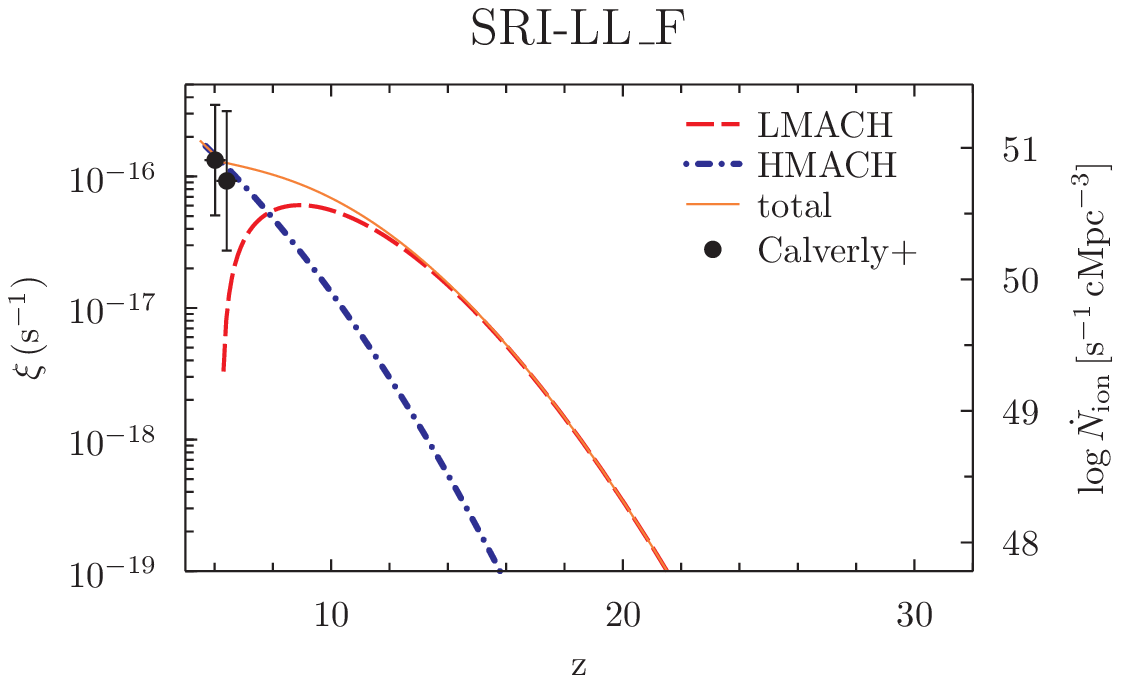}

\caption{
  (top) Ionization rate and ionizing-photon emissivity ($\xi$ in equation
  \ref{eq:dxvol_dt} and $\dot{N}_{{\rm ion}}\equiv\xi n_{b,\,0}$)
  of PLD-favored vanilla and SRI models,
  compared to the $z=6.02$ and $z=6.42$ constraints by
  \citet{Calverley2011}.
 Models are specified in the subtitle.
    For SRI cases, contributions by LMACHs (red, dashed) and HMACHs (blue,
    dot-dashed) are shown separately.
    The net quantities (orange, solid) are also plotted.
  Note that the
  constraint by \citet{Calverley2011} here are values of observed
    $\Gamma$ (section \ref{subsec:Back_UV}) 
  translated into $\dot{N}_{{\rm ion}}$ based on a specific set of $\lambda_{{\rm mfp}}$
  and the spectral hardness of H-ionizing photons: $\lambda_{{\rm mfp}}=10\,{\rm cMpc}$
  and $\alpha_{{\rm s}}=\alpha_{{\rm b}}=2$ in equation (21) of \citet{Bolton2007}.
  \label{fig:Gamma_VSRI}
}

\end{figure*}

\begin{table*}[t!]
\begin{tabular}{cccccccccccc}
\hline 
Model & $g_{\gamma}^{{\rm H}}$ & $g_{\gamma}^{{\rm L}}$ & $M_{{\rm III}}$ & $f_{{\rm esc}}^{{\rm M}}$ & $J_{{\rm LW,th}}$ & $z_{{\rm begin}}$ & $z_{{\rm end}}$ & $\Delta z_{3-97}$ & $\tau_{{\rm es}}$ & $\tau_{{\rm es}}(15-30)$ & $\chi^{2}/\nu$\tabularnewline
\hline 
SRII-L0-100-e0.1-J0.05 & 0.8673 & 0 & 100 & 0.1 & 0.05 & 10.86 & 5.78 & 3.86 & 0.04801 & 3.45E-4 & 0.86\tabularnewline
SRII-L0-100-e0.1-J0.10 & 0.8673 & 0 & 100 & 0.1 & 0.10 & 11.03 & 5.78 & 3.88 & 0.04850 & 6.18E-4 & 0.86\tabularnewline
SRII-L0-100-e0.5-J0.05 & 0.8673 & 0 & 100 & 0.5 & 0.05 & 11.18 & 5.78 & 3.90 & 0.04983 & 1.72E-3 & 0.85\tabularnewline
SRII-L0-100-e0.5-J0.10 & 0.8673 & 0 & 100 & 0.5 & 0.10 & 22.15 & 5.79 & 4.07 & 0.05226 & 3.08E-3 & 0.83\tabularnewline
SRII-L0-100-e1.0-J0.05 & 0.8673 & 0 & 100 & 1.0 & 0.05 & 22.91 & 5.79 & 3.96 & 0.05210 & 3.44E-3 & 0.83\tabularnewline
\cellcolor{lgray}SRII-L0-100-e1.0-J0.10 & 0.8673 & 0 & 100 & 1.0 & 0.10 & 25.38 & 5.79 & 4.51 & 0.05696 & 6.17E-3 & \cellcolor{lgray}0.80\tabularnewline
\cellcolor{lgray}SRII-LL-100-e0.1-J0.05 & 0.8673 & 8.673 & 100 & 0.1 & 0.05 & 14.48 & 6.24 & 6.50 & 0.06168 & 3.75E-4 & \cellcolor{lgray}0.81\tabularnewline
\cellcolor{lgray}SRII-LL-100-e0.1-J0.10 & 0.8673 & 8.673 & 100 & 0.1 & 0.10 & 14.90 & 6.24 & 6.53 & 0.06199 & 6.37E-4 & \cellcolor{lgray}0.81\tabularnewline
\cellcolor{lgray}SRII-LL-100-e0.5-J0.05 & 0.8673 & 8.673 & 100 & 0.5 & 0.05 & 15.26 & 6.24 & 6.54 & 0.06284 & 1.44E-3 & \cellcolor{lgray}0.82\tabularnewline
SRII-LL-100-e0.5-J0.10 & 0.8673 & 8.673 & 100 & 0.5 & 0.10 & 22.14 & 6.24 & 6.76 & 0.06446 & 2.78E-3 & 0.83\tabularnewline
\cellcolor{lgray}SRII-LL-100-e1.0-J0.05 & 0.8673 & 8.673 & 100 & 1.0 & 0.05 & 22.90 & 6.24 & 6.59 & 0.06432 & 2.79E-3 & \cellcolor{lgray}0.82\tabularnewline
SRII-LL-100-e1.0-J0.10 & 0.8673 & 8.673 & 100 & 1.0 & 0.10 & 25.38 & 6.24 & 13.93 & 0.06762 & 5.50E-3 & 0.86\tabularnewline
SRII-LH-100-e0.1-J0.05 & 0.8673 & 86.73 & 100 & 0.1 & 0.05 & 17.11 & 6.93 & 8.35 & 0.08832 & 1.01E-3 & 2.23\tabularnewline
SRII-LH-100-e0.1-J0.10 & 0.8673 & 86.73 & 100 & 0.1 & 0.10 & 17.19 & 6.93 & 8.35 & 0.08847 & 1.15E-3 & 2.24\tabularnewline
SRII-LH-100-e0.5-J0.05 & 0.8673 & 86.73 & 100 & 0.5 & 0.05 & 17.22 & 6.93 & 8.36 & 0.08895 & 1.60E-3 & 2.29\tabularnewline
SRII-LH-100-e0.5-J0.10 & 0.8673 & 86.73 & 100 & 0.5 & 0.10 & 22.05 & 6.93 & 8.41 & 0.08973 & 2.37E-3 & 2.37\tabularnewline
SRII-LH-100-e1.0-J0.05 & 0.8673 & 86.73 & 100 & 1.0 & 0.05 & 17.46 & 6.93 & 8.37 & 0.08976 & 2.35E-3 & 2.36\tabularnewline
SRII-LH-100-e1.0-J0.10 & 0.8673 & 86.73 & 100 & 1.0 & 0.10 & 25.38 & 6.93 & 8.50 & 0.09135 & 3.91E-3 & 2.55\tabularnewline
SRII-HL-100-e0.1-J0.05 & 6.938 & 8.673 & 100 & 0.1 & 0.05 & 14.62 & 8.27 & 4.82 & 0.07586 & 3.48E-4 & 1.17\tabularnewline
SRII-HL-100-e0.1-J0.10 & 6.938 & 8.673 & 100 & 0.1 & 0.10 & 14.71 & 8.27 & 4.83 & 0.07610 & 5.70E-4 & 1.18\tabularnewline
SRII-HL-100-e0.5-J0.05 & 6.938 & 8.673 & 100 & 0.5 & 0.05 & 14.79 & 8.27 & 4.84 & 0.07684 & 1.26E-3 & 1.20\tabularnewline
SRII-HL-100-e0.5-J0.10 & 6.938 & 8.673 & 100 & 0.5 & 0.10 & 22.13 & 8.27 & 4.91 & 0.07807 & 2.40E-3 & 1.26\tabularnewline
SRII-HL-100-e1.0-J0.05 & 6.938 & 8.673 & 100 & 1.0 & 0.05 & 15.35 & 8.27 & 4.86 & 0.07807 & 2.41E-3 & 1.25\tabularnewline
SRII-HL-100-e1.0-J0.10 & 6.938 & 8.673 & 100 & 1.0 & 0.10 & 25.38 & 8.27 & 5.04 & 0.08057 & 4.70E-3 & 1.38\tabularnewline
SRII-HH-100-e0.1-J0.05 & 6.938 & 86.73 & 100 & 0.1 & 0.05 & 17.12 & 8.79 & 6.74 & 0.09325 & 1.01E-3 & 2.96\tabularnewline
SRII-HH-100-e0.1-J0.10 & 6.938 & 86.73 & 100 & 0.1 & 0.10 & 17.19 & 8.79 & 6.75 & 0.09339 & 1.16E-3 & 2.98\tabularnewline
SRII-HH-100-e0.5-J0.05 & 6.938 & 86.73 & 100 & 0.5 & 0.05 & 17.22 & 8.79 & 6.75 & 0.09387 & 1.60E-3 & 3.04\tabularnewline
SRII-HH-100-e0.5-J0.10 & 6.938 & 86.73 & 100 & 0.5 & 0.10 & 22.05 & 8.79 & 6.79 & 0.09464 & 2.35E-3 & 3.15\tabularnewline
SRII-HH-100-e1.0-J0.05 & 6.938 & 86.73 & 100 & 1.0 & 0.05 & 17.43 & 8.79 & 6.76 & 0.09467 & 2.34E-3 & 3.14\tabularnewline
SRII-HH-100-e1.0-J0.10 & 6.938 & 86.73 & 100 & 1.0 & 0.10 & 25.38 & 8.79 & 6.88 & 0.09622 & 3.87E-3 & 3.38\tabularnewline
SRII-0H-100-e0.1-J0.05 & 0 & 86.73 & 100 & 0.1 & 0.05 & 17.11 & 4.10 & 10.12 & 0.08612 & 1.01E-3 & 1.99\tabularnewline
SRII-0H-100-e0.1-J0.10 & 0 & 86.73 & 100 & 0.1 & 0.10 & 17.19 & 4.10 & 10.13 & 0.08627 & 1.15E-3 & 2.00\tabularnewline
SRII-0H-100-e0.5-J0.05 & 0 & 86.73 & 100 & 0.5 & 0.05 & 17.22 & 4.10 & 10.13 & 0.08676 & 1.60E-3 & 2.04\tabularnewline
SRII-0H-100-e0.5-J0.10 & 0 & 86.73 & 100 & 0.5 & 0.10 & 22.05 & 4.10 & 10.18 & 0.08754 & 2.37E-3 & 2.12\tabularnewline
SRII-0H-100-e1.0-J0.05 & 0 & 86.73 & 100 & 1.0 & 0.05 & 17.46 & 4.10 & 10.15 & 0.08756 & 2.35E-3 & 2.11\tabularnewline
SRII-0H-100-e1.0-J0.10 & 0 & 86.73 & 100 & 1.0 & 0.10 & 25.38 & 4.10 & 10.27 & 0.08916 & 3.92E-3 & 2.27\tabularnewline
\hline 
\end{tabular}

\caption{Model parameters and resulting characteristics of SRII, with $M_{{\rm III}}=100\,M_{\odot}$.
Notes: In the table, $M_{{\rm III}}$ and $J_{{\rm LW,\,th}}$ are
in units of $M_{\odot}$ and $10^{-21}\,{\rm erg\,s^{-1}\,cm^{-2}\,Hz^{-1}\,sr^{-1}}$,
respectively. Highlighted in grey are those that fit the observed
$C_{l}^{{\rm EE}}$ of the PLD best.\label{tab:paramsII100}}
\end{table*}

The vanilla model shows a smooth and monotonic evolution of $x$ (Fig.
\ref{fig:typical}). The monotonic behavior of $x(t)$ is easily explained
by the fact that $\xi$ is proportional to the monotonically increasing
$f_{{\rm coll}}$ (F scenario) or $df_{{\rm coll}}/dt$ (dF scenario).
This characteristic makes $\tau_{{\rm es}}$ and $z_{{\rm end}}$
tightly correlated, and thus lacks the ``leverage'' to accommodate
reionization scenarios that are different in $x(z)$ but degenerate
in $\tau_{{\rm es}}$ and $z_{{\rm end}}$, as long as one type of
star formation scenarios are chosen from F or dF. One can of course
make a somewhat more sophisticated variant of this model by e.g. allowing
multiple species of halos with different $f_{\gamma}$'s (mixture
of Pop II and Pop III stars \citealt{2006MNRAS.371..867F}). Nevertheless,
due to the lack of any self-regulation, the resulting reionization
histories of such variants would still remain similar to the original
vanilla model. The duration of reionization is in general more extended
in the dF scenario than the F scenario. This is due to the fact that
$df_{{\rm coll}}/dt$ grows more slowly than $f_{{\rm coll}}$. Therefore,
for given $z_{{\rm end}}$, the dF scenario produces larger $\tau_{{\rm es}}$
than the F scenario. This tendency is clearly presented in figure
\ref{fig:typical} and Table \ref{tab:params}.

\begin{table*}[t!]
\begin{tabular}{cccccccccccc}
\hline 
Model & $g_{\gamma}^{{\rm H}}$ & $g_{\gamma}^{{\rm L}}$ & $M_{{\rm III}}$ & $f_{{\rm esc}}^{{\rm M}}$ & $J_{{\rm LW,th}}$ & $z_{{\rm begin}}$ & $z_{{\rm end}}$ & $\Delta z_{3-97}$ & $\tau_{{\rm es}}$ & $\tau_{{\rm es}}(15-30)$ & $\chi^{2}/\nu$\tabularnewline
\hline 
SRII-L0-300-e0.1-J0.05 & 0.8673 & 0 & 300 & 0.1 & 0.05 & 10.89 & 5.78 & 3.86 & 0.04817 & 4.50E-4 & 0.86\tabularnewline
SRII-L0-300-e0.1-J0.10 & 0.8673 & 0 & 300 & 0.1 & 0.10 & 11.10 & 5.78 & 3.89 & 0.04875 & 7.75E-4 & 0.86\tabularnewline
SRII-L0-300-e0.5-J0.05 & 0.8673 & 0 & 300 & 0.5 & 0.05 & 11.45 & 5.78 & 3.92 & 0.05063 & 2.29E-3 & 0.84\tabularnewline
\cellcolor{lgray}SRII-L0-300-e0.5-J0.10 & 0.8673 & 0 & 300 & 0.5 & 0.10 & 23.26 & 5.79 & 4.14 & 0.05350 & 3.88E-3 & \cellcolor{lgray}0.82\tabularnewline
\cellcolor{lgray}SRII-L0-300-e1.0-J0.05 & 0.8673 & 0 & 300 & 1.0 & 0.05 & 23.50 & 5.79 & 4.02 & 0.05365 & 4.50E-3 & \cellcolor{lgray}0.82\tabularnewline
\cellcolor{lgray}SRII-L0-300-e1.0-J0.10 & 0.8673 & 0 & 300 & 1.0 & 0.10 & 27.15 & 5.79 & 15.28 & 0.05960 & 7.79E-3 & \cellcolor{lgray}0.80\tabularnewline
\cellcolor{lgray}SRII-LL-300-e0.1-J0.05 & 0.8673 & 8.673 & 300 & 0.1 & 0.05 & 14.51 & 6.24 & 6.50 & 0.06177 & 4.49E-4 & \cellcolor{lgray}0.81\tabularnewline
\cellcolor{lgray}SRII-LL-300-e0.1-J0.10 & 0.8673 & 8.673 & 300 & 0.1 & 0.10 & 15.26 & 6.24 & 6.54 & 0.06215 & 7.73E-4 & \cellcolor{lgray}0.82\tabularnewline
\cellcolor{lgray}SRII-LL-300-e0.5-J0.05 & 0.8673 & 8.673 & 300 & 0.5 & 0.05 & 20.00 & 6.24 & 6.55 & 0.06330 & 1.83E-3 & \cellcolor{lgray}0.82\tabularnewline
SRII-LL-300-e0.5-J0.10 & 0.8673 & 8.673 & 300 & 0.5 & 0.10 & 23.26 & 6.24 & 6.86 & 0.06528 & 3.45E-3 & 0.83\tabularnewline
SRII-LL-300-e1.0-J0.05 & 0.8673 & 8.673 & 300 & 1.0 & 0.05 & 23.50 & 6.24 & 6.65 & 0.06527 & 3.57E-3 & 0.83\tabularnewline
SRII-LL-300-e1.0-J0.10 & 0.8673 & 8.673 & 300 & 1.0 & 0.10 & 27.15 & 6.24 & 14.77 & 0.06944 & 6.96E-3 & 0.88\tabularnewline
SRII-LH-300-e0.1-J0.05 & 0.8673 & 86.73 & 300 & 0.1 & 0.05 & 17.12 & 6.93 & 8.35 & 0.08836 & 1.04E-3 & 2.23\tabularnewline
SRII-LH-300-e0.1-J0.10 & 0.8673 & 86.73 & 300 & 0.1 & 0.10 & 17.21 & 6.93 & 8.36 & 0.08856 & 1.23E-3 & 2.25\tabularnewline
SRII-LH-300-e0.5-J0.05 & 0.8673 & 86.73 & 300 & 0.5 & 0.05 & 17.26 & 6.93 & 8.36 & 0.08917 & 1.77E-3 & 2.30\tabularnewline
SRII-LH-300-e0.5-J0.10 & 0.8673 & 86.73 & 300 & 0.5 & 0.10 & 23.22 & 6.93 & 8.42 & 0.09022 & 2.78E-3 & 2.41\tabularnewline
SRII-LH-300-e1.0-J0.05 & 0.8673 & 86.73 & 300 & 1.0 & 0.05 & 17.69 & 6.93 & 8.38 & 0.09024 & 2.74E-3 & 2.40\tabularnewline
SRII-LH-300-e1.0-J0.10 & 0.8673 & 86.73 & 300 & 1.0 & 0.10 & 27.15 & 6.93 & 8.54 & 0.09235 & 4.76E-3 & 2.64\tabularnewline
SRII-HL-300-e0.1-J0.05 & 6.938 & 8.673 & 300 & 0.1 & 0.05 & 14.63 & 8.27 & 4.82 & 0.07594 & 4.10E-4 & 1.17\tabularnewline
SRII-HL-300-e0.1-J0.10 & 6.938 & 8.673 & 300 & 0.1 & 0.10 & 14.75 & 8.27 & 4.83 & 0.07623 & 6.85E-4 & 1.18\tabularnewline
SRII-HL-300-e0.5-J0.05 & 6.938 & 8.673 & 300 & 0.5 & 0.05 & 14.91 & 8.27 & 4.85 & 0.07722 & 1.58E-3 & 1.22\tabularnewline
SRII-HL-300-e0.5-J0.10 & 6.938 & 8.673 & 300 & 0.5 & 0.10 & 23.26 & 8.27 & 4.93 & 0.07875 & 2.97E-3 & 1.28\tabularnewline
SRII-HL-300-e1.0-J0.05 & 6.938 & 8.673 & 300 & 1.0 & 0.05 & 23.50 & 8.27 & 4.88 & 0.07880 & 3.01E-3 & 1.28\tabularnewline
SRII-HL-300-e1.0-J0.10 & 6.938 & 8.673 & 300 & 1.0 & 0.10 & 27.15 & 8.27 & 5.12 & 0.08198 & 5.90E-3 & 1.45\tabularnewline
SRII-HH-300-e0.1-J0.05 & 6.938 & 86.73 & 300 & 0.1 & 0.05 & 17.12 & 8.79 & 6.74 & 0.09329 & 1.05E-3 & 2.97\tabularnewline
SRII-HH-300-e0.1-J0.10 & 6.938 & 86.73 & 300 & 0.1 & 0.10 & 17.20 & 8.79 & 6.75 & 0.09349 & 1.24E-3 & 2.99\tabularnewline
SRII-HH-300-e0.5-J0.05 & 6.938 & 86.73 & 300 & 0.5 & 0.05 & 17.27 & 8.79 & 6.75 & 0.09411 & 1.78E-3 & 3.06\tabularnewline
SRII-HH-300-e0.5-J0.10 & 6.938 & 86.73 & 300 & 0.5 & 0.10 & 23.22 & 8.79 & 6.81 & 0.09513 & 2.77E-3 & 3.21\tabularnewline
SRII-HH-300-e1.0-J0.05 & 6.938 & 86.73 & 300 & 1.0 & 0.05 & 17.58 & 8.79 & 6.77 & 0.09512 & 2.69E-3 & 3.19\tabularnewline
SRII-HH-300-e1.0-J0.10 & 6.938 & 86.73 & 300 & 1.0 & 0.10 & 27.15 & 8.79 & 6.91 & 0.09721 & 4.70E-3 & 3.50\tabularnewline
SRII-0H-300-e0.1-J0.05 & 0 & 86.73 & 300 & 0.1 & 0.05 & 17.12 & 4.10 & 10.12 & 0.08617 & 1.04E-3 & 1.99\tabularnewline
SRII-0H-300-e0.1-J0.10 & 0 & 86.73 & 300 & 0.1 & 0.10 & 17.21 & 4.10 & 10.13 & 0.08637 & 1.24E-3 & 2.01\tabularnewline
SRII-0H-300-e0.5-J0.05 & 0 & 86.73 & 300 & 0.5 & 0.05 & 17.27 & 4.10 & 10.13 & 0.08698 & 1.78E-3 & 2.06\tabularnewline
SRII-0H-300-e0.5-J0.10 & 0 & 86.73 & 300 & 0.5 & 0.10 & 23.22 & 4.10 & 10.19 & 0.08803 & 2.79E-3 & 2.16\tabularnewline
SRII-0H-300-e1.0-J0.05 & 0 & 86.73 & 300 & 1.0 & 0.05 & 17.69 & 4.10 & 10.15 & 0.08802 & 2.71E-3 & 2.14\tabularnewline
SRII-0H-300-e1.0-J0.10 & 0 & 86.73 & 300 & 1.0 & 0.10 & 27.15 & 4.10 & 10.31 & 0.09016 & 4.77E-3 & 2.36\tabularnewline
\hline 
\end{tabular}

\caption{Model parameters and resulting characteristics of SRII, with $M_{{\rm III}}=300\,M_{\odot}$.
Notes: The unit convention is the same as in Table \ref{tab:paramsII100}.
Highlighted in grey are those that fit the observed $C_{l}^{{\rm EE}}$
of the PLD best.\label{tab:paramsII300}}
\end{table*}

The SRI model adds a little more complexity to the characteristics
compared to the vanilla model (Fig. \ref{fig:SR1}). Due to the existence
of self-regulation, the SRI is expected to have more extended reionization
histories than the vanilla model. This has indeed been shown to be
the case for a consistent halo-selection criterion for HMACH and LMACH
(\citealt{Iliev2007}). However, one subtlety in our modelling scheme
complicates such an expectation. Because we use a constant mass criterion
in SRI, while a constant temperature criterion in the vanilla model,
we find that in some cases the duration of SRI models can be shorter
than that of vanilla models. Had we used the same halo selection criterion,
SRI models would have larger $\Delta z$ than vanilla models, which
we actually tested and confirmed. Aside from this complication which
is not essential, the general trend is that (1) the larger the value of
$g_{\gamma}^{{\rm L}}/g_{\gamma}^{{\rm H}}$ is, the larger the duration
of reionization becomes and (2) addition of LMACHs to HMACH-only
  scenarios extends
the duration of reionization. Also, cases with very large $g_{\gamma}^{{\rm L}}/g_{\gamma}^{{\rm H}}$
(e.g. SRI-0H\_F case in Fig. \ref{fig:SR1}) slows down reionization
significantly at the end of reionization, producing histories as symmetric
as the tangent-hyperbolic model that has been used extensively in
the analysis of the CMB data. It is easy to understand this behavior:
$g_{\gamma}^{{\rm L}}/g_{\gamma}^{{\rm H}}$ is a rough measure of
the relative contribution of LMACH to reionization to that of HMACH,
and the self-regulation becomes stronger as $x$ becomes larger. One
very extreme case is SRI-0H\_dF, which never finishes reionization
due to a strong self-regulation. The trend that $\Delta z$ is larger
in the dF scenario than the F scenario is the same as in the vanilla
model.  The general trend of the vanilla and SRI models can also
  be seen in Figure \ref{fig:Gamma_VSRI} in terms of $\xi$.

\begin{figure*}\centering
\includegraphics[width=0.45\textwidth]{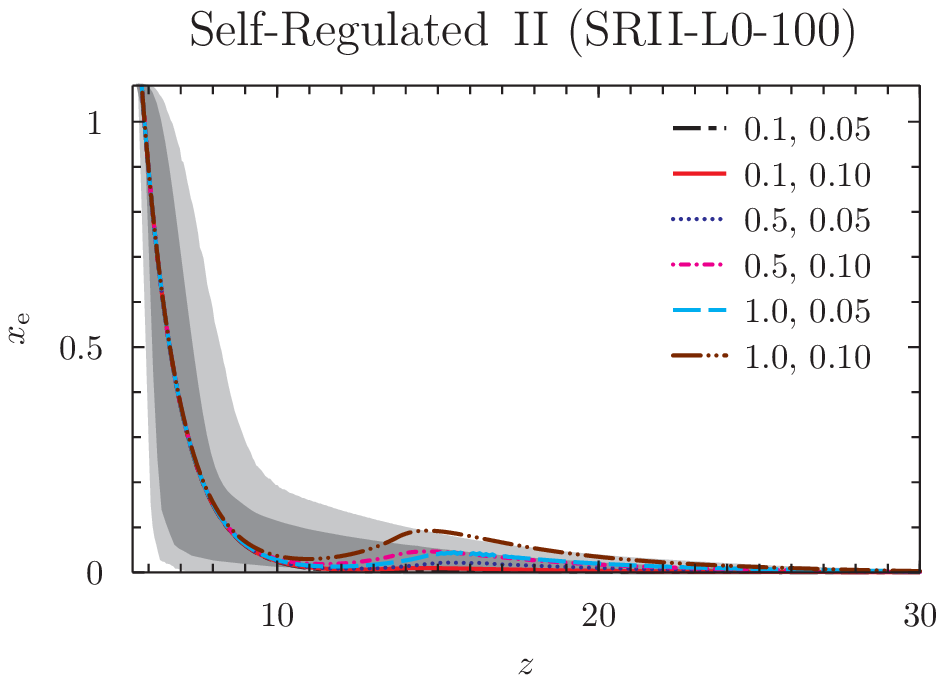}\hspace{0.3cm}\includegraphics[width=0.45\textwidth]{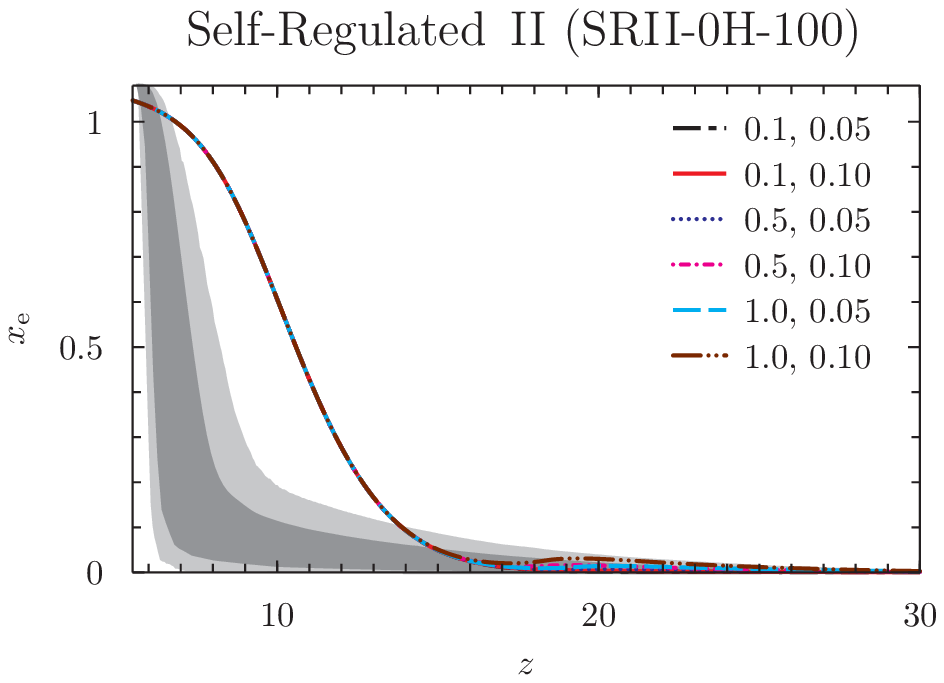}
  \vspace{0.3cm}

\includegraphics[width=0.45\textwidth]{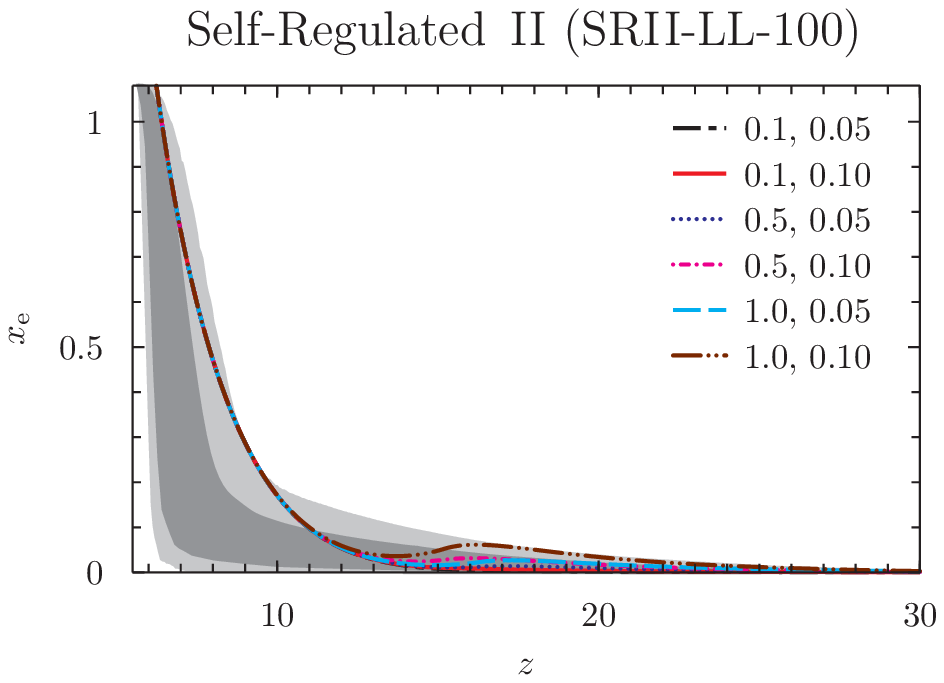}\hspace{0.3cm}\includegraphics[width=0.45\textwidth]{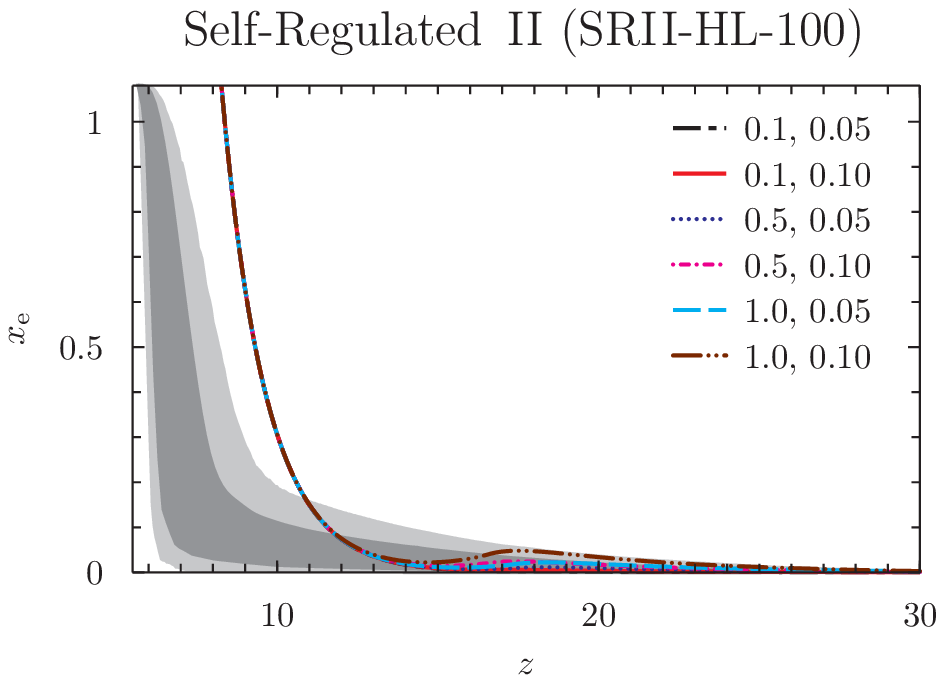}
  \vspace{0.3cm}

\includegraphics[width=0.45\textwidth]{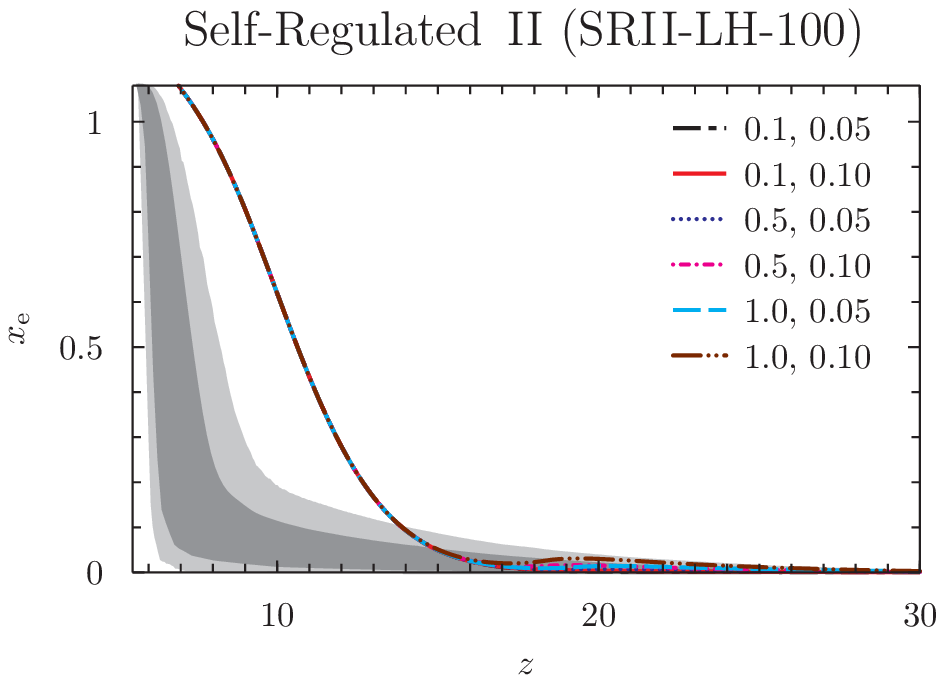}\hspace{0.3cm}\includegraphics[width=0.45\textwidth]{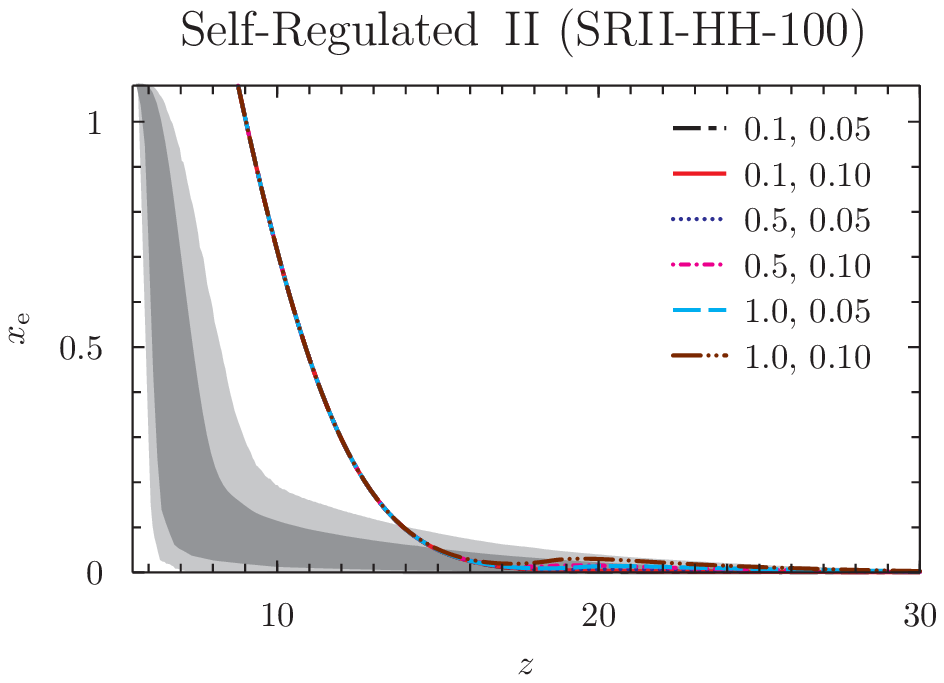}

\caption{Evolution of the global ionized fraction in the model category ``self-regulated
II'', with $M_{{\rm III}}=100\,M_{\odot}$ per MH. Each panel corresponds
to varying sets of $\{g_{\gamma}^{{\rm H}},\,g_{\gamma}^{{\rm L}},\,M_{{\rm III}}\}$,
denoted in the format ``SRII-$g_{\gamma}^{{\rm H}}g_{\gamma}^{{\rm L}}$-$M_{{\rm III}}/M_{\odot}$''.
The letters ``0'', ``L'' and ``H'' denote ``null'', ``low''
and ``high'' values of $g_{\gamma}$, respectively, in relative
sense (listed in Table \ref{tab:params}). In all panels, the set
of $f_{{\rm esc}}^{{\rm M}}$ and $J_{{\rm LW,th}}$ (in units of
$10^{-21}\,{\rm erg\,s^{-1}\,cm^{-2}\,Hz^{-1}\,sr^{-1}}$) are specified
as legends: \{0.1,~0.05\}, black dot-long-dashed; \{0.1,~0.10\},
red solid; \{0.5,~0.05\}, blue dotted; \{0.5,~0.10\}, magenta dot-short-dashed;
\{1.0,~0.05\}, cyan dashed; \{1.0,~0.10\}, brown dot-dot-dashed.
\label{fig:SR2}}
\end{figure*}

\begin{figure*}\centering

\vspace{0.3cm}
\includegraphics[width=0.45\textwidth]{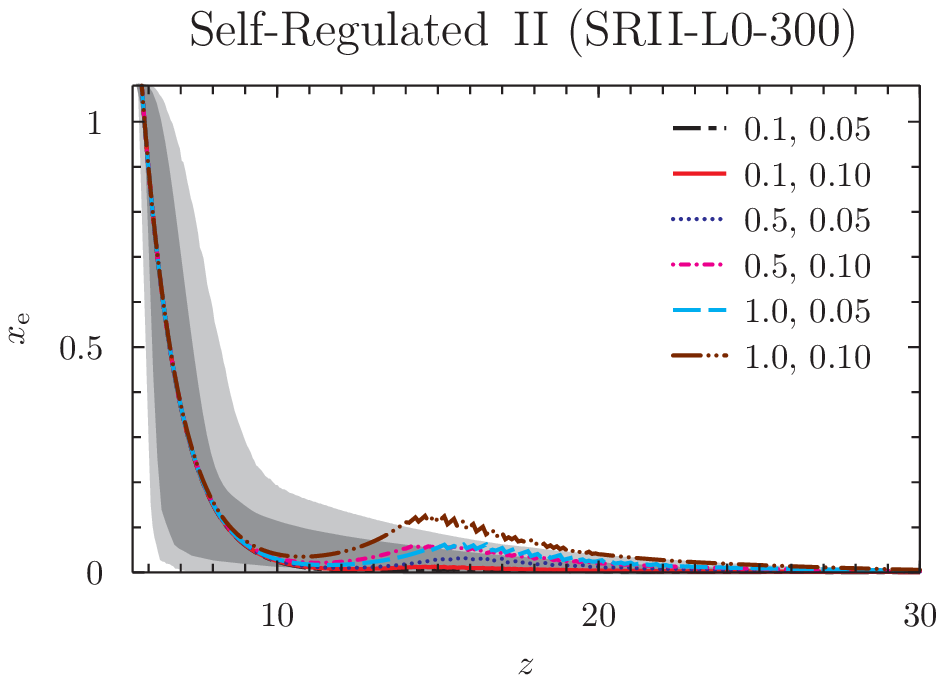}\hspace{0.3cm}\includegraphics[width=0.45\textwidth]{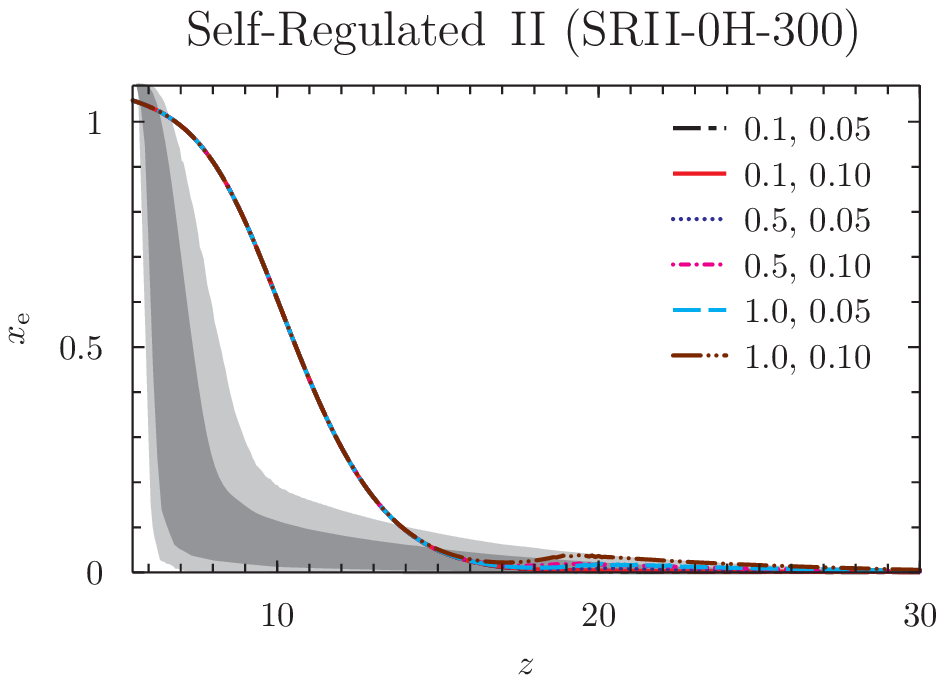}
  \vspace{0.3cm}

\includegraphics[width=0.45\textwidth]{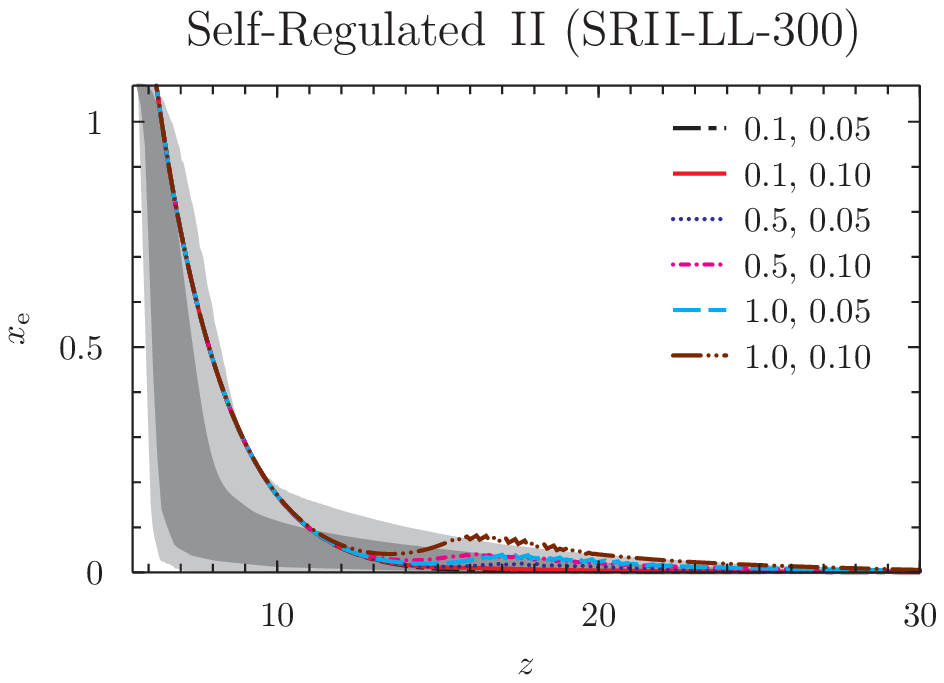}\hspace{0.3cm}\includegraphics[width=0.45\textwidth]{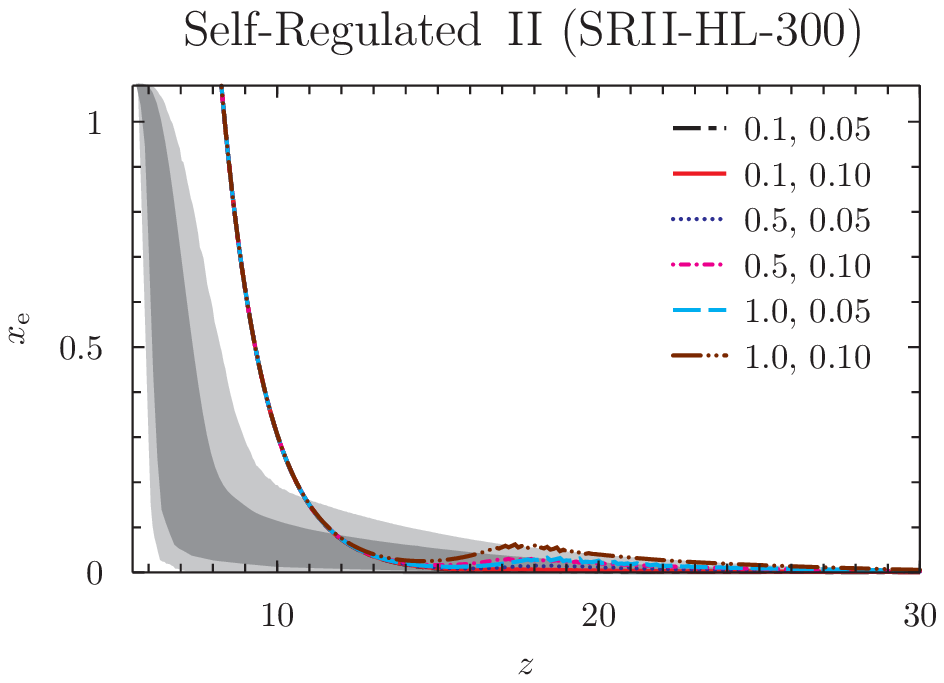}
  \vspace{0.3cm}

\includegraphics[width=0.45\textwidth]{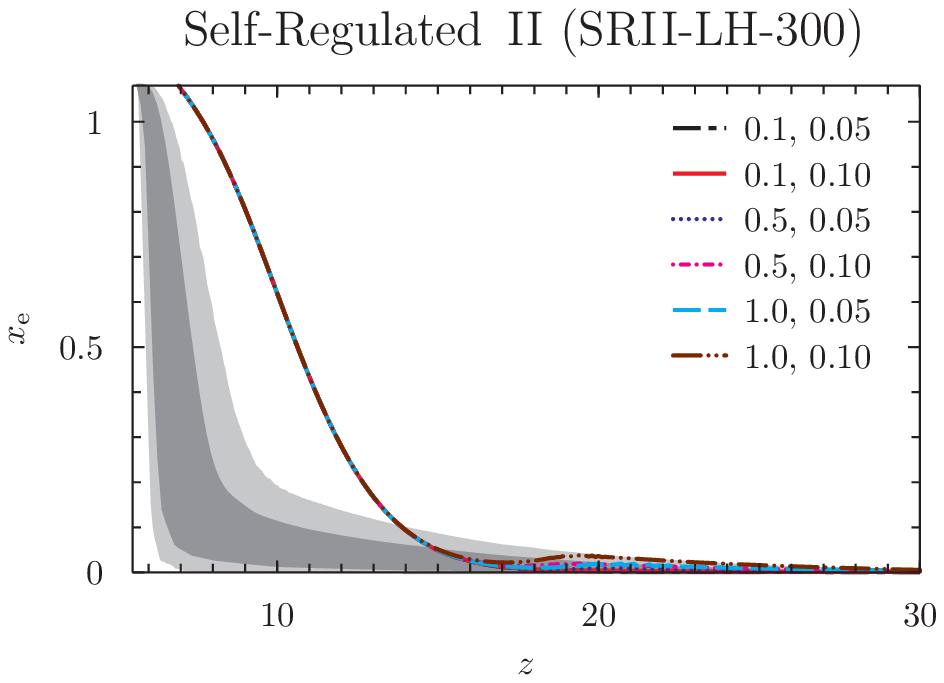}\hspace{0.3cm}\includegraphics[width=0.45\textwidth]{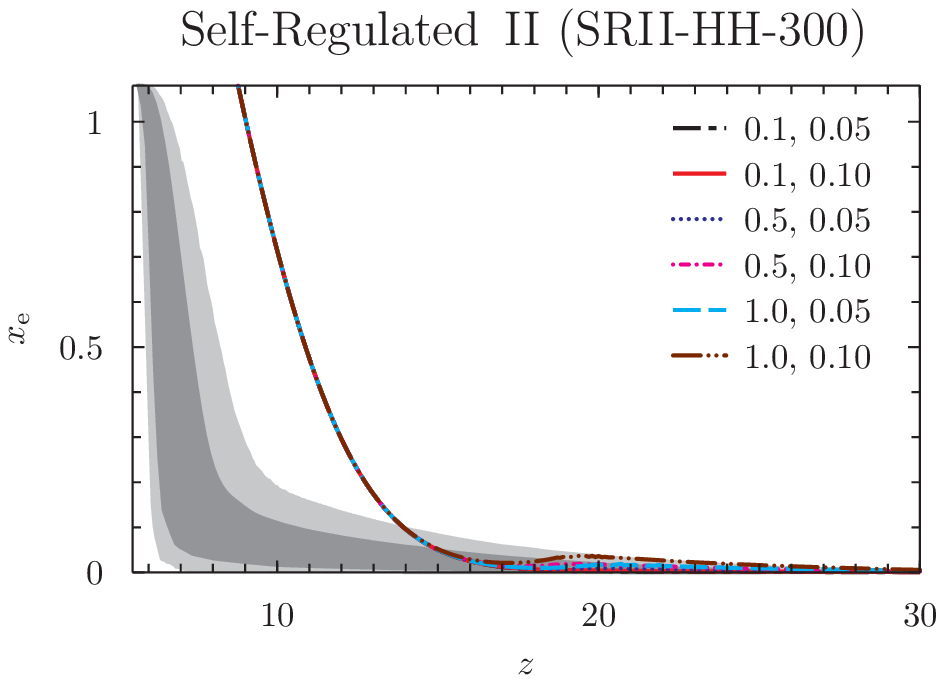}

\caption{Same as Fig \ref{fig:SR2} but with $M_{{\rm III}}=300\,M_{\odot}$.
Each panel corresponds varying $f_{\gamma}^{{\rm H}}$ and $M_{{\rm III}}$
as specified in the title. In all panels, the set of $f_{{\rm esc}}^{{\rm M}}$
and $J_{{\rm LW,th}}$ (in units of $10^{-21}\,{\rm erg\,s^{-1}\,cm^{-2}\,Hz^{-1}\,sr^{-1}}$)
are specified as legends using the same line convention as Figure
\ref{fig:SR2}. \label{fig:SR2-1}}
\end{figure*}

The SRII model has features richer than the vanilla and the SRI models
(Figures \ref{fig:SR2} and \ref{fig:SR2-1}). The most notable feature
is the existence of the early, extended and slowly-increasing phase
in $x$. This is due to the self-regulation of star formation, even
stronger than that in the SRI model, which takes place inside MHs.
\emph{The star formation inside MHs are mainly regulated by the LW
background} $J_{{\rm LW}}$, which quickly builds up to reach $J_{{\rm LW,\,th}}$.
Continuum photons below the Lyman limit and emitted at redshift $z$
travels a cosmological distance ($\sim100\{(1+z)/21\}^{-0.5}\,$cMpc),
in contrast to the hydrogen-ionizing UV photons that travel up to
the ionization front and then absorbed. Therefore, any newly forming
MHs will be under the influence of LW background long before being
exposed to the ionizing photons, and any pre-ionized region would
have been under the over-critical LW intensity ($J_{{\rm LW}}>J_{{\rm LW,\,th}}$).
We find that this is indeed the case: when tested with $\max(x_{{\rm LW}},\,x)$
replaced by $x_{{\rm LW}}$ in equation (\ref{eq:Gamma_mine_F}),
the resulting $x(z)$ was not affected. 

\begin{figure*}\centering
  
\includegraphics[width=0.4\paperwidth]{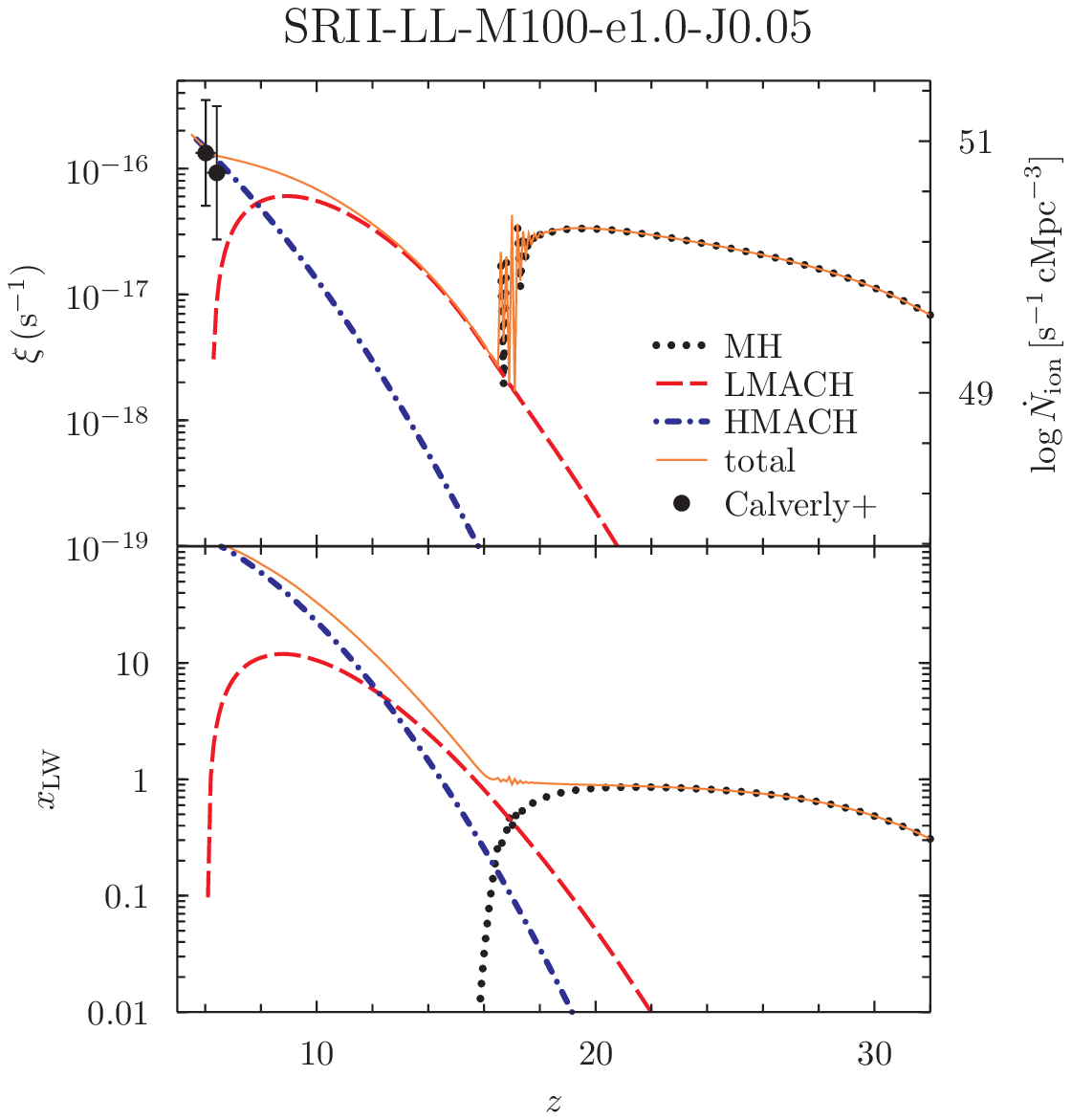}\hspace{0.3cm}\includegraphics[width=0.4\paperwidth]{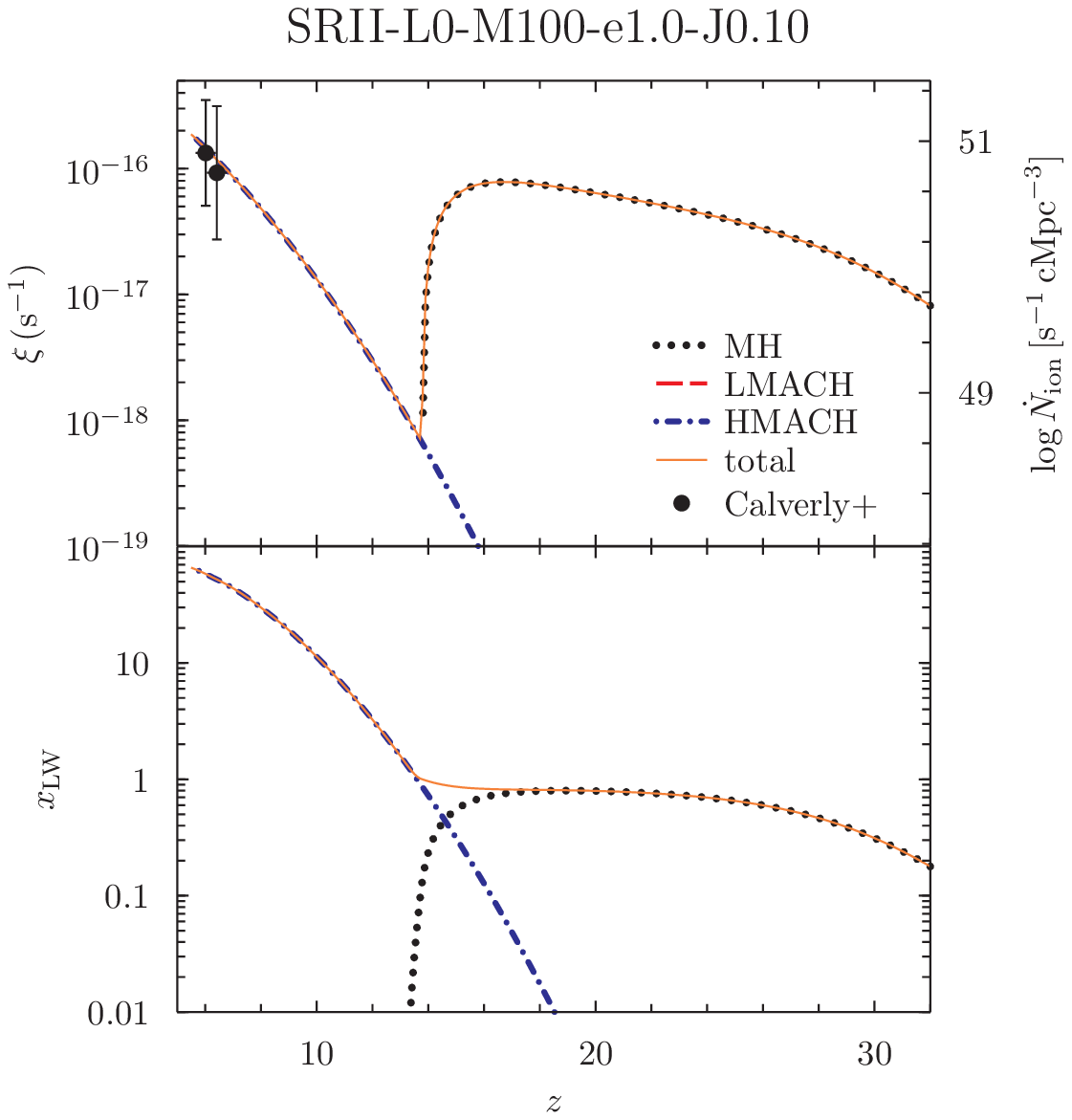}

\caption{ (top) Same as Figure \ref{fig:Gamma_VSRI},
and (bottom) the Lyman-Werner band intensity in terms of $x_{{\rm LW}}\equiv J_{{\rm LW}}/J_{{\rm LW,\,th}}$
(equation \ref{eq:xLW}), for two selected cases among PLD-favored
SRII models. Line convention is identical to Figure
\ref{fig:Gamma_VSRI}, and contributions by MHs (black, dotted) are
shown in addition. 
\label{fig:Gamma_SRII}}

\end{figure*}

Another notable feature of SRII is that, in some cases where the contribution
of ionizing photons by MHs is as significant as to drive $x$ beyond
$\gtrsim10\%$, there exists a phase where $x$ decreases in
time\footnote{Reionization histories of SRII shown in Figs. \ref{fig:SR2}--\ref{fig:SR2-1}
reach smaller values of the midway peak $x$ at $z\simeq15$ and the
recombination is stronger than matching models of \citet{Ahn2012}.
The individual H II regions created by MHs were too small to be numerically
resolved in the simulation of \citet{Ahn2012}, and thus the grid
cells with given resolution were partially ionized before ACHs emerged.
Recombination rate per hydrogen in each grid cell was calculated as
$\alpha C(n_{{\rm H}}+n_{{\rm He}})x^{2}$, even though the rate should
have been $\alpha C(n_{{\rm H}}+n_{{\rm He}})x$ instead (as in Eq
\ref{eq:dxvol_dt}) because UV-driven H II regions are practically
fully ionized and surrounded by neutral IGM. We experimentally calculated
$x(t)$ after changing the sink term in equation \ref{eq:dxvol_dt}
to $\alpha C(n_{{\rm H}}+n_{{\rm He}})x^{2}$, and could recover the
global ionization histories of \citet{Ahn2012} with matching parameters.
Therefore, the quantitative predictions of \citet{Ahn2012} need to
be modified to some extent or considered as models that have more
smooth transition of stars from Pop III to Pop II than SRII models
studied here.}.
This is mainly due to the fact that (1) the LW feedback renders
$J_{{\rm LW}}\simeq J_{{\rm LW\,th}}$ when MHs dominate as the main
radiation sources (Fig. \ref{fig:SR2_LW}) and (2) the large difference
in the number of soft-UV ($h\nu=\sim11-13.6\,{\rm eV}$) photons per
ionizing photon of Pop II and Pop III stars (see the detail in Section
\ref{subsec:Back_21cm}) and (3) the drop of $f_{{\rm esc}}$ from
MH values ($f_{{\rm esc}}\sim1$) to ACHS ($f_{{\rm esc}}\sim0.2$).
Then, as LMACHS (assumed to host Pop III stars) and HMACHs (assumed
to host Pop II stars) start to generate soft-UV photons to make up
$J_{{\rm LW}}$ near $J_{{\rm LW\,th}}$, which is achieved at the
expense of ACHs' putting out much less amount of ionizing photons
than minihalos (assumed to host Pop III stars), the IGM gains a chance
to recombine faster than ionization (Fig. \ref{fig:Gamma_SRII}:  see dips in $\xi$).
In practice,
however, this recombination is slight and not as dramatic as the ``double
reionization'' that has been suggested by \citet{2003ApJ...591...12C}.
Given the same set of $\left\{ g_{\gamma}^{{\rm H}},\,g_{\gamma}^{{\rm L}}\right\} $
as in SRI, the SRII model has $z_{{\rm end}}$ that is practically
identical to that of SRI while $\tau_{{\rm es}}$ and $\Delta z$
that are both boosted from those of SRI (Tables \ref{tab:params}
-- \ref{tab:paramsII300}). This is simply due to the existence of
additional photon sources, or MH stars, that ionize the IGM only to
a limited extent ($x\lesssim15\,\%$ at most under our parameter range
but may be increased if $M_{{\rm III}}$ and $J_{{\rm LW,th}}$ are
pushed to higher values) such that $z_{{\rm end}}$ is not much affected
but can increase $\tau_{{\rm es}}$ and extend the duration of reionization
substantially by strongly regulated ionization history.

\subsection{Comparison with CMB observations\label{subsec:Planck}}

The 2018 PLD provides, among improvements from the 2015 data, so far
the most precise measurement of the large-scale CMB polarization anisotropy.
The large-scale E-mode auto-correlation angular power spectrum, $C_{l}^{{\rm EE}}$,
is strongly affected by the history of reionization. The quadrupole
moment of the CMB anisotropy generates linear polarization after Thomson
scattering from the viewpoint of an electron, and the polarization
signal is observed after being modulated by the relevant wave-modes,
resulting in affecting $C_{l}^{{\rm EE}}$ mostly in the low-$l$
($\lesssim20$) regime (\citealt{hu97,1999ASPC..181..227H,2003moco.book.....D}).

The extended high-redshift ($z\gtrsim15$) ionization tail predicted
by \citet{Ahn2012} and SRII models here has been advocated by \citet{Miranda2017}
and \citet{Heinrich2018}, based on their principal component analysis
(PCA) of the Planck 2015 $C_{l}^{{\rm EE}}$ data observed through
the Low-Frequency Instrument (LFI). They claimed that a specific SRII
model with a substantial high-redshift tail, corresponding roughly
to L2M1J2 case of \citet{Ahn2012}, was favored over the vanilla model
at $\sim1\sigma$ level. Later, \citet{Millea2018} used both the
LFI and the HFI (proprietary at the time) data, with a well-handled
physicality ($x>0$) prior, to claim against too much contribution
from the $z\gtrsim15$ epoch. They constrained the optical depth from
$15\le z\le30$, or $\tau(15,30)$, to $\tau(15,30)<0.015$ at $2\sigma$
level. The Planck 2018 analysis based only on the low-$l$ E-mode
polarization further reduces this value to $\tau(15,30)<0.007$ at
$2\sigma$ level and $\tau_{{\rm es}}=0.0504_{-0.0079}^{+0.0050}$
at $1\sigma$ level (\citealt{PlanckCollaboration2018}).

\begin{figure*}\centering
  
\includegraphics[width=0.45\textwidth]{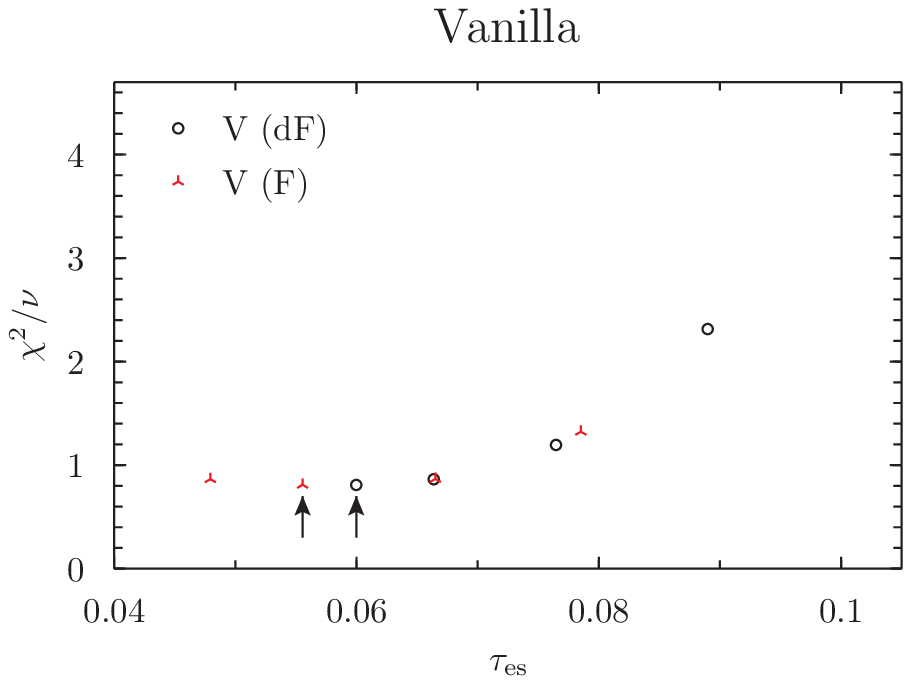}\hspace{0.3cm}\includegraphics[width=0.45\textwidth]{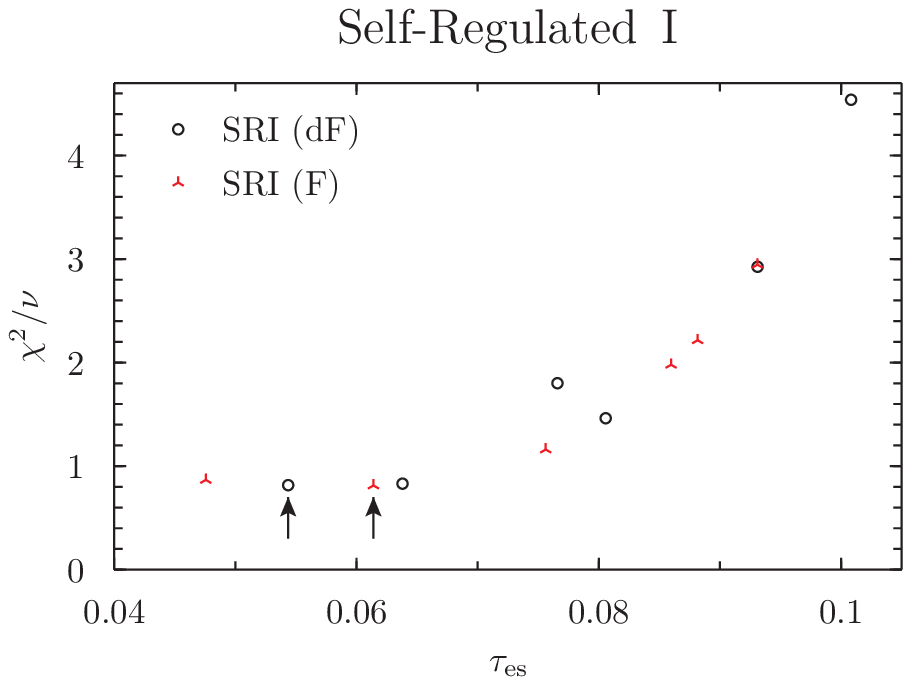}

\includegraphics[width=0.45\textwidth]{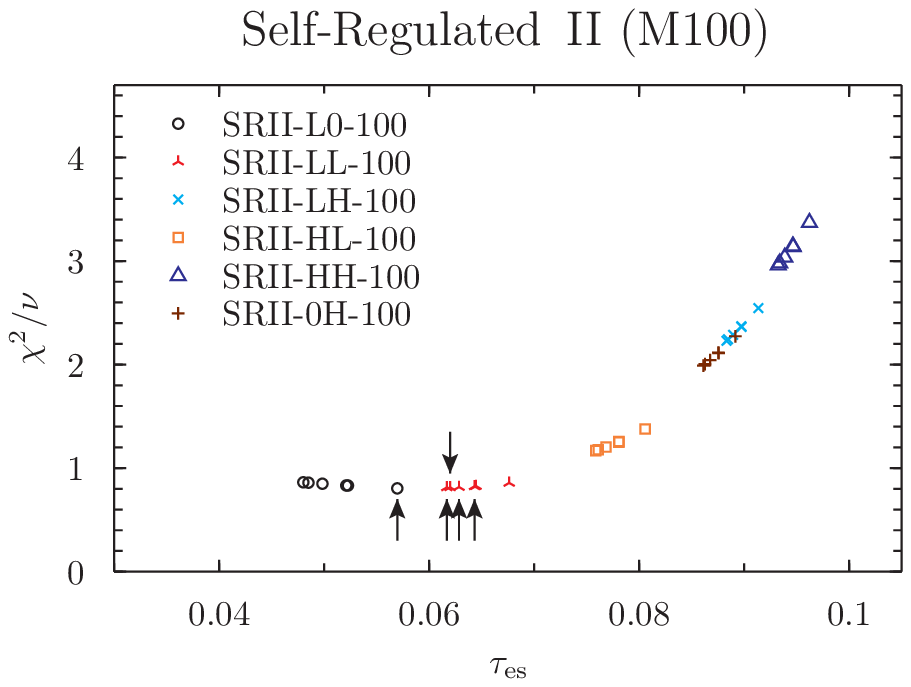}\hspace{0.3cm}\includegraphics[width=0.45\textwidth]{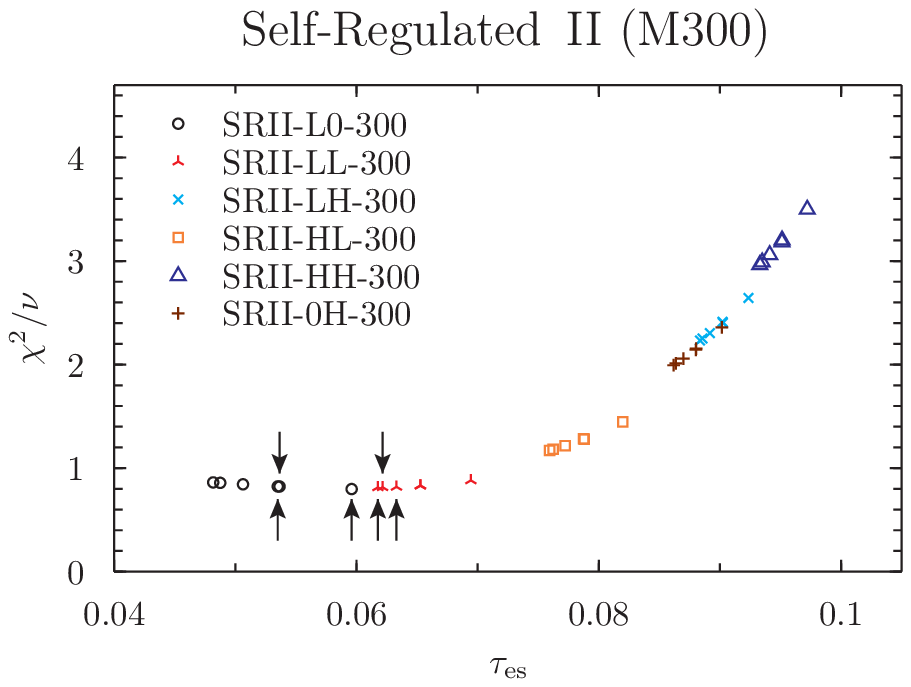}

\caption{$\chi^{2}/\nu$ of the E-mode power spectrum $D_{l}^{{\rm EE}}$ of
each model with respect to the Planck 2018 Legacy Data. In each panel,
plotted points are grouped in sub-categories specified by legends,
and arrows point to the minimum-$\chi^{2}/\nu$ cases in each panel.
Models marked by arrows correspond to those highlighted in Tables
\ref{tab:params} -- \ref{tab:paramsII300}.\label{fig:chi2}}
\end{figure*}

\begin{figure*}\centering
  
\includegraphics[width=0.4\paperwidth]{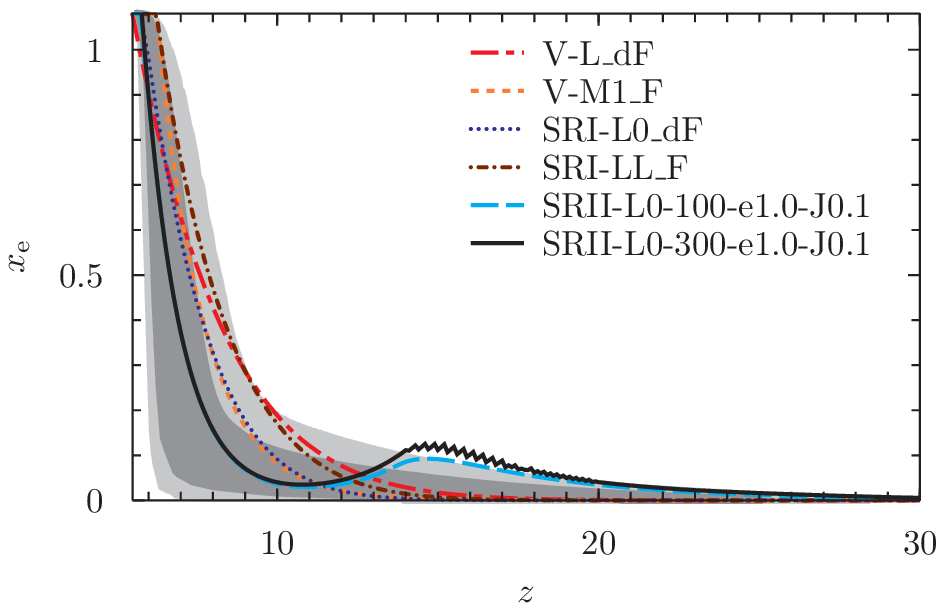}\hspace{0.3cm}\includegraphics[width=0.4\paperwidth]{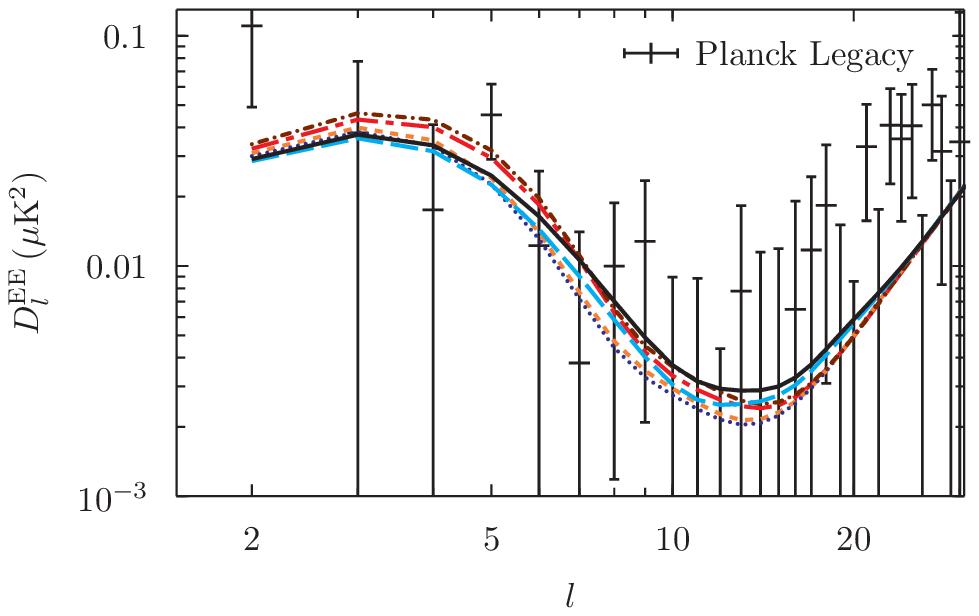}

\includegraphics[width=0.4\paperwidth]{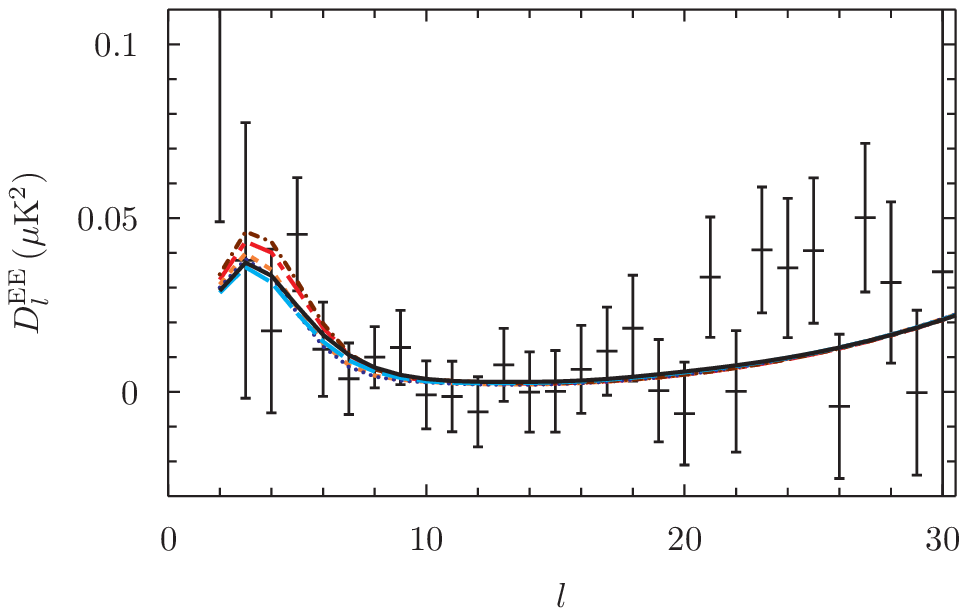}\hspace{0.3cm}\includegraphics[width=0.4\paperwidth]{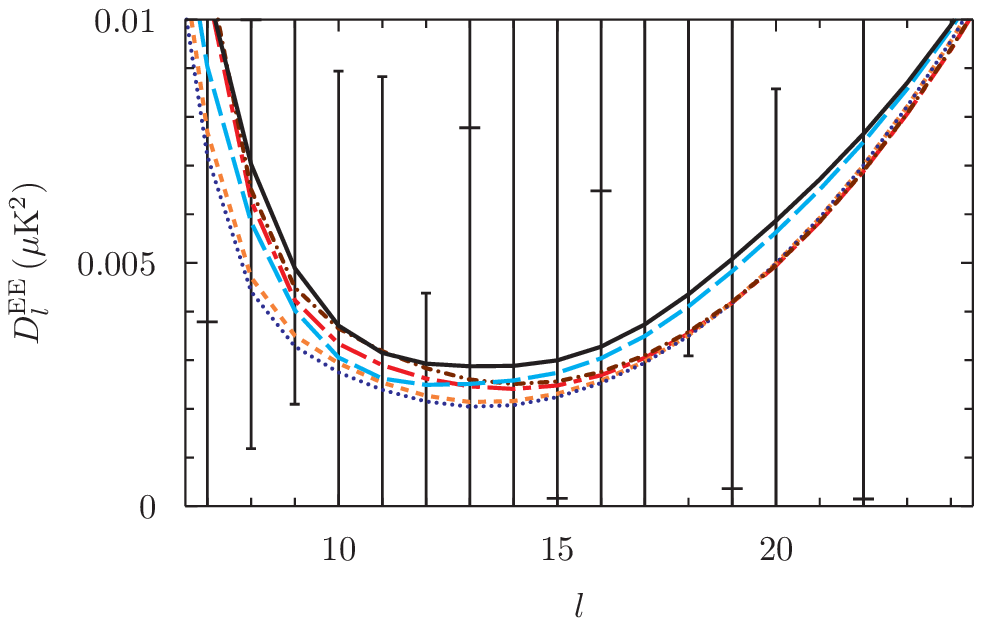}

\caption{Low-multipole E-mode auto-correlation power spectrums predicted by
models (lines with model specification on the top-left panel) with
smallest $\chi^{2}/\nu$ and that of PLD (points with $1\sigma$ error
bar). We select only two models with smallest $\chi^{2}/\nu$ from
SRII models, to avoid crowdedness of lines. CMB lensing is considered
in all cases, even though the net effect in this range of $l$ is
weak. Different plotting schemes (top-right: logarithmic; bottom-left:
linear; bottom-right: linear-zoomed) are used. SRII models with the
high-$z$ ionization tail achieve larger values of $D_{l}^{{\rm EE}}$
for $14\lesssim l\lesssim24$ than vanilla and SRI models, even though
all the models plotted here share about a single value of $\tau_{{\rm es}}$
($\simeq0.06$).\label{fig:ClEE}}
\end{figure*}

In light of the constraints described above, we compare our model $x(z)$'s
from Section \ref{subsec:Result-reion-history} to PLD. The main purpose
of this task is to (1) understand whether any class of our models
are preferred by observation and (2) whether the degeneracy of models
in $\tau_{{\rm es}}$ and $z_{{\rm ov}}$ can be broken. For example,
\citet{Ahn2012} showed that SRII models with an extended tail in
$x(z)$ could be distinguished from the vanilla- or SRI-type models,
even when the models have the same $\tau_{{\rm es}}$ (=0.085) and
$z_{{\rm ov}}$ (=6.8). The pictorial comparison of $x(z)$ to the
Planck constraint ($1\sigma$ and $2\sigma$ constraints shown in
shaded regions) is shown in Figs. \ref{fig:typical} -- \ref{fig:SR2-1}.
We test the relative goodness of several selected models by calculating
the reduced chi square,
\begin{equation}
\chi^{2}/\nu=\frac{1}{29}\sum_{l=2}^{30}\frac{(D_{l}^{{\rm EE}}-\tilde{D}_{l}^{{\rm EE}})^{2}}{\sigma_{l}^{2}},\label{eq:chi2}
\end{equation}
where $D_{l}^{{\rm EE}}\equiv l(l+1)C_{l}^{{\rm EE}}/(2\pi)$ and
$\tilde{D}_{l}^{{\rm EE}}\equiv l(l+1)\tilde{C}_{l}^{{\rm EE}}/(2\pi)$
are the E-mode power spectrums corresponding to a model $x(z)$ and
PLD respectively, and $\sigma_{l}$ is the standard deviation of $\tilde{D}_{l}^{{\rm EE}}$
due to the cosmic variance and the noise\footnote{$\tilde{D}_{l}^{{\rm EE}}$ and $\sigma_{l}$ are from 'COM\_PowerSpect\_CMB-EE-full\_R3.01.txt',
downloadable from the Planck Legacy Archive (\url{https://pla.esac.esa.int}). }. $C_{l}^{{\rm EE}}\equiv \left\langle \left|a_{lm}^{{\rm E}}\right|^{2}\right\rangle $
for given $l$ averaged over $m=[-l,\,l]$, with the spherical-harmonics
decomposition of the E-mode anisotropy $E(\theta,\,\phi)=\sum_{lm}a_{lm}^{{\rm E}}Y_{lm}(\theta,\,\phi)$.
In calculating $D_{l}^{{\rm EE}}$, we use a version of the Boltzmann
solver CAMB that was modified to allow a generic
shape of $x(z)$ (\citealt{Mortonson2008}, downloadable from \url{http://background.uchicago.edu/camb\_rpc/}). For the base cosmology,
we use the best-fit parameter set of PLD. While this is not a full
likelihood analysis including other data products such as the temperature
anisotropy, the value of $\chi^{2}/\nu$ from equation (\ref{eq:chi2})
can indicate the relative goodness of models because the impact of
reionization histories is the strongest in the E-mode (see e.g. an
identical approach by \citealt{Qin2020_CMB}). E-mode power spectrums
of selected models against the PLD are plotted in Fig. \ref{fig:ClEE}.

It is interesting to note that the constraint on reionization by
  the Planck observation provides a good match to the observed $\Gamma$
  (section \ref{subsec:Back_UV}). If the observed $\Gamma$
  (e.g. \citealt{Calverley2011}) is translated into $\dot{N}_{{\rm ion}}$
  with a reasonable set of physical parameters
  ($\lambda_{{\rm mfp}}=10\,{\rm cMpc}$:
  the IGM mean free path to H-ionizing photons at $z\simeq 6$, 
  and $\alpha_{{\rm s}}=\alpha_{{\rm b}}=2$: the spectral hardness of
  H-ionizing photons in equation (21) of \citealt{Bolton2007}),
  PLD-favored models have a good agreement with the observed $\Gamma$
  at  $z\simeq 6$ (Figures \ref{fig:Gamma_VSRI} and \ref{fig:Gamma_SRII}).
  This consistency between the two independent observations, even
  though uncertainties are large, is encouraging.

We now claim that some SRII models with substantial high-redshift
tails are still among those highly favored by PLD. Because $\chi^{2}/\nu$
is a measure of the goodness of a fit, we can use the value of $\chi^{2}/\nu$
to find the PLD-favored models. We indeed find many models can explain
the PLD low-$l$ $C_{l}^{{\rm EE}}$ fairly well even though the variance
in $\tau_{{\rm es}}$ of such models is substantial. Models that fit
the PLD $C_{l\le30}^{{\rm EE}}$ best are marked by arrows in Fig.
\ref{fig:chi2} and highlighted in Tables \ref{tab:params} -- \ref{tab:paramsII300}:
all these models have almost the same likelihood with $\chi^{2}/\nu=0.80-0.82$,
but with a substantial spread on $\tau_{{\rm es}}$ with $\tau_{{\rm es}}\simeq[0.0544,\,0.0643]$.
If allowance is extended to models with $\chi^{2}/\nu\le1$, then
the allowed optical depth becomes $\tau_{{\rm es}}\simeq[0.044,\,0.072]$.
This indicates that some of our SRII models are still well within
the PLD constraint and as much favored as those models without high-redshift
ionization tails. As seen in Tables \ref{tab:params} -- \ref{tab:paramsII300},
the most favored model with the least $\chi^{2}/\nu$, $\chi^{2}/\nu$=0.80,
are indeed SRII-L0-100-e1.0-J0.1 and SRII-L0-300-e1.0-J0.1, which
have a substantial high-redshift tail that reaches maximum $x=0.12$
at $z=13.6$ and $x=0.15$ at $z=13.6$, respectively. Such tails
contribute to $\tau_{{\rm es}}(15-30)$ substantially: $\tau_{{\rm es}}(15-30)=0.0063$
and 0.0078 for SRII-L0-100-e1.0-J0.1 and SRII-L0-300-e1.0-J0.1, respectively.
We also note that these maximum-likelihood models are clustered around
$\tau_{{\rm es}}=0.06$, substantially different from the inferred
value $\tau_{{\rm es}}=0.0504$ by PLD-$C_{l\le30}^{{\rm EE}}$. The
reason why such a difference occurs is unclear; this is nevertheless
a very important issue and a further investigation is warranted.

The model with the strongest ionization tail of all, SRII-L0-300-e1.0-J0.1,
is worthy of a close attention. Compared to the 2$\sigma$ constraint
of PLD, $\tau_{{\rm es}}(15-30)<0.007$, SRII-L0-300-e1.0-J0.1 actually
violates this constraint with $\tau_{{\rm es}}(15-30)=0.0078$ but
is still the best-fit (to PLD $C_{l\le30}^{{\rm EE}}$) of all the
models we tested. This model also has $\tau_{{\rm es}}=0.0596$, which
is about $2\sigma$ away from the E-mode only best-fit estimate by
PLD, $\tau_{{\rm es}}=0.0504$. If we do not consider other CMB observables
and assume a flat prior, we can conclude that this model is as good
as or just slightly better than other tail-less models with $\chi^{2}/\nu=0.81-0.82$.
It is interesting to see that there exists a weak tension between
$\tau_{{\rm es}}$'s estimated by the low-$l$ E-mode polarization
and the CMB lensing: the low-$l$ E-mode data of PLD prefers such
a low $\tau_{{\rm es}}\,(=0.0504_{-0.0079}^{+0.0050})$, while the
CMB lensing of PLD prefers higher $\tau_{{\rm es}}$ at around $\tau_{{\rm es}}\simeq0.08$
(\citealt{PlanckCollaboration2018}). This tendency of the CMB lensing
favoring large values of $\tau_{{\rm es}}$, even though the uncertainty
is large, is in par with favoring two-stage reionization models with
substantial ionization tails. Our findings are in slight disagreement
with the PLD constraint that was constructed using non-parametric
Bayesian inference. Based on our forward modelling and goodness-of-fit
approach, we argue that a family of models with a substantial high-redshift
ionization tail reaching $x_{{\rm max}}\sim0.12$ are still very strong
contenders at the moment just as those tail-less models.

\begin{figure*}\centering
  
\includegraphics[width=0.45\textwidth]{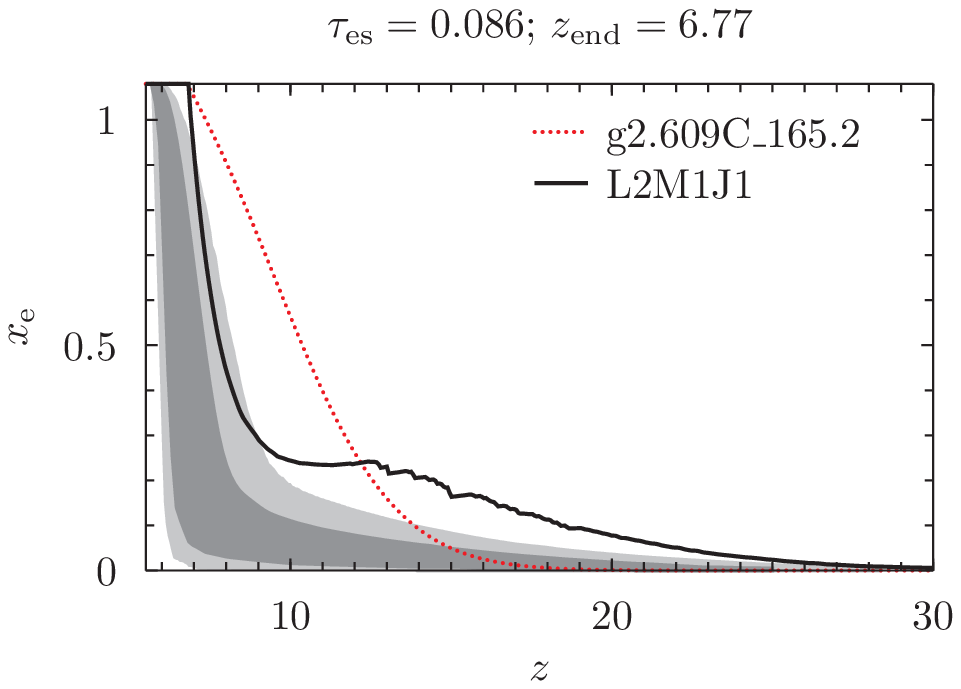}\hspace{0.3cm}\includegraphics[width=0.45\textwidth]{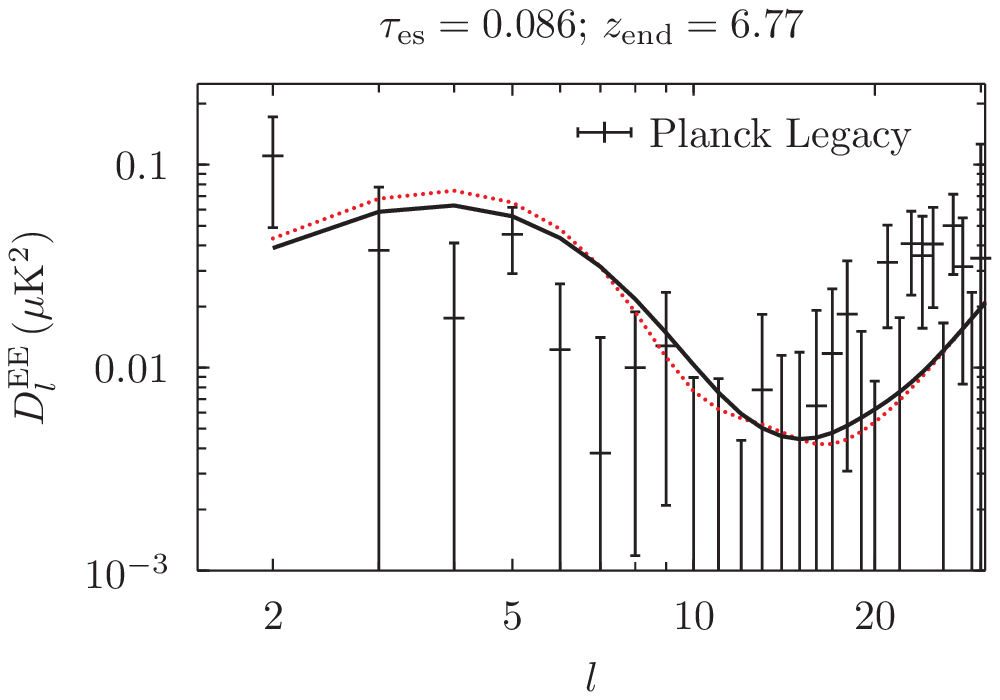}

\caption{Breaking degeneracy in $\tau_{{\rm es}}$ and $z_{{\rm end}}$ by
E-mode polarization observation. (left) $x(z)$'s of two selected
reionization models from \citet{Ahn2012}, sharing the same $\tau_{{\rm es}}$
($=0.086$) and $z_{{\rm end}}$ (=6.77) but are grossly different
in $x(z)$ with (L2M1J1: black, solid) and without substantial high-$z$
ionization tail (g2.609C\_165.2: red, dotted), are plotted against
the PLD constraint (grey shades). (right) E-mode polarization power
spectrums of the two models, against the PLD data (data points with
error bar), showing difference at $l\lesssim24$.\label{fig:break_degeneracy}}

\end{figure*}

Will there be a chance to probe a high-redshift ionization tail in
the future? The high-redshift tail tends to boost $C_{l}^{{\rm EE}}$
at $14\lesssim l\lesssim24$ (e.g. \citealt{Ahn2012,Miranda2017}):
SRII models in Fig. \ref{fig:ClEE} produce $C_{14\le l\le24}^{{\rm EE}}$
larger than that of the rest of models, and especially SRII-L0-300-e1.0-J0.1
has the strongest $C_{l}^{{\rm EE}}$ at $8\le l\le24$. In principle,
models degenerate in $\tau_{{\rm es}}$ can have different $C_{l<30}^{{\rm EE}}$'s
due to the variance in $x(z)$. \citet{Ahn2012}, using a principal
component analysis (PCA), had indeed predicted that high-precision
CMB observation could break the degeneracy in $\tau_{{\rm es}}$ and
probe (or disprove) the existence of the high-redshift ionization
tail. We stress this point again through Fig. \ref{fig:break_degeneracy},
showing two models from \citet{Ahn2012} that are degenerate both
in $\tau_{{\rm es}}$ and $z_{{\rm end}}$ but are clearly different
in $x(z)$, especially in the existence of the high-$z$ ionization
tail, and in the resulting $C_{l}^{{\rm EE}}$. 
From Figures
\ref{fig:ClEE} and \ref{fig:break_degeneracy}, we observe that the
boost of $C_{14\le l\le24}^{{\rm EE}}$ in two-stage reionization models with the high-$z$ ionization
tail against those tail-less models is a universal effect. 
As seen in Fig. \ref{fig:chi2}, the relation between $\tau_{{\rm es}}$ and
$\chi^{2}/\nu$ is not exactly monotonic but instead there exists
some scatter in $\chi^{2}/\nu$ for the same $\tau_{{\rm es}}$ and
vice versa. Such a scatter increases as $\tau_{{\rm es}}$ increases,
which is due to the increased freedom in constructing $x(z)$ for
given $\tau_{{\rm es}}$. However, because the PLD E-mode power spectrum
prefers such a low $\tau_{{\rm es}}$, as of now the leverage of having
a pronounced tail has somewhat diminished from that prediction. Nevertheless,
it is possible that observation by a more accurate apparatus might
find preference for higher $\tau_{{\rm es}}$ than Planck that are
still hampered by the large noise in measuring the polarization anisotropy.
Therefore, we need a better apparatus than Planck to (1) see whether
$\tau_{{\rm es}}$ could get larger than the estimate by PLD to allow
more pronounced two-stage reionization models and (2) break the model
degeneracy in $\tau_{{\rm es}}$ better than Planck to probe the ionization
tail even when the tail is weak.

We also briefly describe another type of constraint from CMB observations.
The kinetic Sunyaev-Zel'dovich effect can arise from the peculiar
motion of H II bubbles during EoR and can affect the small-scale ($l\sim$
a few thousands) temperature anisotropy power spectrum $C_{l}^{{\rm TT}}$.
Measurement of $C_{l}^{{\rm TT}}$ by the South Pole Telescope, especially
$C_{l=3000}^{{\rm TT}}$, (SPT: \citealt{Reichardt2012}) was used
by \citet{Zahn2012} to constrain the duration of reionization $\Delta z\equiv z(x=0.25)-z(x=0.99)$
to $\Delta z<4-7$ at $2\sigma$ level (depending on the assumed correlation
between the thermal Sunyaev-Zel'dovich effect and the cosmic infrared
background; see also the similar assessment by \citealt{Mesinger2012}
and \citealt{Battaglia2013}). Without MH stars, reionization occurs
always in a patchy way and thus any addition of electrons, or equivalently
extension of $\Delta z$, increases $C_{l=3000}^{{\rm TT}}$ monotonically,
as was assumed in \citet{Zahn2012}, \citet{Mesinger2012} and \citet{Battaglia2013}.
However, \citet{Park2013} re-addressed this issue with a variety
of reionization scenarios including the SRII-type, and found that
the added duration of reionization beyond this limit could still be
accommodated by the measured $C_{l=3000}^{{\rm TT}}$. As claimed in
\citet{Park2013}, H II regions by MHs are distributed almost uniformly
(\citealt{Ahn2012}) and thus the increase in $\Delta z$ in SRII
models does not guarantee an increase in $C_{l=3000}^{{\rm TT}}$.
Therefore, the largeness of $\Delta z_{3-97}\equiv z(0.03)-z(x=0.97)$
of many SRII models (see e.g. those highlighted in Tables \ref{tab:paramsII100}
and \ref{tab:paramsII300}) should not be considered as a violation
of such a constraint. Instead, constraining $\Delta z$ using the
small-scale $C_{l}^{{\rm TT}}$ should be restricted to only a limited
set of models without MHs.

\subsection{21 cm background and comparison with EDGES observation\label{subsec:Result-21cm}}

The main variants determining $\delta T_{b}$ are the X-ray heating
efficiency and the Ly$\alpha$ intensity, which determine $T_{{\rm K}}$
and $x_{\alpha}$, respectively. The X-ray efficiency is not a direct
product of the stellar radiation and is thus the main cause of the
uncertainty in $\delta T_{b}$. The Ly$\alpha$ intensity, on the
other hand, is almost solely determined by the stellar radiation and
is closely related to the ionizing PPR and the SED. We do not consider
the creation of Ly$\alpha$ photons due to the excitation of H atoms
by the X-ray-induced electrons, which is a good approximation unless
the X-ray efficiency is extremely high ($f_{x}\gg1$ with $f_{x}$
in equation \ref{eq:fX}). For the X-ray efficiency, we use the common
parameter $f_{X}$ (\citealt{2006MNRAS.371..867F,Mirocha2014}), defined
as the fudge parameter connecting the comoving X-ray luminosity density
$\mathcal{L}_{\nu}$ ($=h\nu\mathcal{N}_{\nu}$; in ${\rm erg\,{\rm s^{-1}\,Hz^{-1}\,cMpc^{-3}}})$
to SFRD (Section \ref{subsec:Gamma_F_dF}): 
\begin{equation}
f_{X}=\frac{\mathcal{L}_{\nu}}{c_{X}{\rm SFRD}},\label{eq:fX}
\end{equation}
where the additional proportionality coefficient $c_{X}$ is fixed
to $c_{X}=3.4\times10^{40}\,{\rm erg\,s^{-1}\,(M_{\odot}\,yr^{-1})}$,
an extrapolation of the 2--10 keV relation between $\mathcal{L}_{\nu}$
and SFRD (or equivalently between $L_{\nu}$ and SFR on average galaxies)
by \citet{Grimm2003} to $h\nu\ge0.2\,{\rm keV}$. Here, we limit
$\mathcal{L}_{\nu}$ to the energy range $h\nu=[0.2,\,30]\,{\rm keV}$
and a power-law SED $\mathcal{L}_{\nu}\propto\nu^{-1.5}$. For the
Ly$\alpha$ intensity and the LW intensity, we take a simple distinction
between Pop II and Pop III stars. Pop III stars are assumed to have
$N_{{\rm ion}}=50000$, $N_{\alpha L}=4800$ and $N_{\beta L}=2130$,
where $N_{\alpha L}$ and $N_{\beta L}$ are the number of photons
emitted by a stellar baryon during the stellar lifetime in the energy
range from Ly$\alpha$ to LL and from Ly$\beta$ to LL, respectively.
Pop II stars are assumed to have $N_{{\rm ion}}=6000$, $N_{\alpha L}=9690$
and $N_{\beta L}=3170$. This makes $N_{\alpha L}/N_{{\rm ion}}$
and $N_{\beta L}/N_{{\rm ion}}$ of Pop III stars about an order of
magnitude smaller than those of Pop II stars, respectively. $N_{\alpha L}/N_{{\rm ion}}$
and $N_{\beta L}/N_{{\rm ion}}$ strongly affect $J_{{\rm LW}}$ (equation
\ref{eq:JLW}) and $N_{\alpha}$ (equation \ref{eq:Nalpha}) for given
PPR. We use the following SED conventions for each category of models:
\begin{itemize}
\item Vanilla model: Pop II SED
\item SRI model: Pop III SED for LMACH; Pop II SED for HMACH
\item SRII model: Pop III SED for LMACH and MH; Pop II SED for HMACH
\end{itemize}
The claimed detection of $\sim500\,{\rm mK}$ absorption dip around
$\nu\simeq78\,{\rm MHz}$ by the EDGES has been a matter of debate,
mainly due to the fact that it is impossible to explain such a large
amplitude in the standard $\Lambda$CDM framework, if the background
after a successful foreground removal is composed only of the CMB
and the 21cm background. In the $\Lambda$CDM universe the kinetic
temperature of the IGM is limited to the adiabatically cooled value
($T_{{\rm K}}\sim10.2\,{\rm K}[(1+z)/21]^{2}$), and even at the maximum
Ly$\alpha$ coupling is limited to $\delta T_{b}\lesssim200\,{\rm mK}$.
Another difficulty faced by the EDGES result is the existence of a
peculiar spectral shape in $\delta T_{b}$, a flat trough of $\delta T_{b}$
from $\nu=72$ to $85\,{\rm MHz}$ and lines connecting to the ends
of the trough from $\nu=65$ and $92\,{\rm MHz}$, which is in contrast
with a smooth dip predicted by models in the $\Lambda$CDM.

We show our model predictions on $\delta T_{b}$ for a selected set
of models, and compare these with the EDGES result. The selection
criterion is the goodness of model fits to the PLD, and we use those
minimum-$\chi^{2}/\nu$ models marked by arrows in Fig. \ref{fig:chi2}.
We also tune $f_{X}$ to produce the largest absorption dip for each
model but under the condition $\delta T_{b}>0$ at $z\lesssim9$,
to (1) comply with the EDGES result with the deepest absorption possible
and (2) compensate for our ignorance of the Ly$\alpha$ heating which
might naturally turn the 21 cm background into emission before the
end of reionization (\citealt{Chuzhoy2007,Ciardi2007,Mittal2020}\footnote{We note
  that \citet{Ghara2020} claims that the Ly$\alpha$ heating 
  is efficient enough to render $\delta T_{b}>0$ before the end
  of reionization. Even though this claim agrees with that of
  \citet{Chuzhoy2007} qualitatively, the heating rate by
  \citet{Ghara2020} is wrongfully
  overestimated and should be reduced by about an order of
  magnitude as clarified by \citet{Mittal2020}. Except for the assumed
  SFRD and the SED, \citet{Mittal2020} is basically identical to
  and a reproduction of \citet{Chuzhoy2007}.})
and some hints
of the IGM heating
at $z\sim9$ (\citealt{Monsalve2017,Singh2018,Mertens2020,Ghara2020a}).
\citet{Chuzhoy2007} first showed that the Ly$\alpha$-recoil heating
can solely increase $T_{{\rm K}}$ beyond $T_{{\rm CMB}}$ before reionization
is completed, correcting the estimate by \citet{2004ApJ...602....1C}.
This way, we show how far off each reionization model is from the
EDGES result even when the maximum absorption is achieved in each
model. One can be more inclusive in model selection because future
CMB observations will probe the CMB polarization with better accuracy;
nevertheless we stick to this choice here.

\begin{figure*}\centering
  
\includegraphics[width=0.44\textwidth]{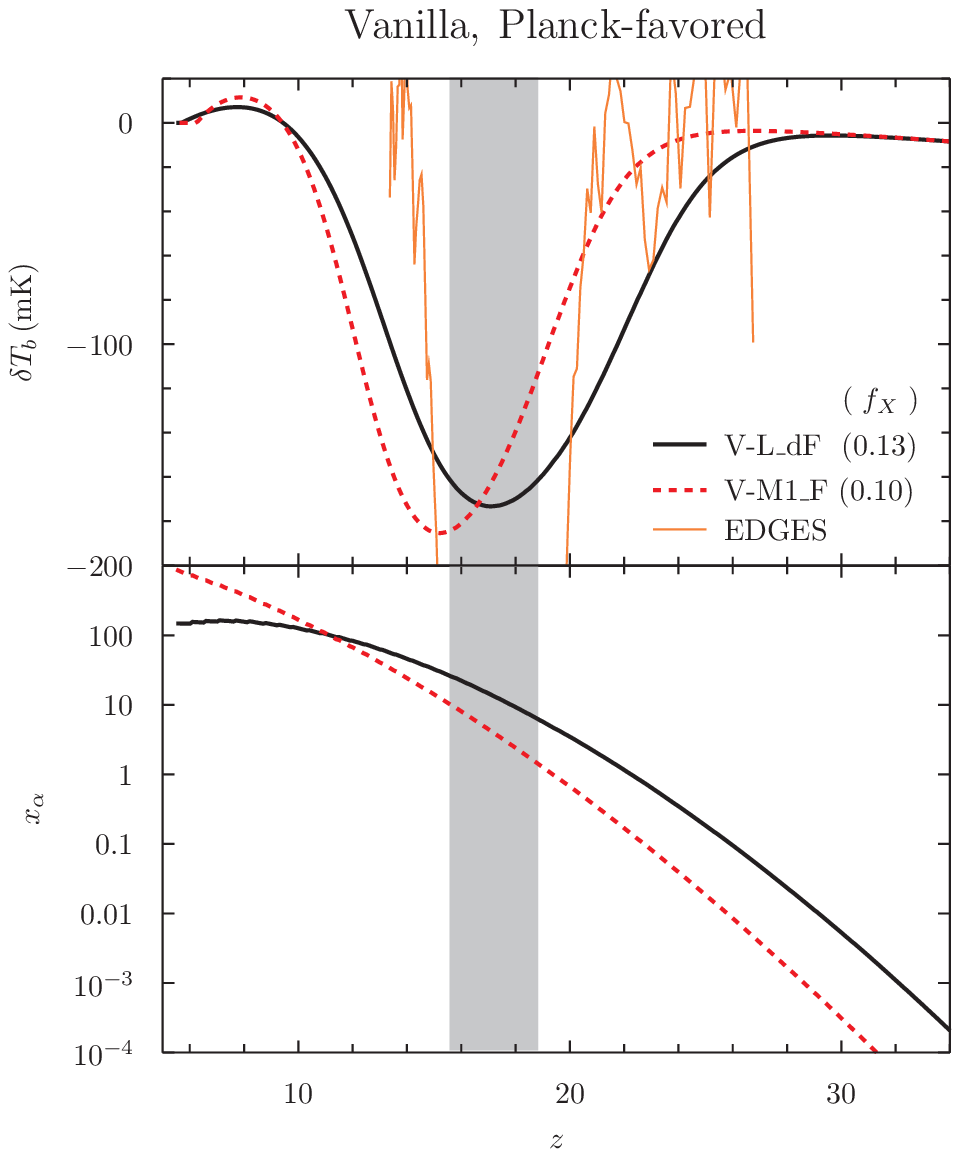}\hspace{0.3cm}\includegraphics[width=0.44\textwidth]{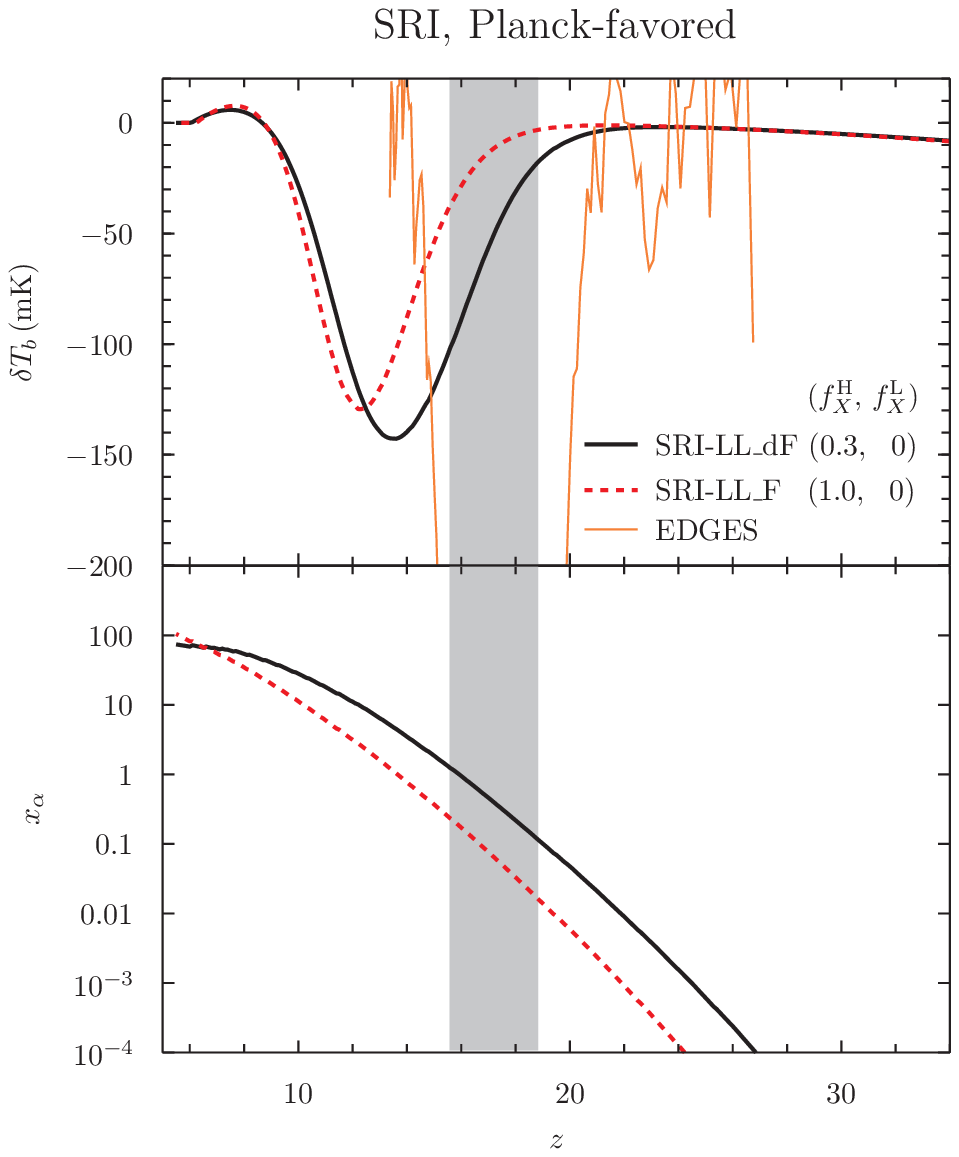}

\caption{21 cm backgrounds (top) and the Ly$\alpha$ coupling coefficients
(bottom) of models that fit PLD best (those marked by arrows in Fig
\ref{fig:chi2}), selected from the vanilla and SRI. The shade indicates
the redshift bin of the 500-mK absorption trough claimed by the EDGES
observation. The EDGES data is shown in orange, thin solid line. \label{fig:21cm}}
\end{figure*}

\begin{figure*}\centering
\includegraphics[width=0.7\textwidth]{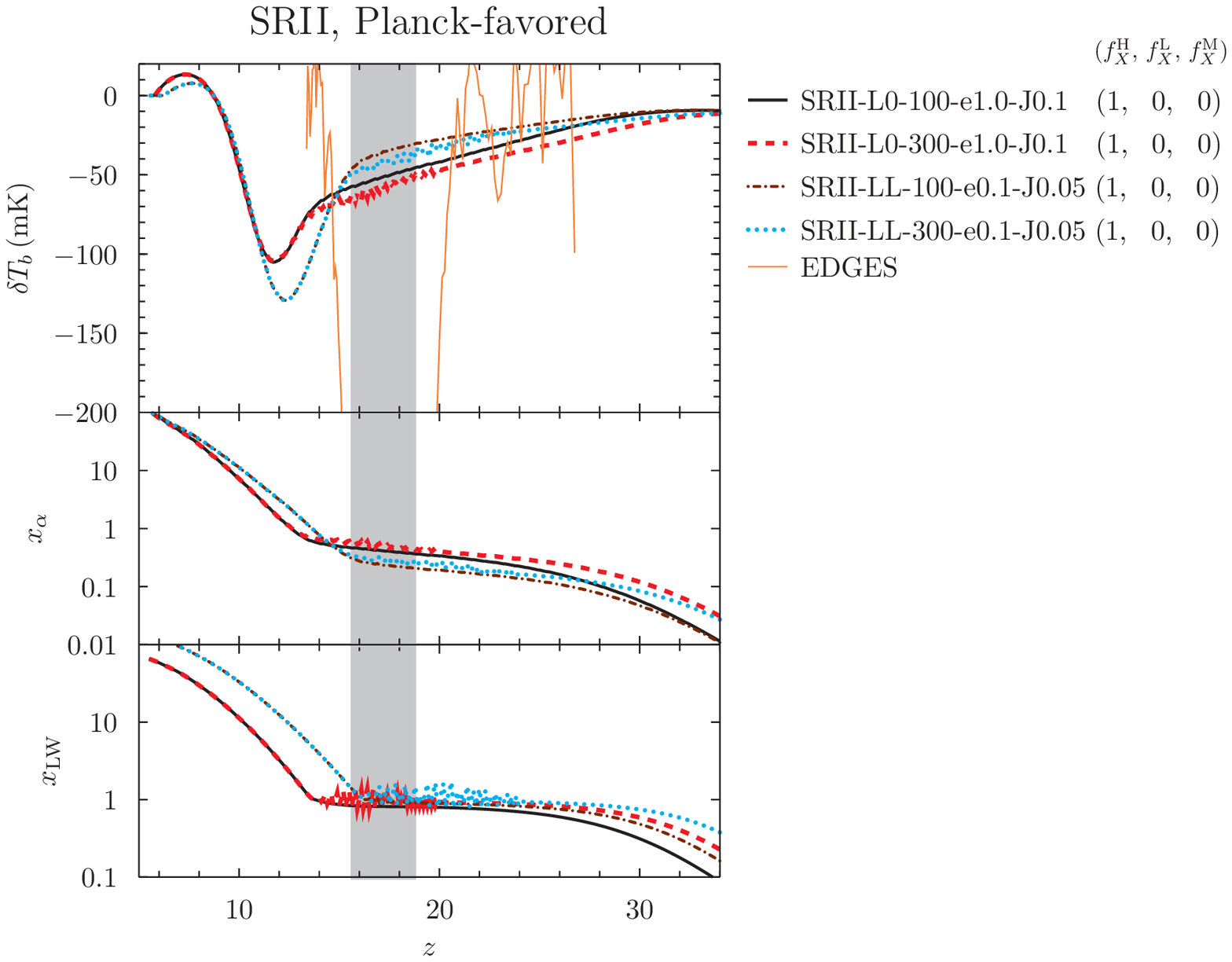}

\caption{21 cm backgrounds (top), the Ly$\alpha$ coupling coefficients (middle)
and $x_{{\rm LW}}$ ($=J_{{\rm LW}}/J_{{\rm LW,th}}$) of models that
fit PLD best (those marked by arrows in Fig \ref{fig:chi2}), selected
from SRII. In order to avoid crowdedness, we take only 4 models among
those that match PLD best. Regardless of the difference in values
of $J_{{\rm LW,th}}$, MHs self-regulate star formation such that
$x_{{\rm LW}}\simeq1$ is maintained after $J_{{\rm LW}}$ reaches
$\sim J_{{\rm LW,th}}$. This tendency continues until ACHs take over
to dominate in contributing to $J_{{\rm LW}}$ by generating stars
without being hindered by the LW feedback. At the same time, this self-regulation
limits $x_{\alpha}\lesssim0.5$, with additive dependence on $f_{{\rm esc}}^{{\rm M}}$.\label{fig:SR2_LW}}
\end{figure*}

Different model categories show distinctive features in $\delta T_{b}$
(Fig. \ref{fig:21cm}, with the shade indicating the redshift bin
of the EDGES absorption trough of $\sim500\,{\rm mK}$), as follows. 
\begin{itemize}
\item The Planck-favored vanilla models show the familiar $\sim180\,{\rm mK}$
absorption dip. The moment of the absorption dip and the start of
the absorption due to the Ly$\alpha$ pumping (to be distinguished
from the absorption due to collisional pumping at $z\gtrsim25$) are
delayed in the dF case compared to the F case. The dip resides at
$z\simeq16$ and $14.5$ for dF and F cases, respectively. The start
of the absorption are at $z\simeq26$ and $21$ for dF and F cases,
respectively.
\item The Planck-favored SRI models show weaker absorption dips, with $\delta T_{b,\,{\rm min}}\simeq[-120,\,-140]\,{\rm mK}$,
than the vanilla models. The dF case shows delays in the moments of
the absorption dip and the start of the absorption compared to the
F case, just as in the vanilla models. Compared to the vanilla models,
both the dip and the start of the absorption are delayed: the dip
is at $z\simeq12$ (F) -- $13$ (dF), and the absorption starts at
$z\simeq18$ (F) -- $20$ (dF). 
\item The Planck-favored SRII models show a very slowly deepening absorption
``slope'' during $28\gtrsim z\gtrsim14$, which is the epoch about
the same as the full EDGES-low observational window, before the absorption
dip at $z\simeq11-12$ occurs. The absorption dip is with $\delta T_{b,\,{\rm min}}\simeq[-90\,({\rm L0}),\,-120\,({\rm LL})]\,{\rm mK}$,
located away from the EDGES trough window. Where the EDGES trough
exists, the models have the limited differential brightness temperature,
$\delta T_{b}\simeq[-30,\,-60]\,{\rm mK}$. 
\end{itemize}
Several details of these results are noteworthy. (1) The $\sim180\,{\rm mK}$
$\left|\delta T_{b,\,{\rm min}}\right|$ of the vanilla model is roughly
the maximum amplitude allowed in the $\Lambda$CDM cosmology (\citealt{2006MNRAS.371..867F,Mirocha2014,Bernardi2015}).
(2) The main cause for the overall delays in the dip and the start
of the absorption for the SRI model relative to the vanilla model
is the fact that LMACHs, which dominate the early phase of reionization,
are assumed to host Pop III stars. Because $(N_{\alpha L}/N_{{\rm ion}})_{{\rm Pop\,III}}$
is much smaller than $(N_{\alpha L}/N_{{\rm ion}})_{{\rm Pop\,II}}$
at a given level of $x$, $N_{\alpha}$ and $x_{\alpha}$ of SRI models,
while LMACHs dominate, also become much smaller than those of the
vanilla models that use the Pop II SED. Such a smallness of $x_{\alpha}$
then renders $T_{{\rm S}}$ to couple only weakly to $T_{{\rm K}}$,
bringing down the amplitude of $\delta T_{b}$ (Fig \ref{fig:21cm}).
(3) Because of the reason same as (2), had we used the Pop III SED
for the vanilla model, we would have gotten delays in the dip and
the start of the absorption just similar to the SRI case. Accordingly,
the amplitude of the absorption dip would have been reduced. \citet{2006MNRAS.371..867F}
tested the vanilla model using both the Pop II and Pop III SEDs to
observe this tendency. (4) The Ly$\alpha$ pumping efficiency of SRII
in the EDGES absorption trough window is almost constant at the restricted
value $x_{\alpha}\simeq0.1-0.5$, and $x_{\alpha}$ is bound to $<1$
at $z\gtrsim14$. This is mainly caused by the combined effect of
the strongly self-regulated SFRD by the LW feedback and the smallness
of $(N_{\alpha L}/N_{{\rm ion}})_{{\rm Pop\,III}}$, both being relevant
to MHs that dominate this epoch.

Compared to the signal interpreted by the EDGES team, none of the
the global $\delta T_{b}$'s of our models can match the EDGES' one
in the amplitude and the spectral shape. Among the tested models,
V-L\_dF provides $\delta T_{b}$ closest to the EDGES data. However,
this is as closest as one can get to the amplitude of the EDGES absorption
trough in the $\Lambda$CDM, because the case is tuned to produce
the coldest $T_{{\rm K}}$ and the largest $x_{\alpha}$, with $x_{\alpha}\simeq5-20$
in the EDGES window, with a reionization history well under the PLD
constraint. The SRI models are terrible in matching the EDGES data,
mainly due to the shift of the absorption dip caused by the nature
of the Pop III SED of LMACH stars dominating this era. The SRII models
are as bad as SRI models in matching the EDGES data within the trough
window, mainly due to the dominance of MH stars with Pop III SED.
Note that all the cases are tuned to produce maximum possible $\left|\delta T_{b}\right|$
in absorption with $f_{X}^{{\rm L}}=0$ (SRI) and $f_{X}^{{\rm L}}=f_{X}^{{\rm M}}=0$
(SRII) with non-zero $f_{X}^{{\rm H}}$. Any addition of non-zero
X-ray heating will increase $T_{{\rm K}}$ from these null-heating
cases to reduce $\left|\delta T_{b}\right|$ and worsen the mismatch
between the EDGES' interpretation and the theory.

The SRII models produce $\delta T_{b}$ that is the most peculiar
in the spectral shape, because across the full EDGES window ($14\lesssim z\lesssim24$)
$\delta T_{b}(z)$ is almost featureless without much variation. It
would be even possible that the process of foreground-removal, utilizing
the spectral smoothness of the foreground, from the observed signal
could completely remove the true EoR (or Dark Ages) signal at $z\gtrsim14$
or $\nu\lesssim95\,{\rm MHz}$ to yield only a null result. The limited
amplitude of the absorption depth, $\left|\delta T_{b}(z\gtrsim14)\right|\lesssim60\,{\rm mK}$,
and the featureless spectral shape place our SRII model category as
the one that disagrees with the EDGES data most. Considering the excellent
agreement of many two-phase reionization models in the SRII category
with the PLD polarization data, rather extreme alternative explanations
to the standard model are required to explain the EDGES data in order
to accept SRII. If SRII were the right model, the excess radio background
or other alternatives should (1) almost solely contribute to the absorption
trough of $\sim500\,{\rm mK}$ because even the small ($\lesssim60\,{\rm mK}$)
absorption signal is likely to be removed in the foreground removal
process and (2) offset the possible absorption depth of $\delta T_{b}\sim-100\,{\rm mK}$
at $z\sim12$ in case X-ray heating is inefficient. Obviously, independent
observations such as the 21 cm intensity mapping by radio interferometers
will help to settle this issue.

We note that the limited amplitude and the featureless spectral shape
of the global $\delta T_{b}(z)$ in SRII models are a novel result,
and such features are in large disagreement with other studies that
also implement the LW feedback on Pop III stars inside MHs to investigate
its impact on $\delta T_{b}$ (\citealt{Mirocha2018,Mirocha2019,Mebane2020,Qin2020_tt_I,Qin2021_tt_II}).
Let us explain the major reason for such a discrepancy. A big difference
lies among this work and others in the LW feedback is implemented.
As described in Section \ref{subsec:SRII}, we assume that there exists
a threshold value of $J_{{\rm LW}}$ such that star formation inside
MHs are fully suppressed as long as $J_{{\rm LW}}>J_{{\rm LW,\,th}}$,
based on the observed abrupt change in $M_{{\rm min}}$ (the minimum
mass of star-forming halos) and $T_{{\rm vir,\,min}}$ (the minimum
virial temperature of star-forming halos) of star-forming MHs as $J_{{\rm LW}}$
varies across the value $J_{{\rm LW}}\sim0.1\times10^{-21}\,{\rm erg\,s^{-1}\,cm^{-2}\,Hz^{-1}\,sr^{-1}}$
in \citet{Yoshida2003} and \citet{OShea2008}. In contrast, other
studies usually adopt a prescription where $M_{{\rm min}}$ and consequently
$T_{{\rm vir,\,min}}$ are smooth functions of $J_{{\rm LW}}$. While
numerical coefficients vary somewhat in the literature (e.g. see difference
between \citealt{Fialkov2013} and \citealt{Schauer2020}), the commonly
used functional form for $M_{{\rm min}}$ is either a redshift-independent
one (advocated by \citealt{Machacek2001} and \citealt{Wise2007}),
\begin{equation}
M_{{\rm min}}/M_{\odot}=2.5\times10^{5}\left[1+6.8\left(4\pi J_{21}\right)^{0.47}\right]\label{eq:Mmin}
\end{equation}
where $J_{21}\equiv J_{{\rm LW}}/(10^{-21}\,{\rm erg\,s^{-1}\,cm^{-2}\,Hz^{-1}\,sr^{-1}})$,
or a redshift-dependent one (\citealt{Fialkov2013,Qin2021_tt_II,Visbal2020}),
\begin{align}
  &M_{{\rm min}}/M_{\odot} \nonumber\\
  &=2.5\times10^{5}\left(\frac{1+z}{26}\right)^{-1.5}f(v_{{\rm bc}};\,z)\left[1+6.96J_{21}^{0.47}\right]\label{eq:Mmin_z}
\end{align}
where $f(v_{{\rm bc}};\,z)\ge1$ is an additional factor accounting
for the suppression of star formation due to the baryon-dark matter
streaming velocity (\citealt{Tseliakhovich2010,Ahn2016}). If one
uses one of equations (\ref{eq:Mmin}) and (\ref{eq:Mmin_z}), the
effective $J_{{\rm LW,\,th}}$ is much larger than $J_{{\rm 21}}=0.1$
and thus star formation inside MHs will become much stronger than
our prescription at a given $J_{{\rm LW}}$. Therefore, studies adopting
equation (\ref{eq:Mmin_z}), e.g. \citet{Fialkov2013}, \citet{Qin2021_tt_II}
and \citet{Visbal2020}, find Ly$\alpha$ intensity much stronger
than our prediction, $x_{\alpha}(z\gtrsim14)\lesssim0.5$, in SRII
models. Consequently, these studies find that MH stars bring the absorption
dip occur much earlier at $z\sim16-18$ than in cases without MH stars
(\citealt{Mebane2020,Qin2020_tt_I}) and with amplitude easily reaching
$\left|\delta T_{b}\right|\gtrsim100\,{\rm mK}$. However, we stress
that if one focuses on $T_{{\rm vir,\,min}}$, \citet{OShea2008}
clearly shows $T_{{\rm vir,\,min}}\simeq8000\,{\rm K}$ when $J_{21}\simeq0.1$.
If we take this result and extrapolate to any other redshift, one
can instead conclude that $J_{{\rm LW,\,th}}$ should lie around $J_{21}=0.1$.
In this paper, we parameterize $J_{{\rm LW,\,th}}$ but to a limited
value of $J_{{\rm LW,\,th,\,21}}\le 0.1$, respecting the results of
\citet{Yoshida2003} and \citet{OShea2008}. Of course, one cannot
exclude one LW feedback scheme against another at the moment, because
there is practically no observational constraint on ultra high-$z$
($z\gtrsim14)$ radiation sources.

The success of our SRII models in producing two-stage reionization
models, which are somewhat favored by PLD against vanilla and SRI
models, can be taken as a hint that SRII models may represent the
reality including how the LW feedback operates in nature. If this
were true, the gross disagreement of the global $\delta T_{b}(z)$
of SRII models with the EDGES result would be hardly conceivable in
the standard model or the EDGES result could be an incorrect claim.
Further study is warranted.

\section{Summary and Conclusion\label{sec:Discussion}}

We studied three types of reionization models semi-analytically: the
vanilla, SRI and SRII models. SRI and SRII models implement the negative
feedback effects on star formation inside LMACHS (SRI, SRII) and MHs
(SRII only), namely the Jeans-mass filtering due to photoionization
(LMACH, MH) and the LW feedback (MH). As long as LMACHs and MHs host
Pop III stars and dominate the era of $z\gtrsim14$, we find that
$\delta T_{b}$ is constrained to $-50\,{\rm mK}\lesssim\delta T_{b}<0$
in both SRI and SRII models that is in stark contrast with the $\sim500\,{\rm mK}$
absorption trough claimed by the EDGES. $\delta T_{b}$'s predicted
by SRII models are almost featureless in its spectral shape for $z\gtrsim14$
($\nu\lesssim95\,{\rm MHz}$) due to the strong self-regulation of
star formation inside MHs by the LW background built up my MHs.

At this stage with the PLD being the most accurate large-scale-CMB
anisotropy observation, we find that all three models can provide
acceptable reionization scenarios. Especially, we find that SRII models
with substantial high-redshift ($z\gtrsim15$) ionization tails are
as favored as those models without such tails, if corresponding $C_{l}^{{\rm EE}}$
is analyzed against the $C_{l}^{{\rm EE}}$ of the PLD. SRII models
have the two-stage ionization feature, the high-redshift slow-ionization
stage and the low-redshift fast-ionization stage, which seems more
favored by PLD than tail-less reionization models even though this
tendency is largely uncertain in PLD. In conclusion, SRII models with
substantial high-$z$ ionization tails should NOT be ruled out as
claimed by \citet{PlanckCollaboration2018}. Our $C_{l}^{{\rm EE}}$-only
analysis favors models with $\tau_{{\rm es}}\simeq0.055-0.064$, which
is relatively larger than the $C_{l}^{{\rm EE}}$-only inference by
\citet{PlanckCollaboration2018}, $\tau_{{\rm es}}=0.0504_{-0.0079}^{+0.0050}$.
This issue needs to be investigated further.

In light of the peculiarity of $\delta T_{b}$ but the good agreement
with PLD-$C_{l}^{{\rm EE}}$ and the hint of two-stage reionization,
we stress that SRII models should be considered more seriously. Such
a disagreement with the EDGES data requires a change in the standard
$\Lambda$CDM model or the ``interpretation'' of the EDGES data.
Even though many statistical analyses have been already carried out
by accepting the full result (e.g. \citealt{Mebane2020}) or
a part of the result (\citealt{Qin2020_tt_I}: only the redshift window
of the absorption trough is taken as the possible location for the
absorption dip) of EDGES, we question the foreground-removal scheme
of the EDGES team as \citet{Hills2018} and \citealt{Tauscher2020}
did. The claimed signal is indeed a result of extracting an arbitrary
smooth signal from the residual signal (Fig 1b in \citealt{Bowman2018})
after the foreground removal. There is no guarantee that a combination
of log-power-law spectral curves could completely remove the foreground,
and such an additional arbitrary removal seems even more dubious.
It would be even harder to probe the EoR signal if nature were indeed
described by our SRII models, because the featureless spectral shape
at $z\gtrsim14$ of the models is going to be removed by any foreground-removal
scheme utilizing the spectral smoothness. In this regard, probing
the dipole anisotropy of $\delta T_{b}$ seems very promising (\citealt{Deshpande2018,Trombetti2020}),
because the dipole moment is caused by the monopole moment seen by
an observer with a peculiar motion and thus can be an independent probe
of the global $\delta T_{b}$.

It is therefore crucial to carry out higher-precision CMB observation
and the 21-cm intensity mapping to further constrain the reionization
history. A superb CMB polarization apparatus limited only by the cosmic
variance (for all $2\le l\lesssim200$), LiteBIRD (Lite (Light) satellite
for the studies of B-mode polarization and Inflation from cosmic background
Radiation Detection), is under way, which will significantly sharpen
the constraint on the reionization history. Many different reionization
histories can share the same $\tau_{{\rm es}}$ and $z_{{\rm ov}}$
(see e.g. Fig. 4b in \citealt{Ahn2012}), and breaking this degeneracy
seems elusive with the PLD at this low-$\tau_{{\rm es}}$ era but
will become more feasible with such a high-precision CMB observation.
The boost of $C_{8\lesssim l\lesssim24}^{{\rm EE}}$ of SRII models
relative to that of the vanilla and SRII models is of particular interest.
The 21-cm intensity mapping will be able to probe large-scale fluctuation
of HI density when large H II regions are produced during EoR. The
high-redshift regime with $z\gtrsim15$ is more in the Dark Ages without
too much ionization in typical reionization scenarios, while some
SRII models produce a substantial amount of ionization. Nevertheless,
SRII models predict only small-size H II regions by MHs (\citealt{Ahn2012})
that might not be imaged individually by radio interferometers. If
future CMB observations favored two-stage reionization but radio interferometry
did not reveal any noticeable H II bubbles at high redshift, SRII
models would be the strongest candidate to explain both. We will investigate
detailed observational prospects in the future.

\acknowledgements{This work was supported by Korea NRF grant NRF-2016R1D1A1B04935414
and a research grant from Chosun University (2016).}


\end{document}